\documentclass[]{aa}
\usepackage{graphicx}
\usepackage[varg]{txfonts}
\usepackage{url}
\newcommand{\kms}{km\,s$^{-1}$}
\usepackage{CJK}
\DeclareRobustCommand{\cjkKTW}{\begin{CJK*}{Bg5}{bkai}\CJKchar[Bg5]{"B6}{"C0}\CJKchar[Bg5]{"B9}{"C5}\CJKchar[Bg5]{"B9}{"46}\end{CJK*}}
\usepackage{multirow}
\usepackage[dvipsnames]{xcolor}
\usepackage[normalem]{ulem}
\begin{document} 
\title{Molecular remnant of Nova 1670 (CK Vulpeculae)}
\subtitle{I. Properties of the gas and its enigmatic origin}
   \author{Tomek Kami\'nski \inst{1,2}
           \and
          Karl M.\ Menten\inst{3}
          \and
          Romuald Tylenda\inst{1}
          \and
          Ka Tat Wong ({\cjkKTW})\inst{3,4}
          \and 
          Arnaud Belloche\inst{3}
           \and
          Andrea Mehner\inst{5}
          \and 
          Mirek R.\ Schmidt\inst{1}
          \and 
          Nimesh A. Patel\inst{2}
          }

\institute{\centering
Nicolaus Copernicus Astronomical Center, Polish Academy of Sciences, Rabia{\'n}ska 8, 87-100 Toru{\'n}, \email{tomkam@ncac.torun.pl}\label{inst1}
\and
Center for Astrophysics $|$ Harvard \& Smithsonian , 60 Garden Street, Cambridge, MA 02138 \label{inst2}
\and
Max Planck Institut f\"ur Radioastronomie, Auf dem H\"ugel 69, D-53121 Bonn, Germany \label{inst3}
\and
Institut de Radioastronomie Millim{\'e}trique, 300 rue de la Piscine, 38406 Saint-Martin-d'H{\`e}res, France\label{inst4}
\and
ESO -- European Organisation for Astronomical Research in the Southern Hemisphere, Alonso de Cordoba 3107, Vitacura, Santiago, Chile\label{inst5}
}
\authorrunning{T. Kami\'nski et al.}
 
\abstract{CK\,Vul erupted in 1670 and is considered a Galactic stellar-merger candidate. Its remnant, observed 350\,yr after the eruption, contains a molecular component of surprisingly rich composition, including polyatomic molecules as complex as methylamine (CH$_3$NH$_2$). 
We present interferometric line surveys with subarcsec resolution with ALMA and SMA. The observations provide interferometric maps of molecular line emission at frequencies between 88 and 243\,GHz 
that allow imaging spectroscopy of more than 180 transitions of 26 species. We present, classify, and analyze the different morphologies of the emission regions displayed by the molecules. We also perform a non-LTE radiative-transfer analysis of emission of most of the observed species, deriving the kinetic temperatures and column densities in five parts of the molecular nebula. 
Non-LTE effects are clearly seen in complex species including methanol absorption against the cosmic microwave background. 
The temperatures are about 17\,K in the inner remnant and 14\,K in the extended lobes, both higher than excitation temperatures estimated earlier in an LTE approach and based on single-dish spectra. We find total (hydrogen plus helium) densities in the range of $10^4-10^6$\,cm$^{-3}$. The column densities provide rough relative abundance patterns in the remnant which currently are not understood. Attempts to derive elemental abundances within the assumption of a chemical equilibrium give only loose constraints on the CNO elements. 
That the formation of many of the observed molecules requires a major involvement of circumstellar shocks remains the preferred possibility. The molecular gas could have formed 350\,yr ago or more recently. The molecules are well shielded from the interstellar radiation field by the circumstellar dust. Their presence alone indicates that the unobservable central star cannot be a hot object such as a white dwarf. This excludes some of the proposed scenarios on the nature of CK\,Vul. The general characteristics of the molecular environment of CK\,Vul derived in this study 
resemble quite well those of some pre-planetary nebulae and asymptotic giant branch stars, most notably that of OH231.8+4.2.}

\keywords{Astrochemistry -- Shock waves -- stars: individual: CK Vulpeculae -- circumstellar matter --
                techniques: interferometric}
\maketitle

\section{Introduction}\label{intro}
The year 2020 marks the 350\,yr anniversary of the discovery of the eruption of Nova 1670 (or CK\,Vul) made by European astronomers \citep{shara85}. Their observations, predominantly performed with a naked eye, traced the object's evolution on the sky in 1670--1672. From the archive records, we know that the eruption was rather unusual, in particular it was very much unlike classical novae. The light curve of CK\,Vul displayed three peaks and the star was described as reddish in the later stages of the eruption \citep{Hevelius,shara85}. These characteristics resemble closest the behavior often observed in (luminous) red novae \citep{kato,blg360}, a modern category of eruptive stars known from our and other galaxies \citep[e.g.][]{pastorello}. Red novae are recognized as manifestations of on-going mergers of non-compact stars such as main-sequence dwarfs, sub-giants, or red giants \citep{ST03,TS06,tylenda2011,pastorello}. While the number of known red novae, mainly extragalactic ones, is quickly rising \citep[e.g.,][]{stritzinger}, we know only a few red-nova remnants that are decades old \citep{submmRN}. The remnant of the 1670 eruption of CK\,Vul, as a candidate post-merger site, could be the oldest (counting from the onset of the eruption) known object of this type and as such offers the opportunity to investigate a merger aftermath centuries after the stellar coalescence. The nature of the progenitor system of CK\,Vul has been debated. \cite{Eyres} proposed that the seventeenth-century merger took place between a white dwarf and a brown dwarf, but there is little quantitative evidence to support this. Based on the analysis of the source's chemical and isotopic composition, including the unique presence of the radioactive isotope of $^{26}$Al, \cite{kamiAlF} found that the progenitor system of CK\,Vul included at least one red-giant-branch (RGB) star with a fully developed helium core.

The remnant of CK\,Vul was recovered for modern astrophysics in the 1980s \citep{shara82,shara85} and was long considered to consist only of a very weak optical nebula with a large (71\arcsec) hourglass component and several brighter knots located near the center of the large bipolar structure \citep{hajduk2007,hajduk2013}. The stellar component has never been observed directly, but \cite{hajduk2007} found a radio source near the center of the hour-glass structure, which, as we know today, coincides with the location of the only energy source of the remnant. First studies of the atomic emission suggested an anomalous gas composition, which was quantified only recently by \cite{xshooter} based on deep Very Large Telescope spectra of the brightest knot ("the northern jet"). Helium is found to be twice as abundant and the nitrogen-to-oxygen ratio ten times higher than in the Sun. These findings are consistent with nucleosynthesis products of the RGB progenitor. The plasma within the optical knot has an electron temperature of 10--15\,kK and, based on multiple line ratios, is most likely excited by radiative shocks which may be younger than the remnant itself. Fast ($>100$\,\kms) and active shocks are also required to explain the longevity of emission of some of the observed ions (e.g., [\ion{Ne}{iii}] and [\ion{O}{iii}]), which otherwise would recombine on time scales shorter than the age of the remnant \citep{xshooter}. Direct 
evidence for shocks has so far only been available for the central part of the optical nebula, within about 10\arcsec.

The discovery that the CK\,Vul remnant contains a remarkably rich molecular component with peculiar composition came as a surprise. Molecular features were identified at millimeter (mm) and submillimeter (submm) wavelengths and were found to originate from a region overlapping with the central shocked parts of the optical nebula \citep{nature}. The molecular emission arises from a bipolar structure of a size of 13\arcsec\ and is centered on the low-frequency radio source. Emission of molecules, including lines of H$_2$, was later found in infrared spectra \citep{evans2016,xshooter}. A comprehensive investigation of the molecular component of the remnant was presented by \cite{kami-singledish} (hereafter KMT17) based on single-dish spectra obtained with the Atacama Pathfinder Experiment (APEX) and the IRAM 30\,m telescopes and with a dense coverage of the 70--500\,GHz atmospheric windows. The molecular emission lines are generally weak and typically display broad
profiles, with full-widths of up to 300\,\kms, and full widths (to zero power) up to a remarkable 450\,\kms, which is perhaps the reason why such emission had not been detected earlier at mm wavelengths. The molecular inventory includes now over 28 different species ranging from simple di-atomics to organic molecules consisting of up to seven atoms  (CH$_3$NH$_2$). The list of molecules combines species typically seen only in oxygen-rich stars (e.g., SO$_2$) as well as those only seen in carbon stars (e.g., HC$_3$N). This is related to the anomalous ratios of the CNO elements which we assign to the nucleosynthesis in the progenitor stars. Complex species such as methanol (CH$_3$OH) are very rarely observed in evolved stars and with several similarly complex molecules, CK\,Vul definitely stands out among known stellar mm-wave sources. Lacking a better interpretation, the more complex species are typically interpreted as signatures of shocks in the circumstellar environment \citep[e.g.,][]{olofsson,oh231}. The mm and submm spectra of CK\,Vul are particularly rich in molecular features because most of the molecules show strong emission in their different isotopic forms. The presence of these features allows a determination of the isotopic composition of the remnant, linking it to the nucleosynthesis history of the matter that is now dispersed in the remnant. Another surprising aspect of the molecular component is its low temperature, roughly constrained to be $\lesssim$12\,K by KMT17, while the eruption 350\,yr ago involved gas temperatures of the order of 10\,kK. The cool state is most straightforwardly interpreted as a result of effective cooling during an adiabatic expansion that followed after the eruption. 

Here, we discuss new interferometric observations that allow us to better study the molecular component of the ancient remnant.  Being first in a series of papers, this study presents the survey and focuses on the molecular composition in different parts of the nebula. The observations and their analysis are presented in Sect.\,\ref{sec-obs} and a summary of the observed molecular inventory is given in  Sect.\,\ref{sec-summary}. The morphology of the distribution of different molecules is discussed in Sect.\,\ref{sec-morph}, while Sect.\,\ref{sec-rad-transf} describes our radiative transfer modeling of the source's different components. Sect.\,\ref{sec-CE} presents a general view that has emerged on the nature of the chemical processes at work in CK\,Vul and attempt to determine elemental abundances in the molecular gas. Sects.\,\ref{sec-shocks} and \ref{sec-surv}, respectively, discuss the importance of shocks and the survival of molecules, while, in Sect.\,\ref{sec-final} we present our summary.  

\section{Observations}\label{sec-obs}
We report interferometric observations from a spectral line survey obtained mainly in 2017 and 2018 with the Atacama Large Millimeter/submillimeter Array (ALMA). The ALMA data are supplemented by observations with the Submillimeter Array (SMA) and NSF's Karl G. Jansky Very Large Array (VLA) at lower sensitivities but at comparable angular resolutions. These interferometric data were combined to have the most comprehensive view on the molecular emission in CK\,Vul. The spectral and spatial coverage was optimized for the sake of an excitation analysis of different parts of the molecular nebula. Continuum observations obtained with these data sets will be reported and analyzed in a separate study (in prep.) We also briefly present here our near-infrared observations of H$_2$ emission of CK\,Vul which are important in the context of shocks.  

\subsection{ALMA survey}
The survey was intended to obtain interferometric maps of molecular emission at a sub-arcsec resolution over ALMA bands 3, 4, and 5, in which molecular emission lines are brightest and most numerous for CK\,Vul.  Archival data extend the  spectral coverage up to band 6. Individual observing runs are characterized in Table\,\ref{tab-alma-setups}. Each frequency setup was executed in two different array configurations of 12-m antennas to cover a wide range of spatial frequencies and in particular to minimize the interferometric filtering effect of the emission with the largest angular scales (missing ``zero spacings''). For instance, the observations at the lowest frequencies were obtained with projected baselines 13--3200\,m. The shortest projected baselines are the consequence of a low elevation of CK\,Vul at the ALMA site. The nominal angular resolution and largest angular scales are listed in Table\,\ref{tab-alma-setups}, following the ALMA technical handbook.\footnote{\url{https://almascience.nrao.edu/documents-and-tools/latest/alma-technical-handbook}} The actual values depend on choices made in the imaging procedure.  Roughly, the survey allows us to study structures larger than about 0\farcs3 and should recover emission smaller than about 15\arcsec. This latter limit is larger than the typical size of emission regions except for those of some transitions of CO. Because we wished to fully cover bands 3 and 4, some of the 25 setups overlap in frequency. Our native spectral resolution of a few \kms\ is sufficiently small to resolve the narrowest lines found in CK\,Vul. The field center for the ALMA and SMA surveys was set to ICRS coordinates RA=19:47:39.07 and Dec=27:28:45.16.  Spatial offsets here are given with respect to that position.

We supplemented the dedicated survey data with all earlier ALMA observations of CK\,Vul which extend the observational material to band 6. These observations were described in detail in \citet{kamiAlF} and \citet{Eyres}. The three extra setups are included at the bottom of Table\,\ref{tab-alma-setups}. Note that the observations from April 2017 (PI: Evans) have a much poorer spectral resolution than the other data. The archival data are also limited to a single array configuration and are more prone to large-scale filtering effects.

The visibility data were calibrated with the default ALMA data reduction pipeline in the Common Astronomy Software Applications (CASA) package \citep{casa}.\footnote{\url{https://casa.nrao.edu/}} Most data were further self-calibrated using the strong continuum emission of CK\,Vul.
The continuum contribution was subtracted in the calibrated visibility ($uv$) domain. Visibility data overlapping in frequency were combined. The continuum-subtracted data were used to create data cubes employing the {\tt clean} and {\tt tclean} implementations of the CLEAN algorithm in CASA versions 5.4--5.6. 
Different weighting schemes of visibilities were used in the imaging process. For most of the data cubes discussed in this paper, we used the optimal Briggs weighting with the robust parameter of 0.5. Most cubes were imaged at a channel spacing of 
a few MHz, resulting in channel spacings (mostly) between 3 and 5\,\kms, see Table \ref{tab-alma-setups}.  

Owing to the frequency overlaps, different numbers of antennas and weather conditions on different dates, variations in sky transparency with wavelength (especially in band 5), and other technical issues -- the line sensitivities in the calibrated data are not uniform, especially when different bands are compared. Roughly, however, the typical rms noise values are 0.04\,Jy/beam in band 3 and 0.01\,Jy/beam in bands 4 and 5 (at 5\,MHz binning). See also Fig.\,\ref{fig-sma-alma} for a visual comparison of data quality with wavelength. This sensitivity was required to map the weakest emission from complex molecules reported in KMT17.

\begin{sidewaystable*}\footnotesize
\caption{Observing runs of the ALMA survey towards CK\,Vul.}\label{tab-alma-setups}
\centering
\begin{tabular}{ccc cccc ccc cccc} 
\hline\hline
            &Integr.&    & &&&                                          &\multicolumn{2}{c}{Resolution}&\multicolumn{2}{c}{Sensitivity}&     &  Field   & Largest\\
Observation &time   &    & &&&                                          &spectral & angular            &\multicolumn{2}{c}{(mJy/beam)} & PWV &	of view & angular\\
date        &(s)    &Band&\multicolumn{4}{c}{Frequency coverage (GHz)} &(\kms) &(\arcsec)             &line$_{\rm 10km/s}$& continuum    & mm  & (\arcsec)& scale (\arcsec) \\
\hline\hline
03-Dec-17    &    1391  &  3	&   88.32--90.19   & 90.19--92.07   & 100.32--102.19 & 102.19--104.07 & 5.6  &	 0.24  &  0.67  &  0.015 &  2.3  &   60.5  &  4.1   \\  
05-Dec-17    &    1361  &  3	&   92.07--93.94   & 93.94--95.82   & 104.07--105.94 & 105.94--107.82 & 5.4  &	 0.25  &  0.65  &  0.015 &  2.5  &   58.3  &  4.0   \\  
07-Dec-17    &    1361  &  3	&   99.57--101.44  & 101.44--103.32 & 111.57--113.44 & 113.55--115.42 & 5.1  &	 0.24  &  0.67  &  0.017 &  3.0  &   54.2  &  3.9   \\  
07-Dec-17    &    1391  &  3	&   84.57--86.44   & 86.44--88.31   & 96.57--98.44   & 98.44--100.32  & 5.8  &	 0.28  &  0.78  &  0.016 &  3.1  &   63.0  &  4.4   \\  
10-Dec-17   &    1331  &  3	&   95.82--97.69   & 97.69--99.57   & 107.82--109.69 & 109.69--111.57 & 5.2  &	 0.30  &  0.55  &  0.013 &  0.9  &   56.2  &  5.1   \\  
14-Dec-17   &    1210  &  4	&   130.22--132.09 & 132.09--133.97 & 142.22--144.09 & 144.09--145.97 & 4.0  &	 0.25  &  0.81  &  0.021 &  3.5  &   42.2  &  4.7   \\  
14-Dec-17   &    1149  &  4	&   126.47--128.34 & 128.34--130.22 & 138.47--140.34 & 140.34--142.22 & 4.1  &	 0.27  &  0.84  &  0.021 &  3.3  &   43.3  &  5.2   \\  
31-Dec-17   &    968   &   4	&   146.27--148.15 & 148.17--150.04 & 158.22--160.09 & 160.17--162.04 & 3.6  &	 0.23  &  0.78  &  0.021 &  1.7  &   37.8  &  4.3   \\  
31-Dec-17   &    1119  &  4	&   137.72--139.59 & 139.59--141.47 & 149.72--151.59 & 151.59--153.47 & 3.8  &	 0.24  &  0.63  &  0.017 &  1.8  &   40.0  &  4.4   \\  
31-Dec-17   &    1179  &  4	&   133.97--135.84 & 135.84--137.72 & 145.97--147.84 & 147.84--149.72 & 3.9  &	 0.26  &  0.65  &  0.017 &  1.7  &   41.1  &  5.0   \\  
04-Jan-18    &    1210  &  4	&   141.47--143.34 & 143.34--145.22 & 153.47--155.34 & 155.34--157.22 & 3.7  &	 0.27  &  0.77  &  0.021 &  2.7  &   39.0  &  5.2   \\  
21-Mar-18   &    544   &   3	&   99.57--101.44  & 101.44--103.32 & 111.57--113.44 & 113.55--115.42 & 5.1  &	 1.07  &  1.35  &  0.033 &  2.3  &   54.2  &  13.6  \\  
21-Mar-18   &    544   &   3	&   95.82--97.69   & 97.69--99.57   & 107.82--109.69 & 109.69--111.57 & 5.2  &	 1.14  &  1.05  &  0.024 &  2.1  &   56.2  &  14.4  \\  
04-Apr-18    &    5262  &  5	&   185.67--187.54 & 187.56--189.43 & 197.67--199.54 & 199.56--201.43 & 11.6 &	 0.82  &  0.48  &  0.016 &  0.3  &   30.1  &  10.2  \\  
09-Apr-18    &    484   &   4	&   130.22--132.09 & 132.09--133.97 & 142.22--144.09 & 144.09--145.97 & 4.0  &	 1.25  &  1.26  &  0.032 &  3.6  &   42.2  &  14.2  \\  
09-Apr-18    &    726   &   4	&   126.47--128.34 & 128.34--130.22 & 138.47--140.34 & 140.34--142.22 & 4.1  &	 1.29  &  1.09  &  0.027 &  3.6  &   43.3  &  14.7  \\  
23-Apr-18   &    1089  &  3	&   92.07--93.94   & 93.94--95.82   & 104.07--105.94 & 105.94--107.82 & 5.4  &	 1.23  &  1.00  &  0.022 &  2.6  &   58.3  &  21.0  \\  
13-May-18   &    635   &   5	&   164.57--166.44 & 166.44--168.32 & 176.77--178.64 & 178.44--180.32 & 3.2  &	 1.63  &  0.68  &  0.022 &  0.4  &   33.8  &  15.1  \\  
25-May-18   &    605   &   4	&   141.47--143.34 & 143.34--145.22 & 153.47--155.34 & 155.34--157.22 & 3.7  &	 1.90  &  0.68  &  0.019 &  0.5  &   39.0  &  19.1  \\  
25-May-18   &    575   &   4	&   137.72--139.59 & 139.59--141.47 & 149.72--151.59 & 151.59--153.47 & 3.8  &	 2.00  &  0.70  &  0.019 &  0.5  &   40.0  &  20.1  \\  
25-May-18   &    605   &   4	&   133.97--135.84 & 135.84--137.72 & 145.97--147.84 & 147.84--149.72 & 3.9  &	 2.13  &  0.74  &  0.019 &  0.5  &   41.1  &  21.5  \\  
26-May-18   &    484   &   4	&   146.27--148.15 & 148.17--150.04 & 158.22--160.09 & 160.17--162.04 & 3.6  &	 1.90  &  0.95  &  0.025 &  1.3  &   37.8  &  19.0  \\  
27-Aug-18   &    575   &   3	&   88.32--90.19   & 90.19--92.07   & 100.32--102.19 & 102.19--104.07 & 5.6  &	 1.40  &  0.94  &  0.020 &  0.7  &   60.5  &  18.6  \\  
27-Aug-18   &    575   &   3	&   84.57--86.44   & 86.44--88.31   & 96.57--98.44   & 98.44--100.32  & 5.8  &	 1.47  &  0.98  &  0.020 &  0.7  &   63.0  &  19.5  \\  
19-Sep-18   &    1240  &  5	&   164.57--166.44 & 166.44--168.32 & 176.77--178.64 & 178.44--180.32 & 3.2  &	 0.38  &  0.60  &  0.021 &  0.9  &   33.8  &  6.0   \\[5pt]  
\multicolumn{14}{l}{archival:}\\
02-Jun-16    &    2419  &  6	&   230.57--232.44 & 233.51--235.38 & 246.07--247.94 & 248.12--249.99 & 9.4  &   0.47  &  0.76  &  0.026 &  1.4  &   24.2  &  5.6   \\  
13-Oct-16   &    1391  &  6	&   222.99--224.98 & 224.99--226.98 & 238.99--240.98 & 240.99--242.98 & 38.6 &	 0.13  &  0.52  &  0.017 &  0.5  &   25.0  &  1.9   \\  
24-Apr-17   &    605   &   6	&   222.98--224.96 & 224.98--226.96 & 238.98--240.96 & 240.98--242.96 & 38.6 &	 0.84  &  0.88  &  0.029 &  0.7  &   25.0  &  7.7   \\  
\hline
\end{tabular}
\tablefoot{Angular resolutions and sensitivities are from the ALMA archive and are only approximate. Line sensitivity is given for a 10\,\kms\ bin. Weather conditions are characterized by the precipitable water vapor (PWV).}
\end{sidewaystable*}

\subsection{SMA survey}
A dedicated and complementary extension of the ALMA survey towards higher frequencies was done in 2018 with the SMA and the SWARM correlator \citep{swarm}. The observing runs are listed and characterized in Table\,\ref{tab-sma-setups}. Two frequency setups were chosen to cover 199.3--231.5 and 241.0--273.0\,GHz in band 6. Each setup was observed in compact and extended SMA configurations with projected baselines of 17--227\,m. SMA also observed CK\,Vul in one tuning in band 7, covering 328.8--361.2\,GHz. To fill the {\it uv} plane, the observations were typically obtained over a few hours in each run. The SMA field of view for band 6 and 7 observations is about 52\arcsec\ and 35\arcsec, respectively. The data were collected at a spectral resolution of 140\,kHz but were degraded to a channel binning of 1.12\,MHz before calibration for a computational efficiency.

\begin{table*}\footnotesize
\caption{Observing runs of the SMA survey towards CK\,Vul.}\label{tab-sma-setups}
\centering
\begin{tabular}{ccc cccc} 
\hline\hline
            & Frequency & Number   &               &                 &              & Continuum \\    
            & range &    of        & SMA           &                 &              & sensitivity \\
Date        &  (GHz)    & antennas & configuration & Beam, PA (\degr) & $\tau_{225}$ & (mJy/beam)\\
\hline\hline
20-May-18 & 329.0--361.0                   & 8                    & extended       & 1\farcs06$\times$0\farcs61, 79                 &  0.06--0.11       & 1.9 \\
13-Jun-18 & 329.0--361.0                   & 7                    & compact        & 2\farcs31$\times$1\farcs77, 9                   &  0.06--0.10       & 2.2 \\
12-Aug-18 &  328.8--360.8                   & 6--7                 & subcompact   & 3\farcs35$\times$1\farcs81, 37                 &  0.10--0.20       & 8.0\\
02-Oct-18 &  328.8--360.8                   & 7                     & subcompact   & 3\farcs04$\times$1\farcs73, 25                 &  0.04--0.11       & 3.0\\
\multirow{2}{*}{13-Apr-18} & 199.3--207.3,215.3--223.3, & \multirow{2}{*}{7}                     & \multirow{2}{*}{extended}      & \multirow{2}{*}{1\farcs24$\times$0\farcs88, 70}& \multirow{2}{*}{0.15--0.30} & \multirow{2}{*}{1.3} \\ 
            & 241.0--249.0257.0--265.0                  &                       &               &                                                 & \\
27-Apr-18 &  (as above)                      & 7                     & extended      & 1\farcs07$\times$0\farcs89, 68                  & 0.05--0.14        & 0.6 \\
26-May-18 & (as above)                      & 7                     & compact       & 2\farcs63$\times$2\farcs39, 59                  & 0.16--0.18        & 1.5 \\
\multirow{2}{*}{18-Apr-18} &   207.3--215.3,223.3--231.3, & \multirow{2}{*}{7} & \multirow{2}{*}{extended}     & \multirow{2}{*}{1\farcs08$\times$0\farcs85, 86}&\multirow{2}{*}{0.17--0.24} & \multirow{2}{*}{0.6} \\
                          &249.0--257.0,265.0--273.0 &&&&&\\ 
25-May-18 &  (as above)                     & 6                     & compact       & 3\farcs17$\times$2\farcs63, 5                    & 0.16--0.18        & 1.6 \\[5pt]
\multicolumn{7}{l}{archival:}\\
\multirow{2}{*}{30-Jul-14}  & 216.9--218.8,219.9--220.8, & \multirow{2}{*}{7} & \multirow{2}{*}{subcompact} &\multirow{2}{*}{7\farcs6$\times$4\farcs1, 69}& \multirow{2}{*}{0.07--1.10} & \multirow{2}{*}{1.9} \\
                            & 228.9--230.8,230.9--232.8 &                    &                             &                                             &                             & \\
\multirow{2}{*}{03-Jul-14} & 330.2--332.2,335.2--337.2,  & \multirow{2}{*}{8} & \multirow{2}{*}{compact}  &  \multirow{2}{*}{2\farcs0$\times$1\farcs3, 270} &\multirow{2}{*}{0.06--0.10}  &\multirow{2}{*}{0.8} \\ 
                           & 345.2--347.2,350.2--352.2 &                     &                           &                                                 &                   &                  \\                   
\hline
\end{tabular}
\tablefoot{The weather conditions are characterized by the atmosphere opacity, $\tau_{225}$, measured at 225\,GHz.}
\end{table*}

The SMA data were calibrated using standard procedures in the MIR package.\footnote{\url{https://www.cfa.harvard.edu/~cqi/mircook.html}}  The calibrated visibilities were exported to CASA. Self-calibration in phase was applied to data for which the continuum signal-to-noise was sufficiently high. Visibilities from different configurations were combined. The continuum was subtracted from the calibrated visibilities. Line data were imaged in CASA at a reduced channel spacing of 23\,MHz and with Briggs weighting at the robust parameter of 0.5. 

The complex gains were calibrated by observations every 10--15\,min of two calibrators, the quasar 2015+371 and the star MWC\,349A. A band-pass calibration was performed by observations of the quasar 3C\,279 and Solar system objects. Careful calibration to absolute flux was obtained by observations of Callisto, Neptune, or Titan. The same flux calibration was applied to gain and band-pass calibrators and compared to their independent flux measurements in monitoring programs at ALMA and SMA. The flux calibration is consistent to within 20\%.

The 2018 SMA data in band 6 overlap partially with SMA observations of CK\,Vul made in 2014 and described in detail in \citet{nature}. These older data were acquired in the subcompact (i.e., most compact) configuration which provides more short baselines and thus gives a better access to larger spatial scales than the rest of the 2018 SMA data in bands 6 and 7. To investigate how the missing short spacings influence our observations, we combined and imaged all band 6 data that  include sub-compact baselines. Examination of the distribution of the CO 2--1 emission in maps obtained with and without the subcompact data show that emission associated with CK\,Vul is somewhat affected by the lack of short baselines but for other species, including $^{13}$CO 2--1, the effect is negligible. We also found that the surrounding narrow emission of $^{12}$CO and $^{13}$CO, originating from diffuse interstellar clouds (see KMT17), is recovered from visibilities at the subcompact configuration. Especially a component near the local-standard-of-rest (LSR) velocity of 20\,\kms\ is readily visible north-east of CK\,Vul and partially overlaps with the molecular remnant of CK\,Vul.

\begin{figure*}[!ht]
  \includegraphics[trim=0 0 0 50, width=.99\textwidth]{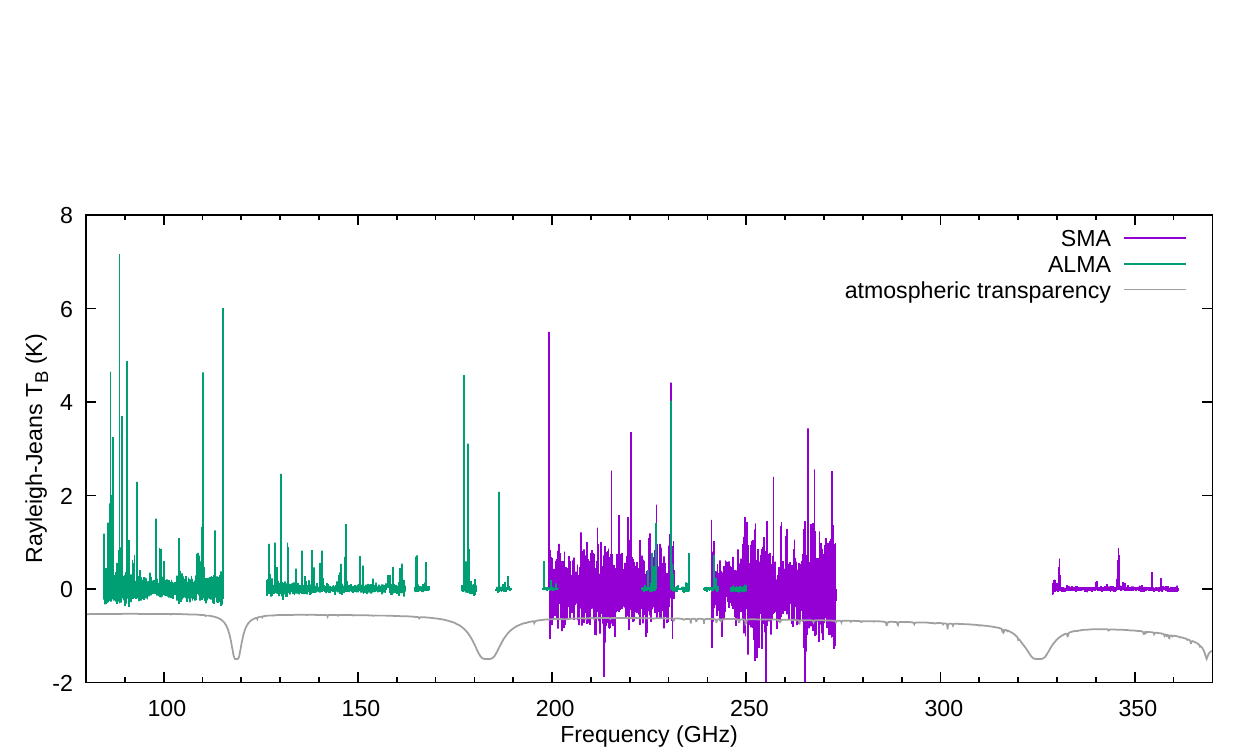}
\caption{The coverage of the ALMA and SMA surveys. Spectra were extracted for the central beam of each data-cube. Atmospheric transparency is shown for reference.}\label{fig-sma-alma}
\end{figure*}

The coverage and relative quality of the ALMA and SMA data are illustrated in Fig.\,\ref{fig-sma-alma}. In the spectral regions where SMA and ALMA data overlap, we used the more sensitive ALMA data for the analysis that follows.

\subsection{VLA observations}
We observed CK\,Vul on 18 and 22 Jan 2016 with the VLA in the hybrid array configuration DnC (PI: K. Menten). We covered three frequency ranges between 18 and 49\,GHz in the $K$, $K_a$, and $Q$ radio bands.  Data were flagged, calibrated, and imaged using the standard procedures in CASA. The beam sizes of the observations are in the range from about 1\farcs6$\times$1\farcs1 to about 3\arcsec$\times$1\farcs6. Part of the VLA observations were reported in \citet{kamiAlF} and their full description will be given in an associated paper on continuum emission. In the current paper, we only present a map of NH$_3$ emission (for completeness, in Sect.\,\ref{sec-morph}) and used the $Q$-band spectra to achieve a better convergence in modeling the excitation of SiO (Sect.\,\ref{sec-rad-transf}). 

\subsection{H$_2$ emission in X-shooter near-infrared spectra}\label{sec:xshooter}
In the following analysis and discussions, we often refer to the near-infrared spectrum of H$_2$ obtained with the X-shooter spectrograph at the Very Large Telescope. The observations are described in detail in \cite{xshooter}, where visual parts of the spectrum ($\lesssim$1\,$\mu$m) were fully analyzed. These very deep long-slit (11\arcsec)  observations probe the brightest optical clump of the nebula, called here "the northern jet". 

The near-infrared X-shooter spectrum  covering the spectral range 1--2.48\,$\mu$m has not been analyzed. The applied observation strategy was optimized for visual features and in consequence the near-infrared part of the spectrum is heavily affected by sky emission and telluric lines. 
Nevertheless, we were able to identify and measure fluxes of about 15 $S$ and $Q$ transitions from the vibrational bands $\varv$=1--0, 2--0, and 2--1. A few strongest lines readily show that the emission is extended. 

The line fluxes, corrected for the reddening $E_{B-V}$ with a standard extinction law, were used to construct a rotational diagram for both ortho and para transitions. The measurements are consistent with an ortho-to-para H$_2$ ratio of 3. By performing linear fits to the measurements, we find that the vibrational excitation temperature of 2613$\pm$173\,K (1$\sigma$ error) is within errors the same as the rotational temperature of 2781$\pm$212\,K. Those estimates assumed optically thin emission produced under local-thermodynamic-equilibrium (LTE) conditions. Both assumptions are consistent with the observed emission as the linear fits reproduce the observations well. The rotational temperature was derived from $\varv$=1--0 transitions, while the fit to all measured lines is interpreted as the vibrational temperature.

\section{Survey summary}\label{sec-summary}
In a single report, it is impossible to fully describe the rich and peculiar molecular composition and morphology of the envelope of CK\,Vul imaged by ALMA and give justice to all the intriguing details. Here, we attempt to give a general overview. About 180 spectral features, as counted in spectra extracted near the central source, were attributed to molecular transitions following the identification presented in KMT17. There, a detailed identification was based on single-dish spectra in a broader spectral coverage and with only slightly lower signal-to-noise ratio. The identifications were validated in all common spectral regions and  only minor revisions had to be made for the weakest features reported in KMT17. In the newly explored frequency ranges (e.g., 178.5--180.3\,GHz in ALMA band 5) that were not covered by previous single-dish spectra, we used the CASSIS toolbox\footnote{\url{http://cassis.irap.omp.eu}} \citep{Vastel2015} and LTE models to identify lines. We do not present the detailed identification in this paper because it is almost unchanged. Also, in the spatially-resolved images of the nebula, the presence of spectral lines and their line-of-sight velocities depend on the position, making a presentation of line identification more ambiguous than in single-dish spectra \citep[cf.][]{kamiSurv}. We refer the reader to Appendices B and C of KMT17 for source-integrated (single dish) spectra and tabulated lists of transitions. The inventory of molecules detected by ALMA and SMA includes: AlF, CN, CO, CS, NO (very weak lines), NS, PN, SO, SiO, SiS, CCH, H$_2$S, HCN, HCO$^{+}$, HNC, N$_2$H$^{+}$, SO$_2$, H$_2$CO, HNCO, H$_2$CS (weak lines), CH$_2$NH, HC$_3$N, CH$_3$CN, CH$_3$OH, CH$_3$NH$_2$, and many of their isotopologues. One tentatively new detected species is OCS but its transitions are too weak for a detailed analysis here. We find no evidence for the presence of NH$_2$CHO claimed by \citet{Eyres} and consider it a mis-identification. 

In the current paper, we focus on exploring the molecular content and excitation in the different parts of the nebula which could not have been performed with the single-dish data due to insufficient angular resolution. In forthcoming papers, we will explore in detail ({\it i}) the spatio-kinematical structure of the nebula in 3D, ({\it ii}) its isotopic composition, and ({\it iii}) the continuum emission of CK\,Vul.

\section{Morphology of the molecular emission}\label{sec-morph}
The most crucial and novel information delivered by our interferometric imaging spectroscopy survey consists of high-angular-resolution maps of emission of almost all molecules detected in CK\,Vul. We present the total-intensity maps in Figs.\,\ref{fig-gallery1} and \ref{fig-lowres}. Maps of emission of a given molecule were created by combining the contributions of all clearly detected and unblended transitions of its various isotopologues. This averaging, which optimizes signal-to-noise ratio (S/N), is justified because the emission morphology typically does not change much with transition owing to the narrow range of the upper-level energies ($E_u$) of the contributing transitions. 
While details in the emission distributions of lines from different isotopologues 
of a molecule may change owing to opacity effects and potential variations in isotopic ratios, for our analyses such combined maps give a good representation of the general morphology for each molecule. 

In the combination procedure, different maps of a given species were smoothed to the resolution of the image with the largest beam. Additionally, for the presentation here we used circular beams with a FWHM being the geometric mean of the major and minor axes of the data-informed synthesized beam. For species with a large number of transitions imaged with a very high S/N, e.g., SiO, HNC, and HCN, we excluded the maps obtained with the largest beams to preserve a high resolution of the combined images without significantly compromising the overall S/N of the final map. In the combination procedure, we calculated a weighted mean with weights set by the respective noise levels in the smoothed maps. For comparison and discussion, in Fig.\,\ref{fig-lowres} we present the continuum maps for bands 3 and 4 of the ALMA survey and also $K$-band ammonia emission imaged with the VLA (all hyperfine-structure transitions of NH$_3$ averaged). 

The gallery presented in Figs.\,\ref{fig-gallery1} and \ref{fig-lowres} constitutes a unique set of detailed 
images of an eruptive star with a molecular remnant. Few other stellar objects have a molecular envelope of comparable richness that is so well observed and has such a peculiar composition.
We now briefly summarize the morphological structures of the emission regions, emphasizing the most characteristic features of the remnant.

\begin{figure*}[!ht]
  \includegraphics[trim=0 35 35 0, width=.225\textwidth]{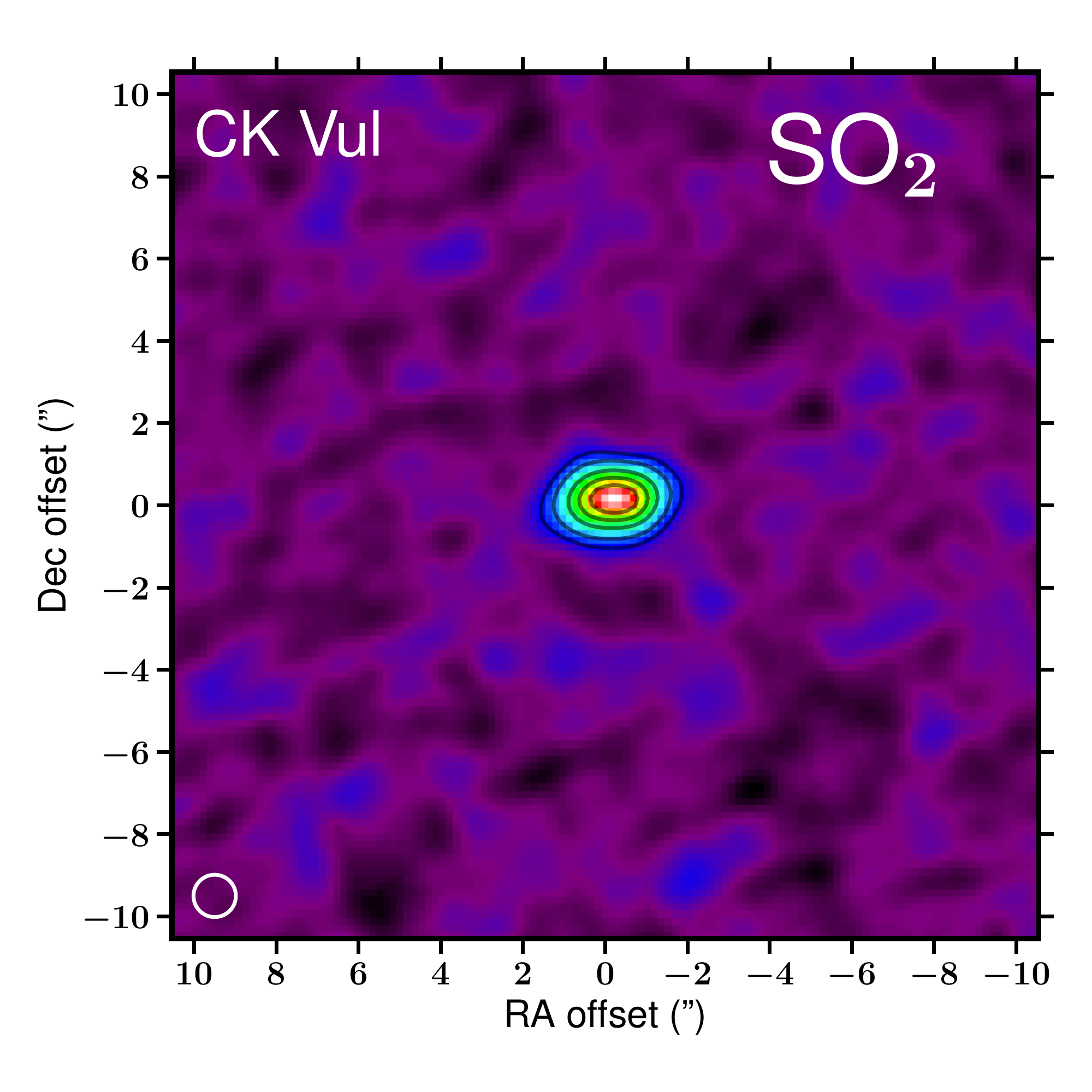}
  \includegraphics[trim=0 35 35 0, width=.225\textwidth]{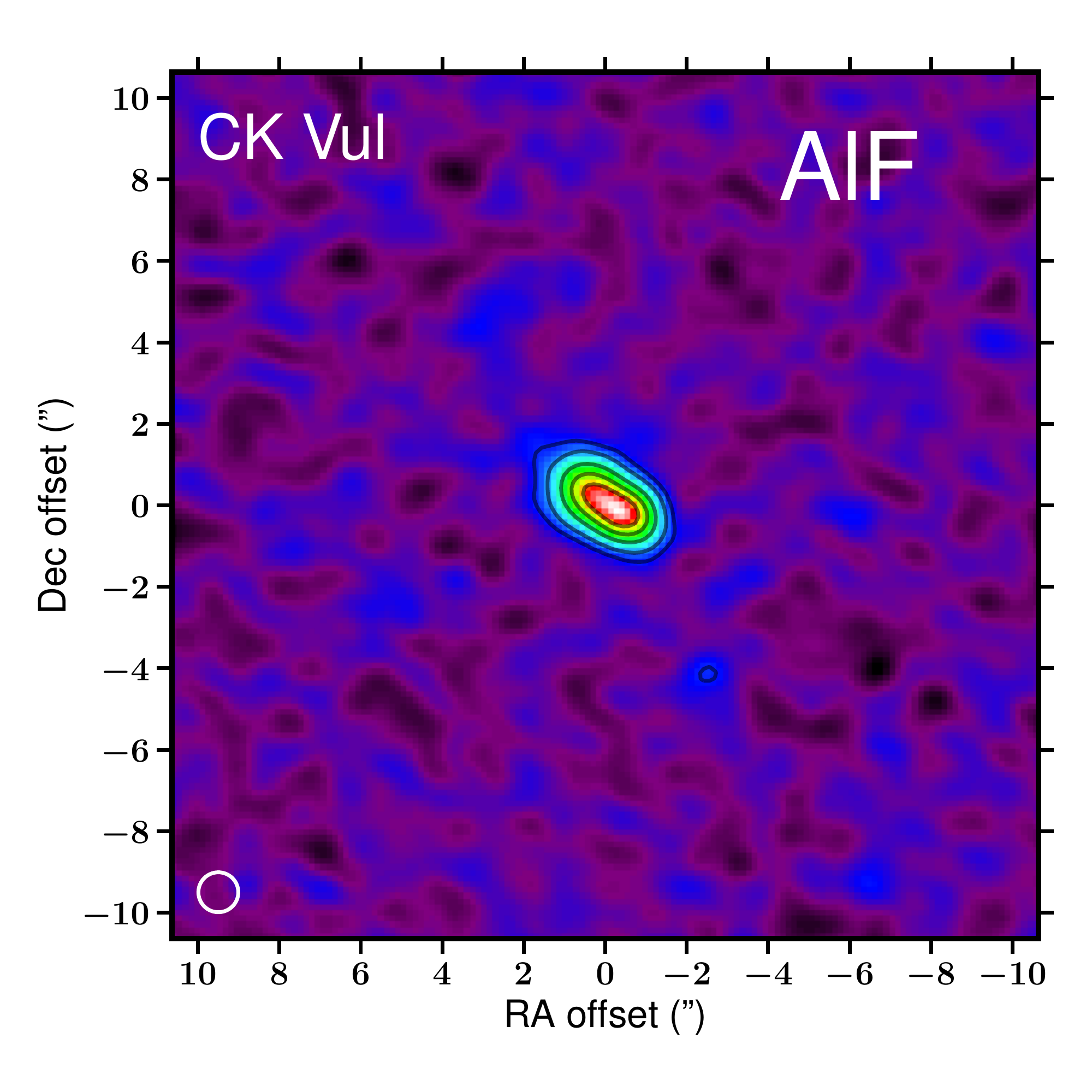}
  \includegraphics[trim=0 35 35 0, width=.225\textwidth]{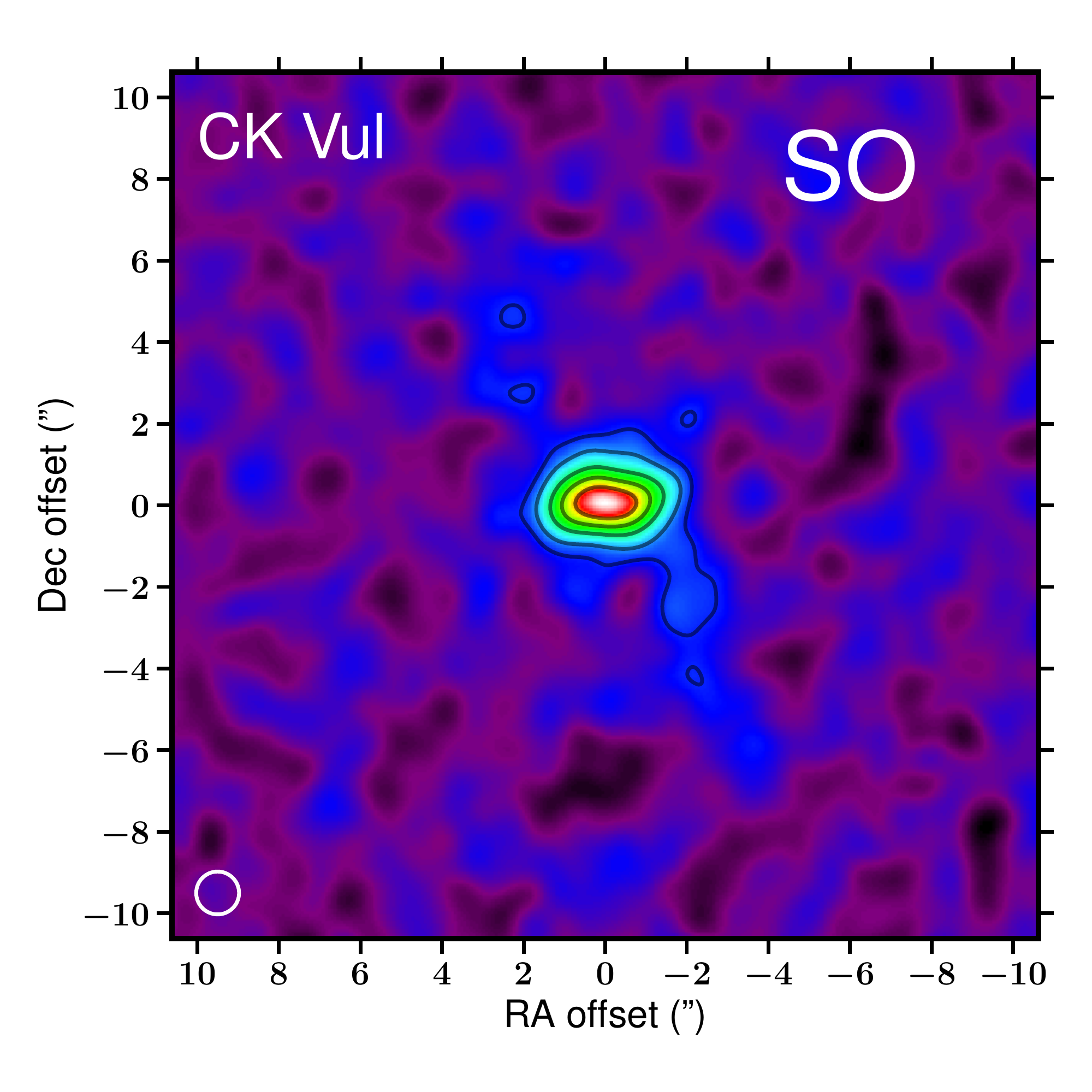}
  \includegraphics[trim=0 35 35 0, width=.225\textwidth]{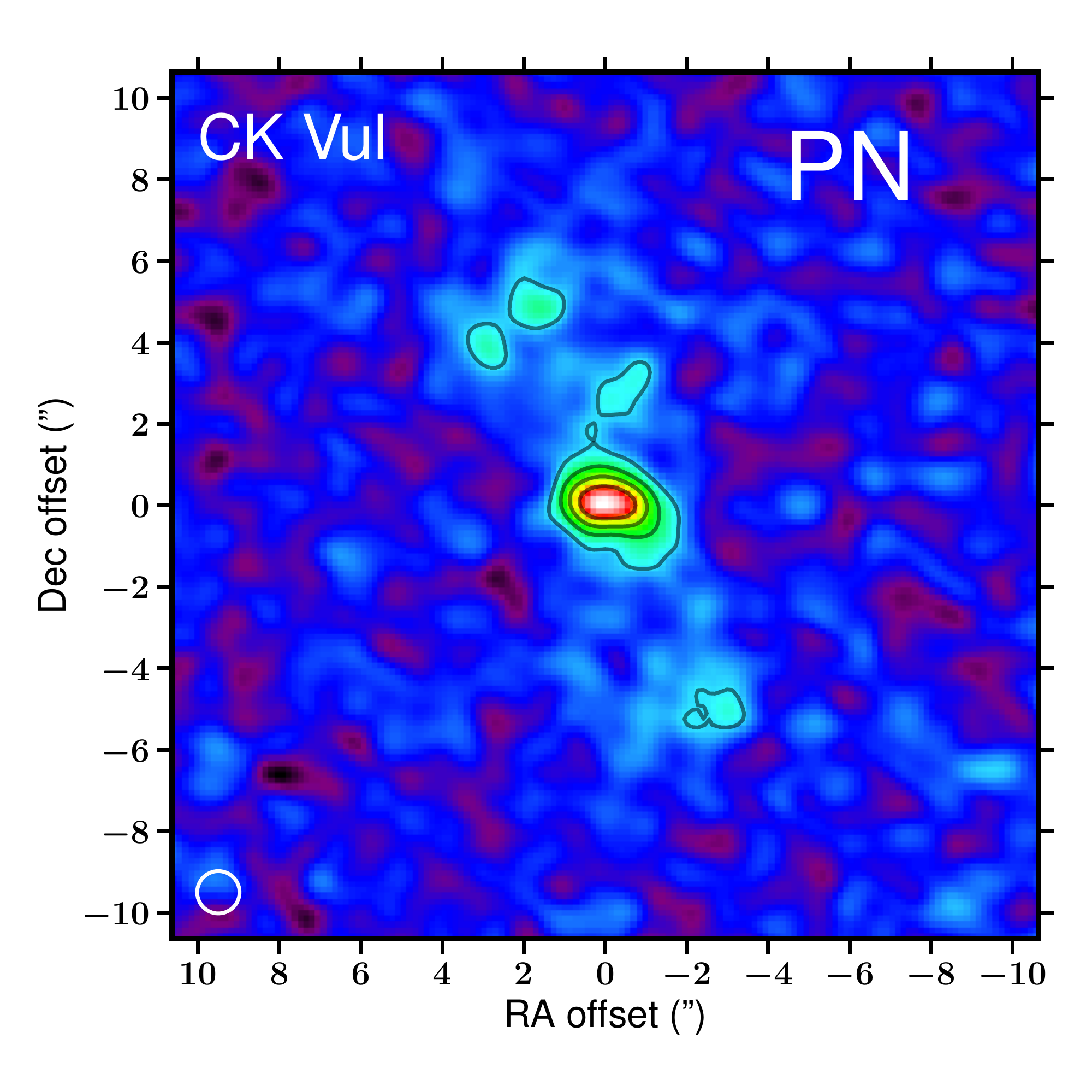}\\
  \includegraphics[trim=0 35 35 0, width=.225\textwidth]{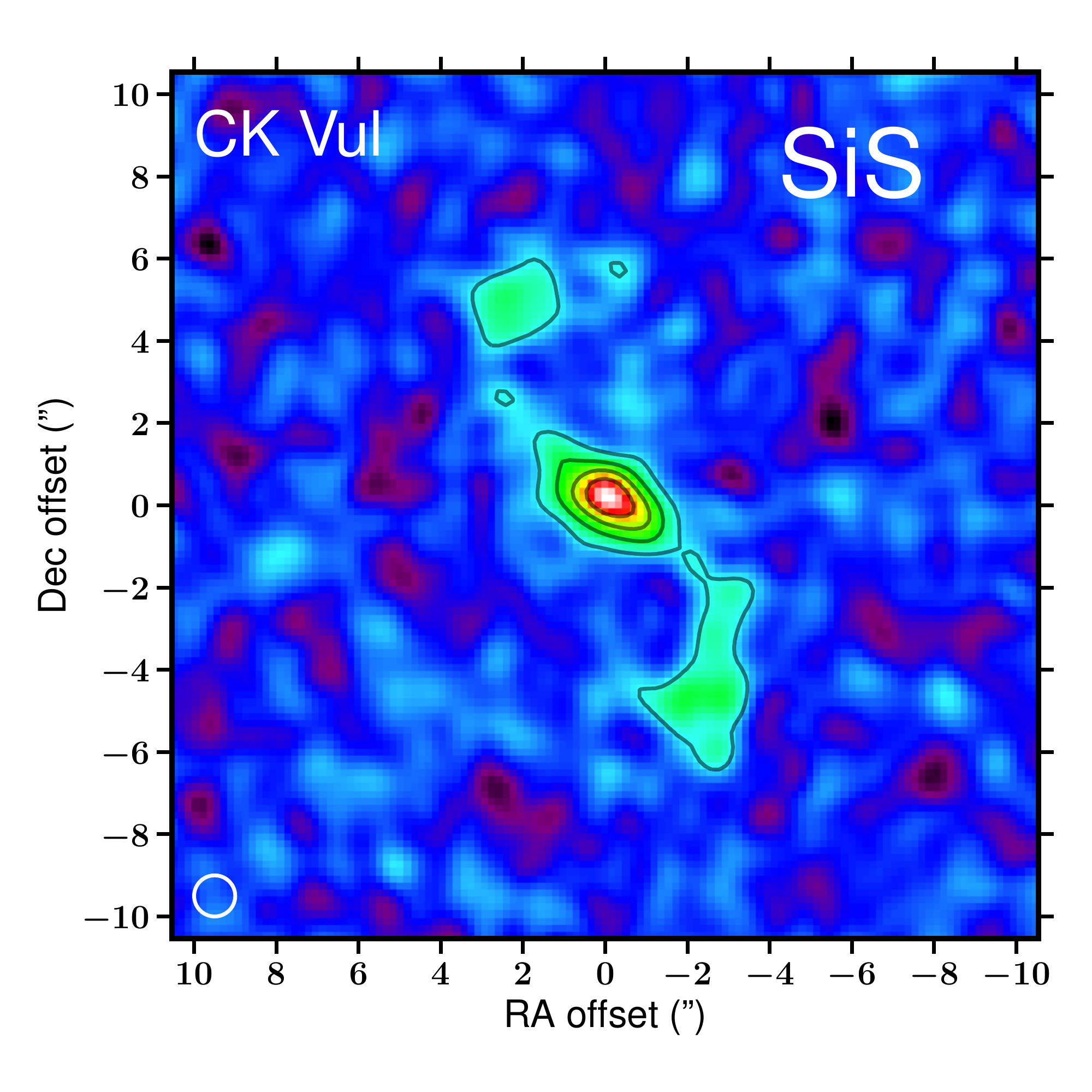}
  \includegraphics[trim=0 35 35 0, width=.225\textwidth]{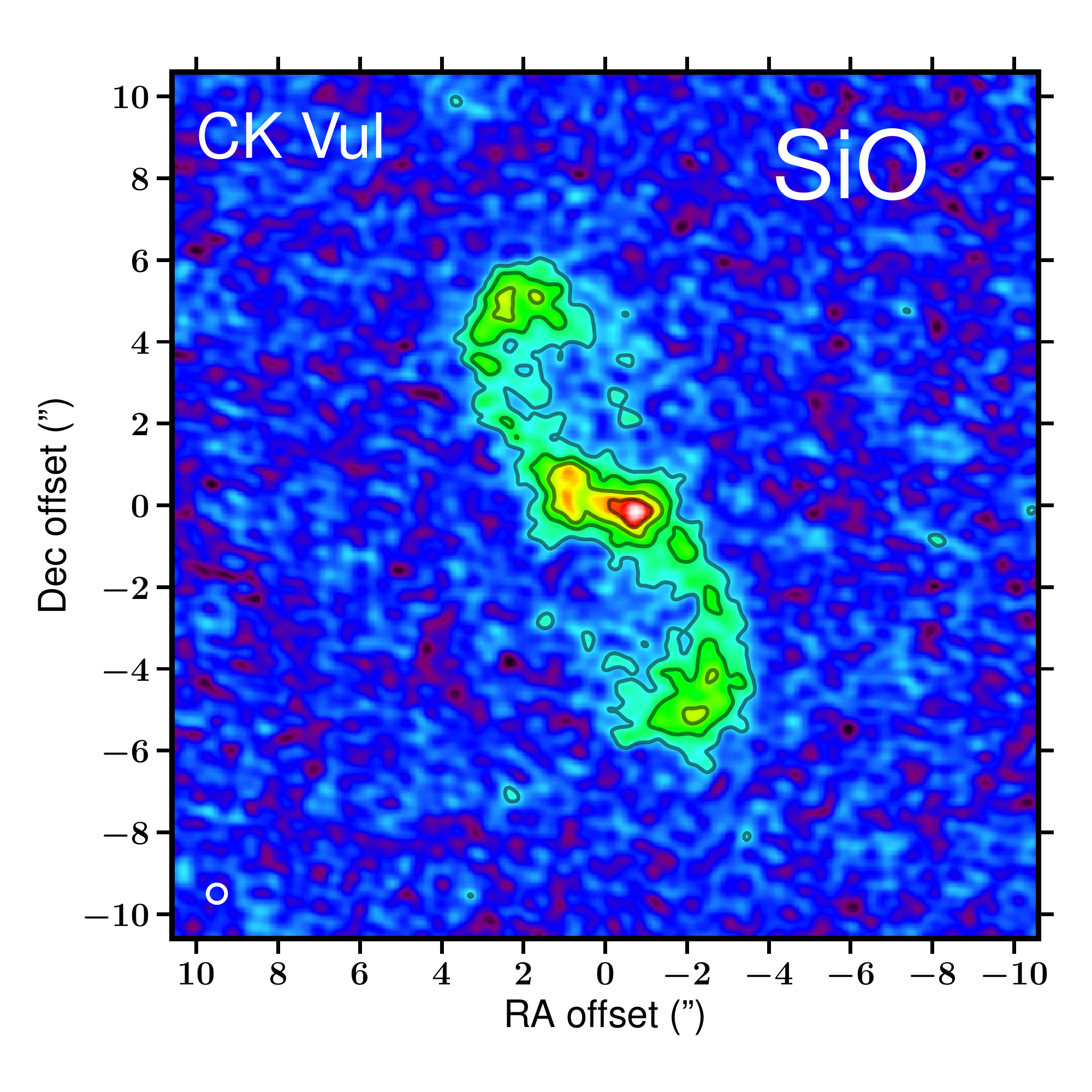}
  \includegraphics[trim=0 35 35 0, width=.225\textwidth]{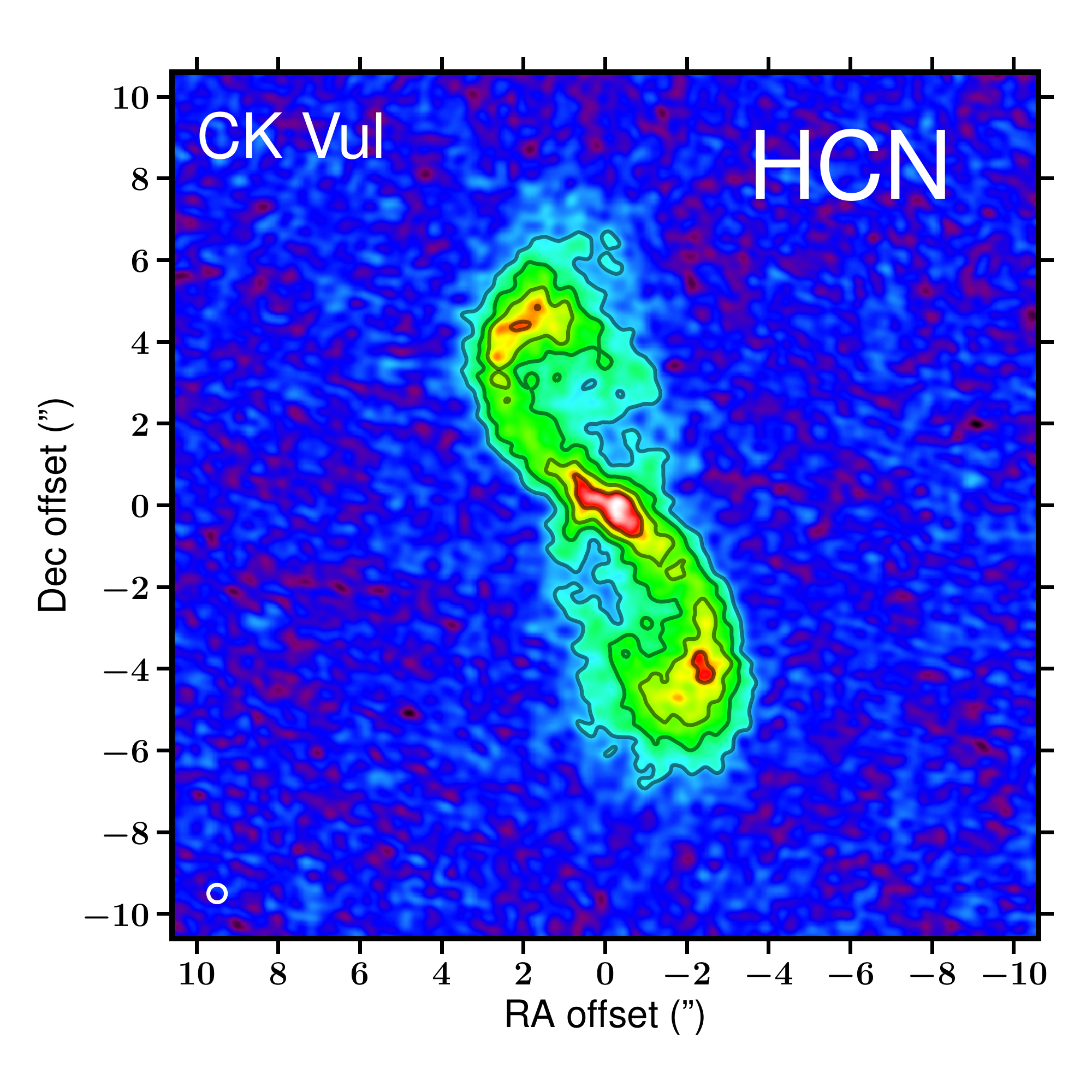}
  \includegraphics[trim=0 35 35 0, width=.225\textwidth]{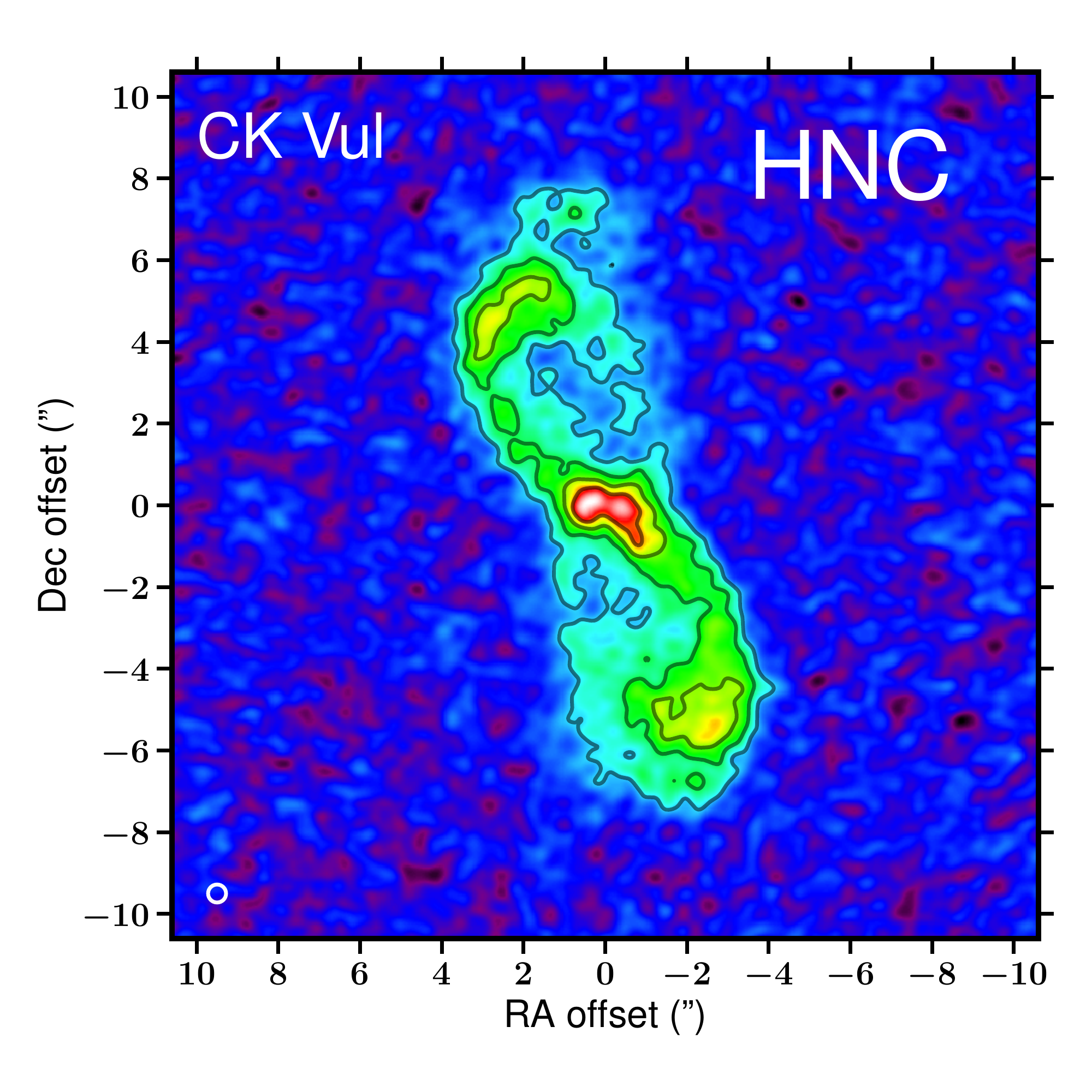}\\
  \includegraphics[trim=0 35 35 0, width=.225\textwidth]{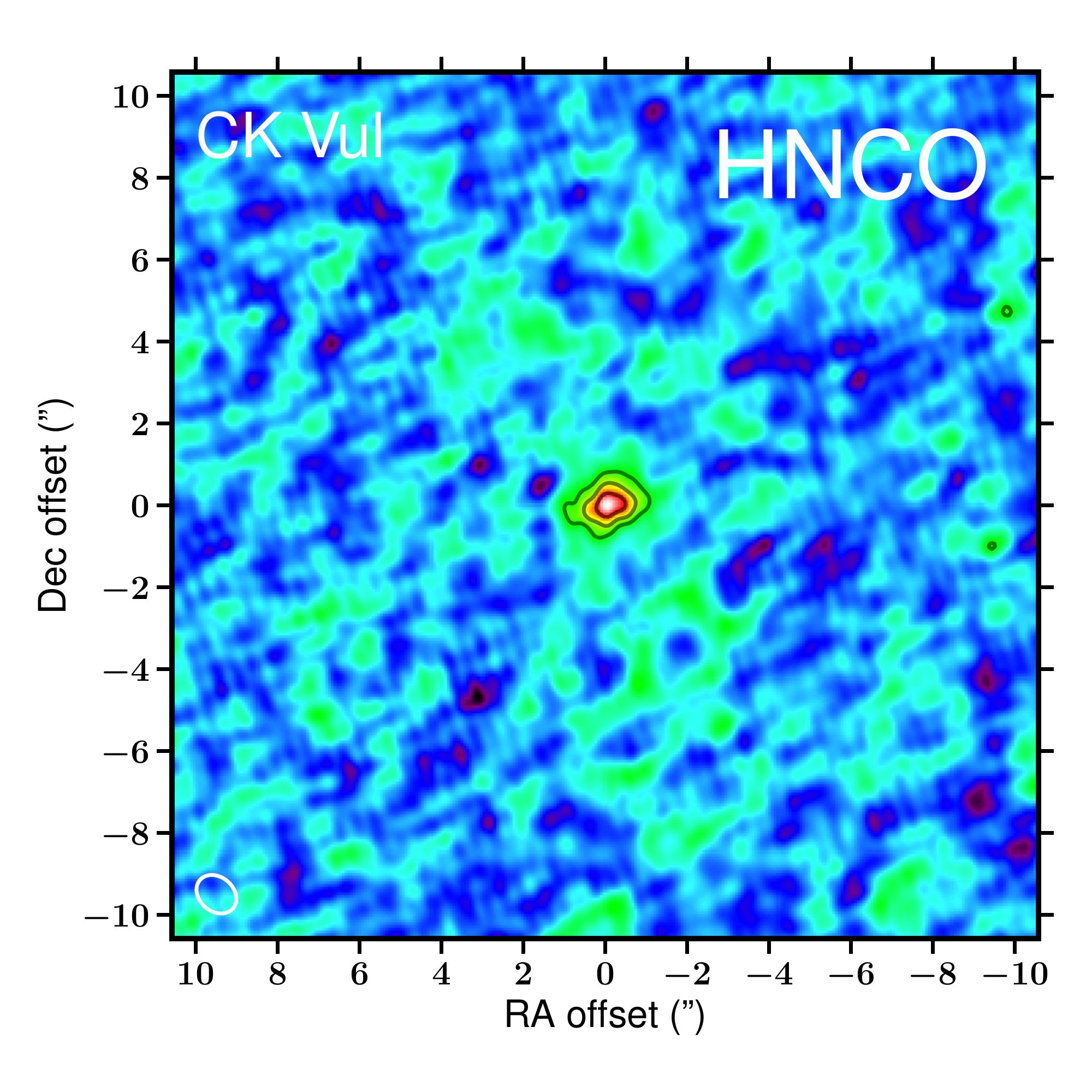}
  \includegraphics[trim=0 35 35 0, width=.225\textwidth]{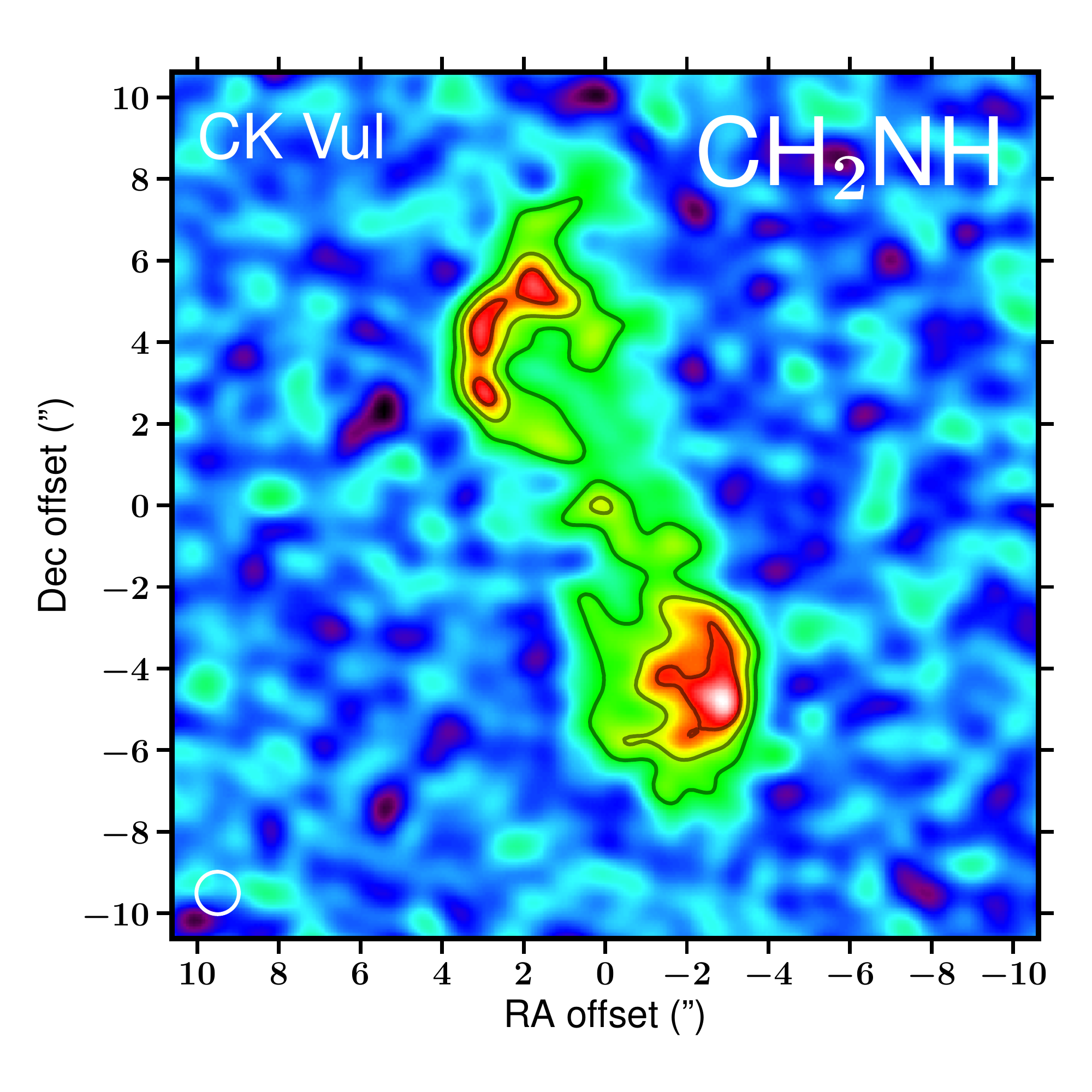}
  \includegraphics[trim=0 35 35 0, width=.225\textwidth]{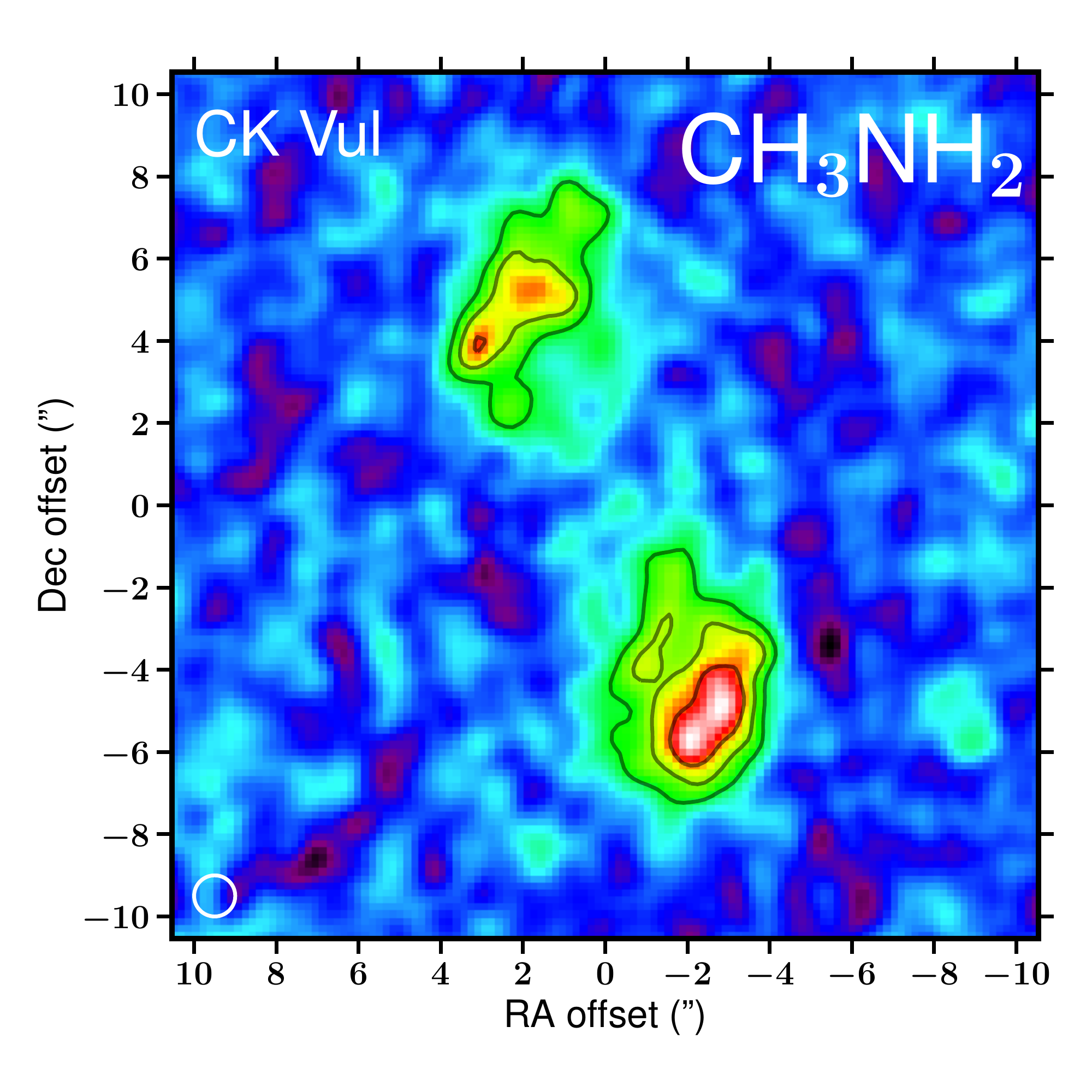}
  \includegraphics[trim=0 35 35 0, width=.225\textwidth]{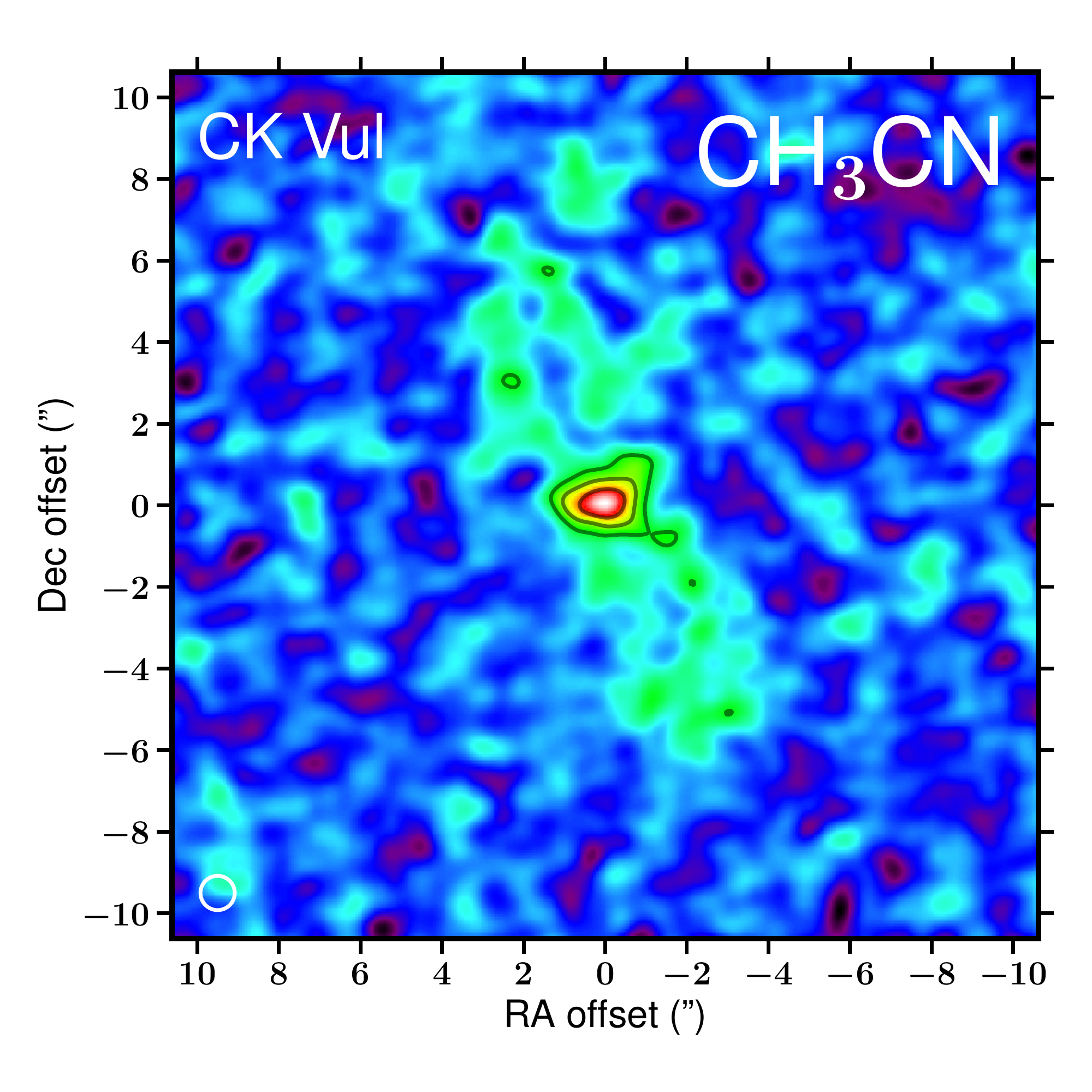}\\
  \includegraphics[trim=0 35 35 0, width=.225\textwidth]{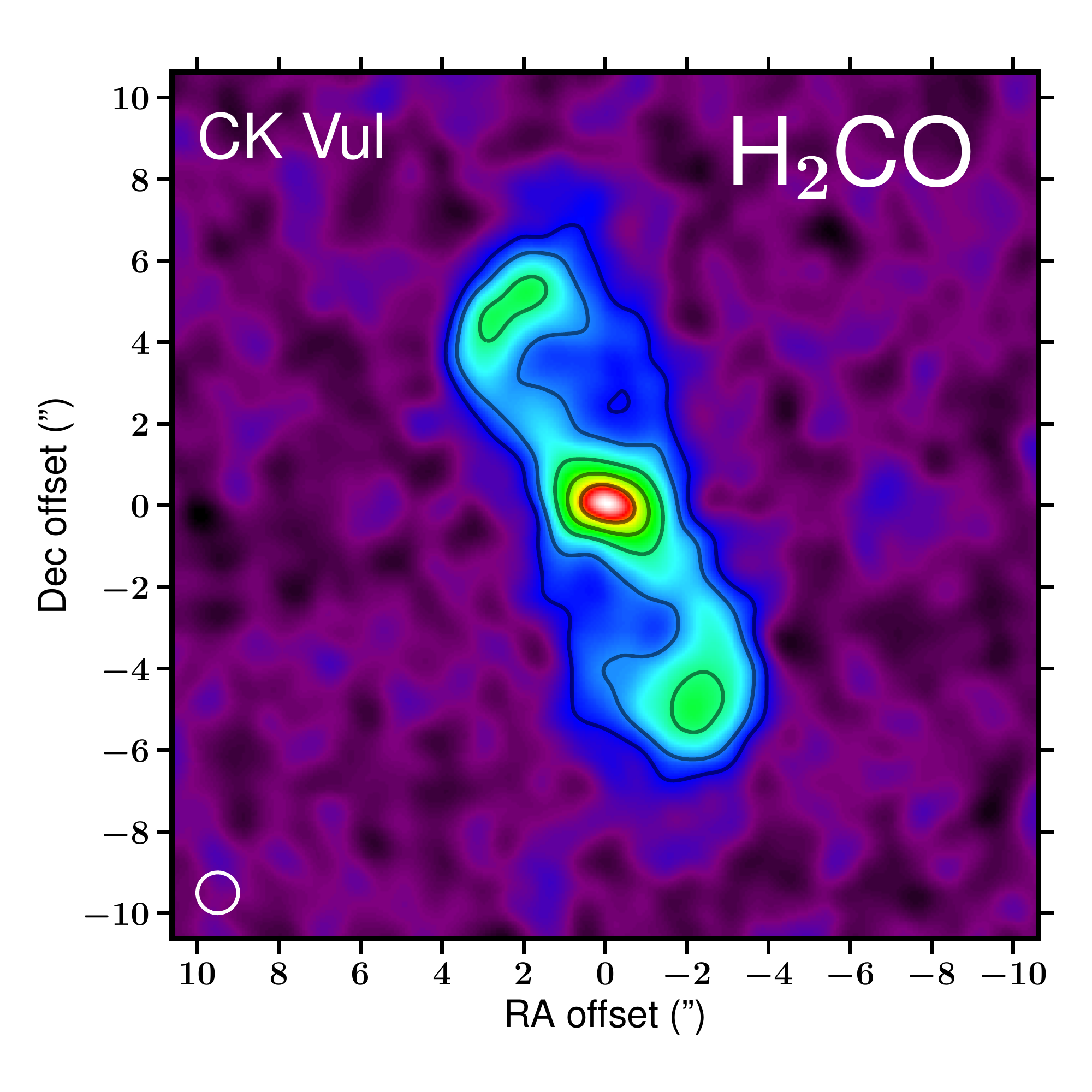}
  \includegraphics[trim=0 35 35 0, width=.225\textwidth]{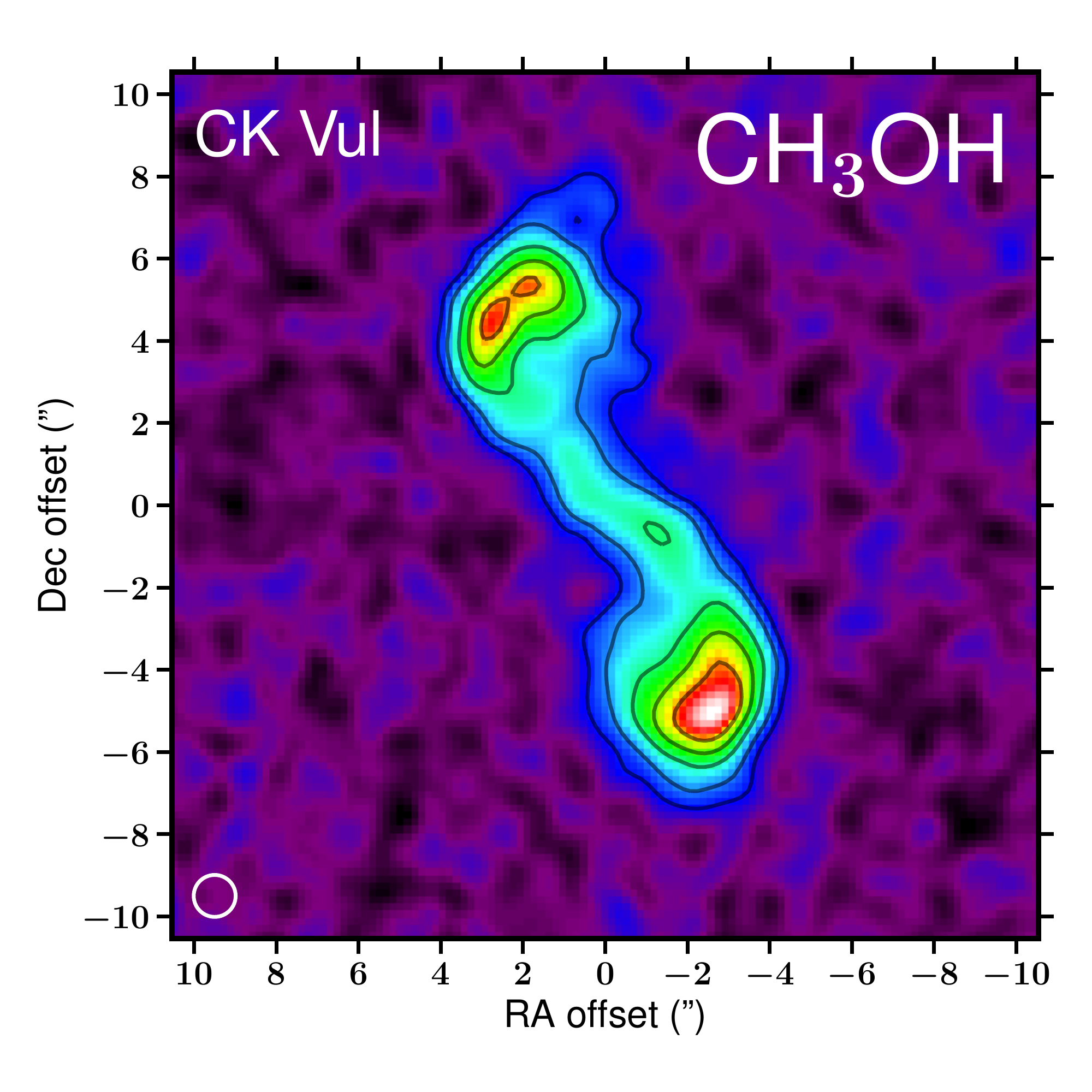}
  \includegraphics[trim=0 35 35 0, width=.225\textwidth]{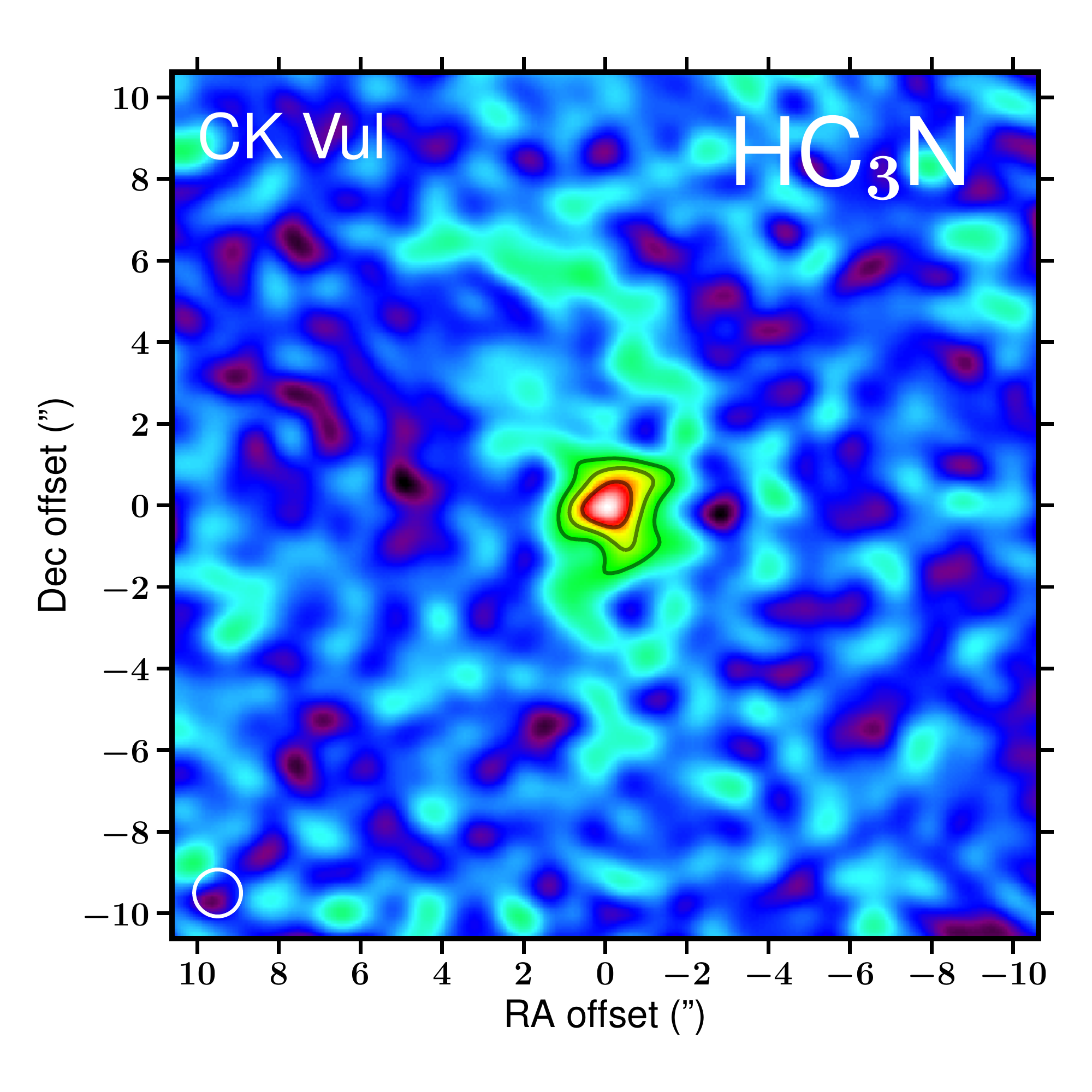}
  \includegraphics[trim=0 35 35 0, width=.225\textwidth]{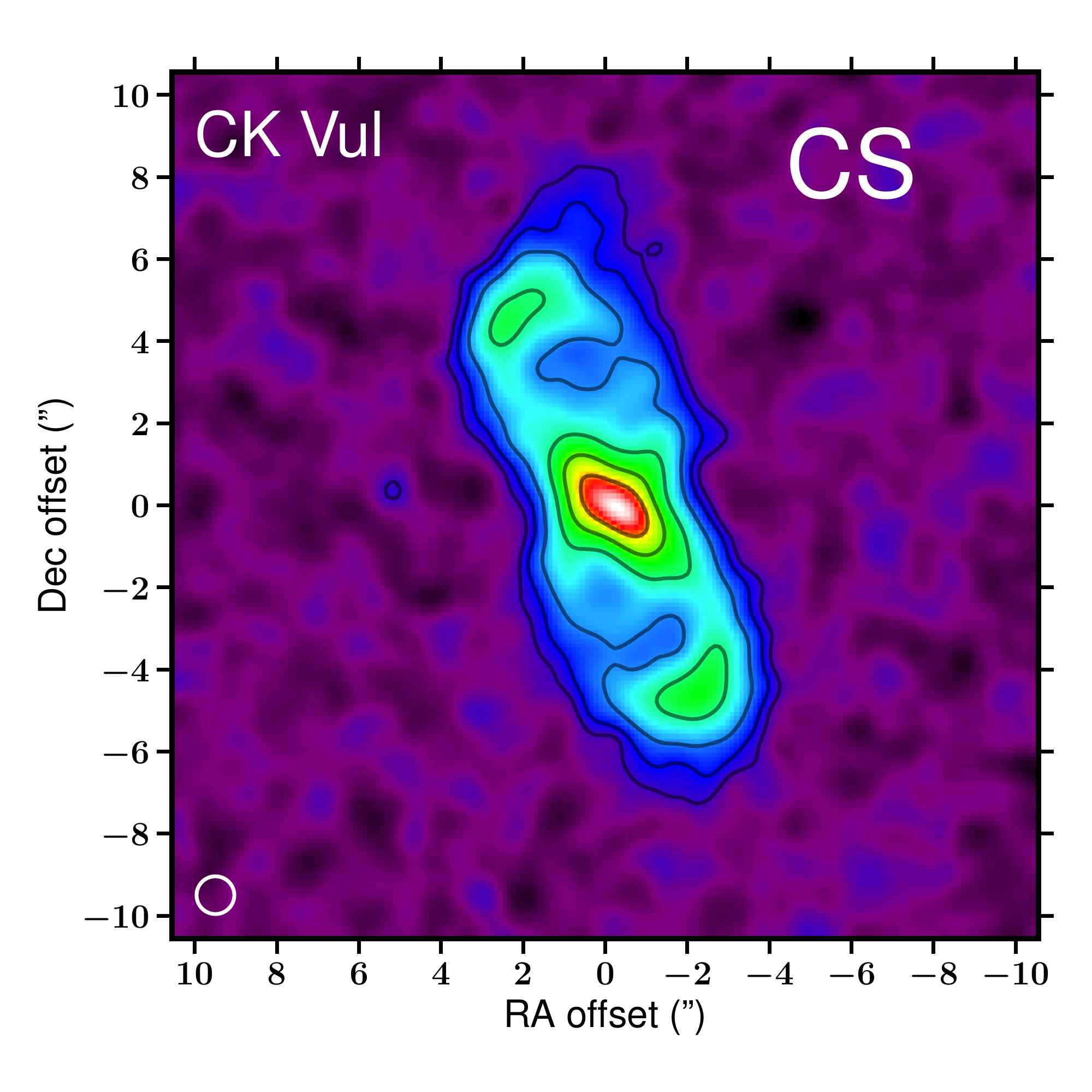}\\
  \includegraphics[trim=0 35 35 0, width=.225\textwidth]{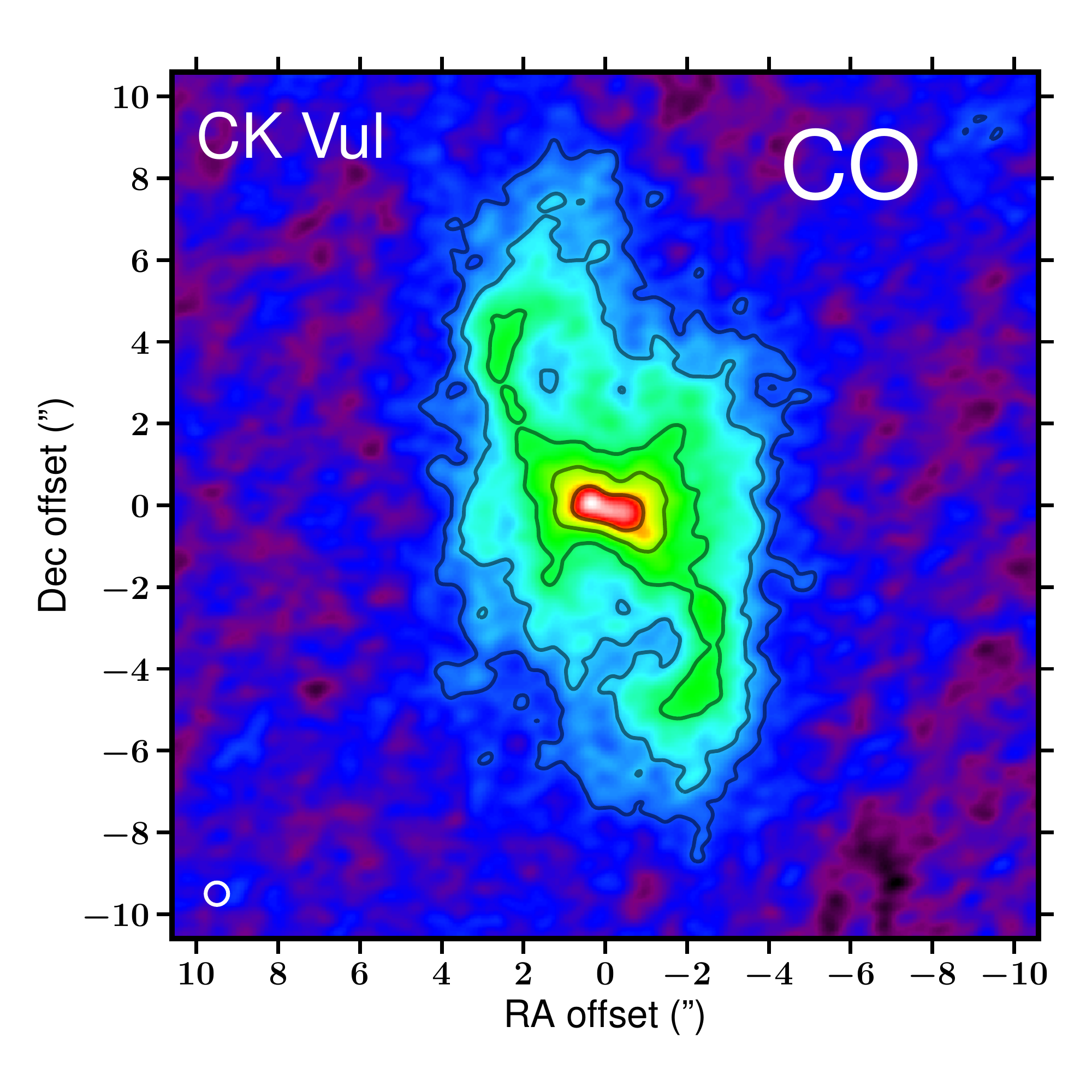}
  \includegraphics[trim=0 35 35 0, width=.225\textwidth]{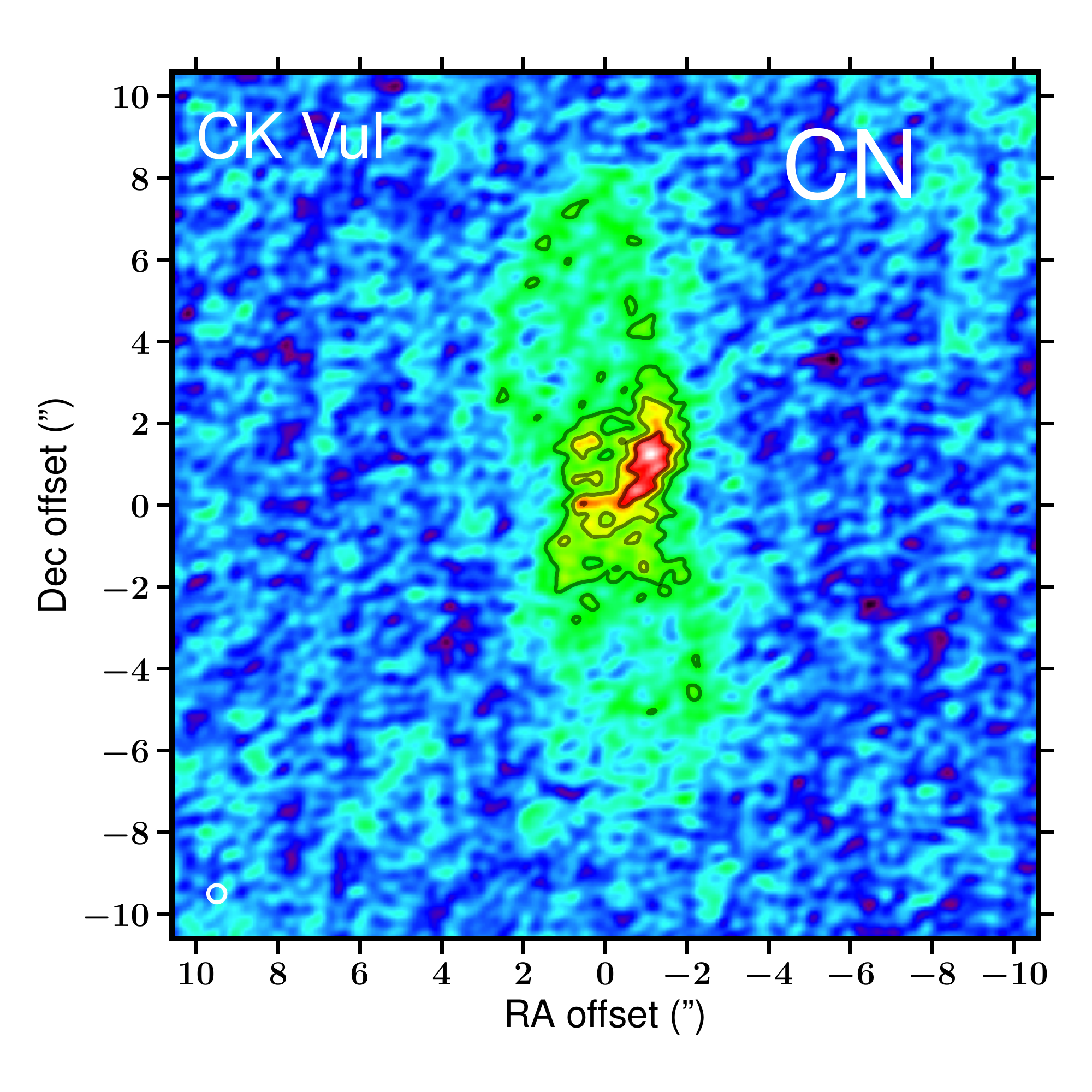}
  \includegraphics[trim=0 35 35 0, width=.225\textwidth]{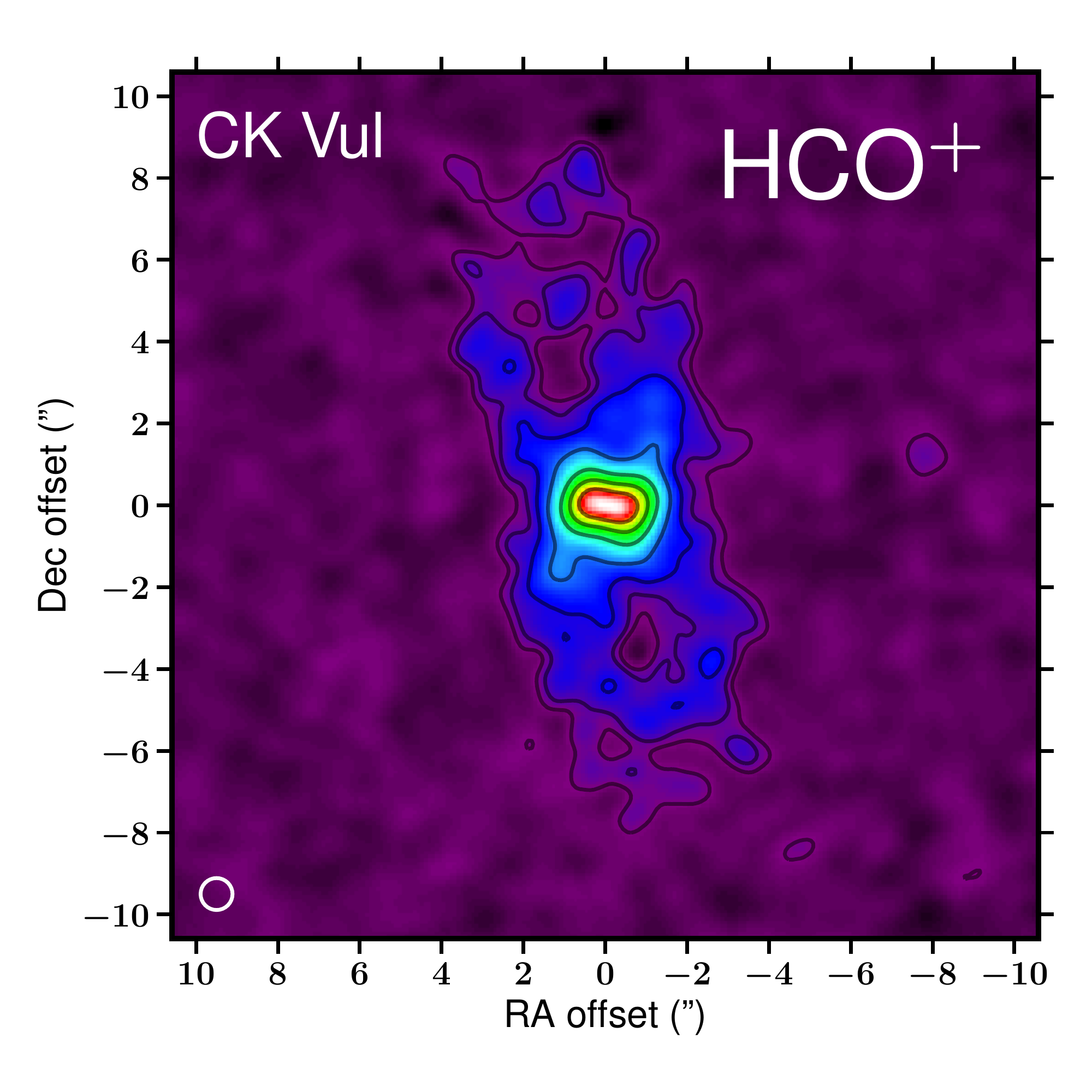}
  \includegraphics[trim=0 35 35 0, width=.225\textwidth]{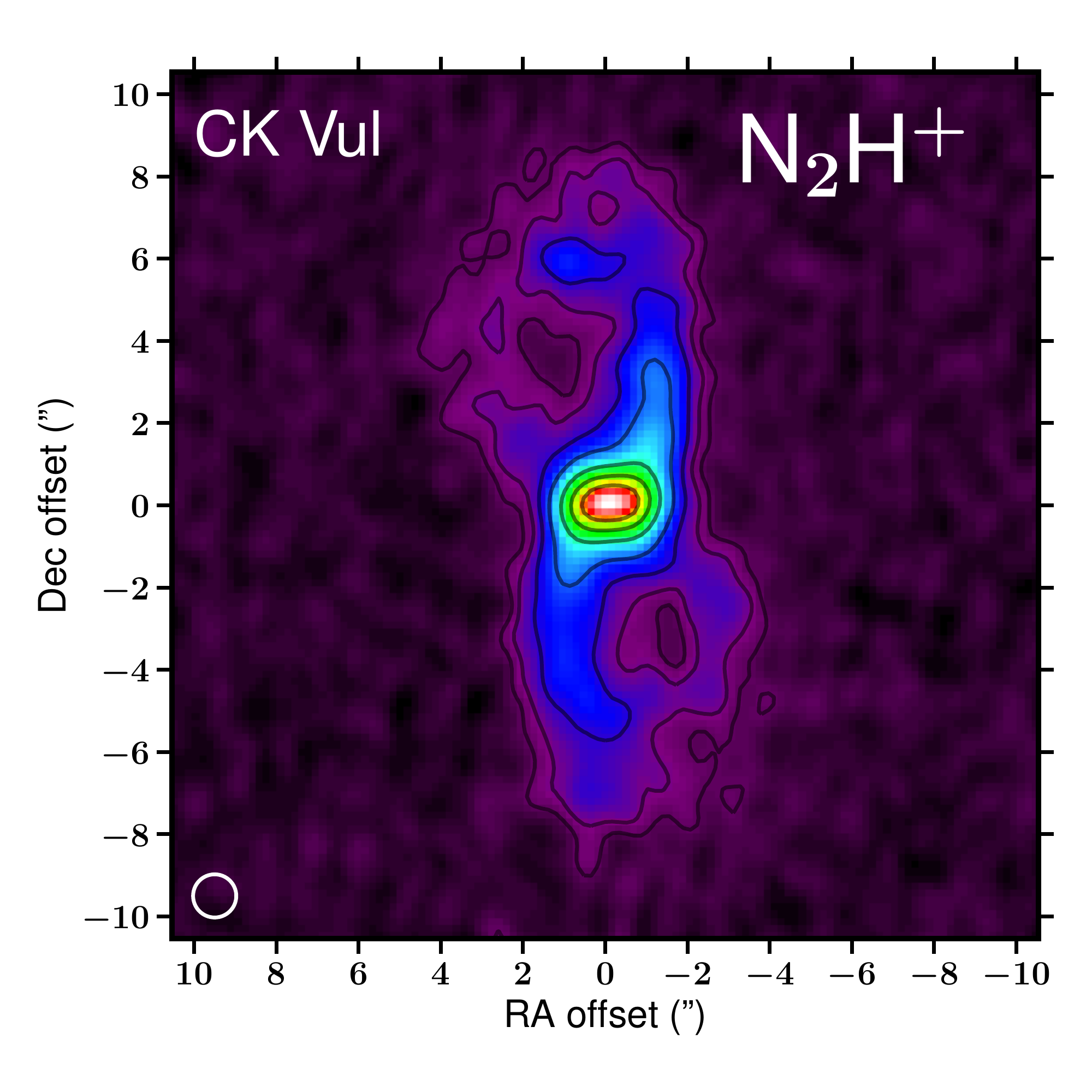}\\
\caption{Maps of molecular and continuum emission in CK\,Vul. Each map shows the weighted mean of the emission in multiple transitions of the given species and its isotopologues. Contours are drawn at 2.5, 5, 10, 20, 40, 80\% of the emission peak in each map starting from a level which is above the respective 3$\times$rms noise. All images use the same linear color scale.}\label{fig-gallery1}
\end{figure*}

\begin{figure*}[!ht]
  \includegraphics[trim=0 35 35 0, width=.225\textwidth]{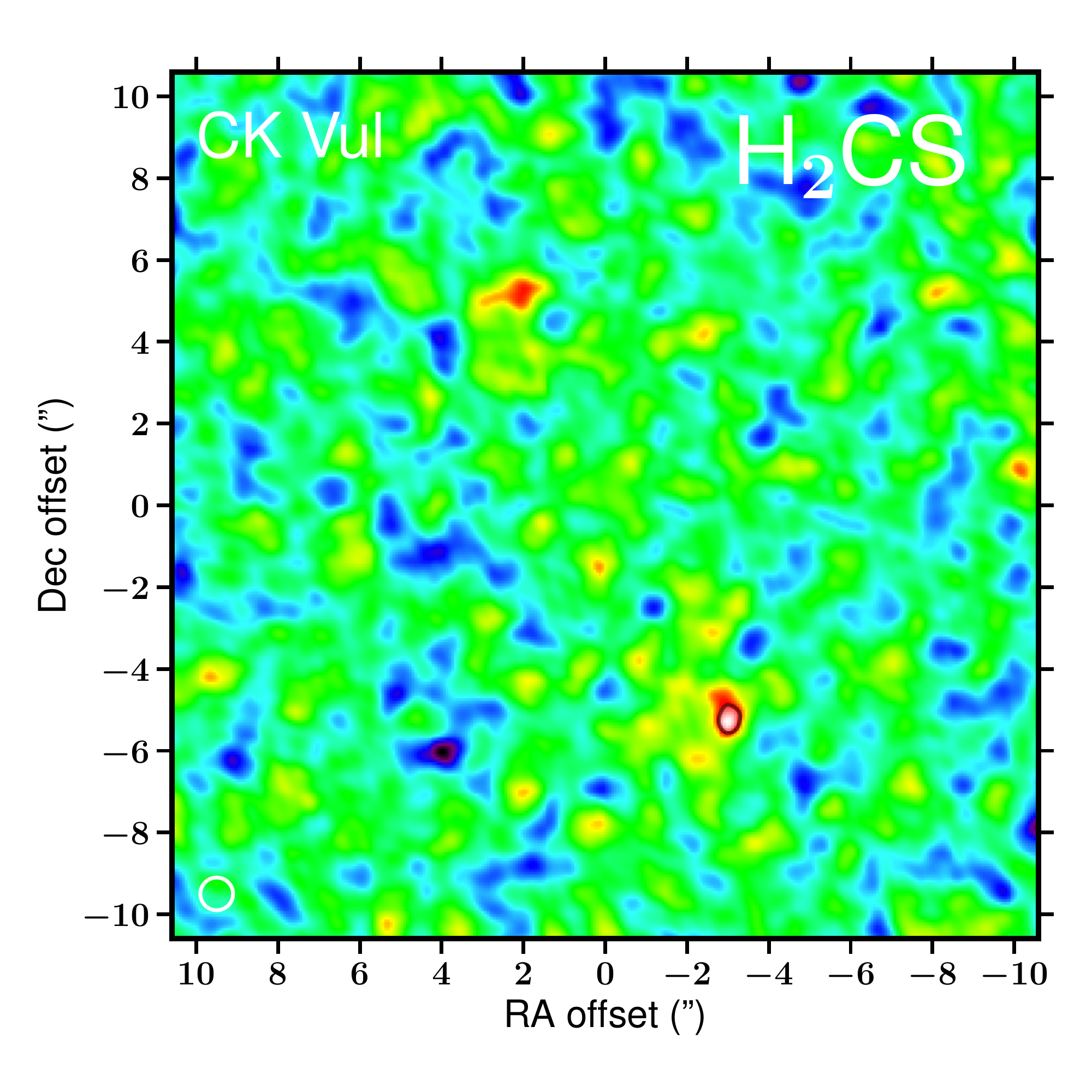}
  \includegraphics[trim=0 35 35 0, width=.225\textwidth]{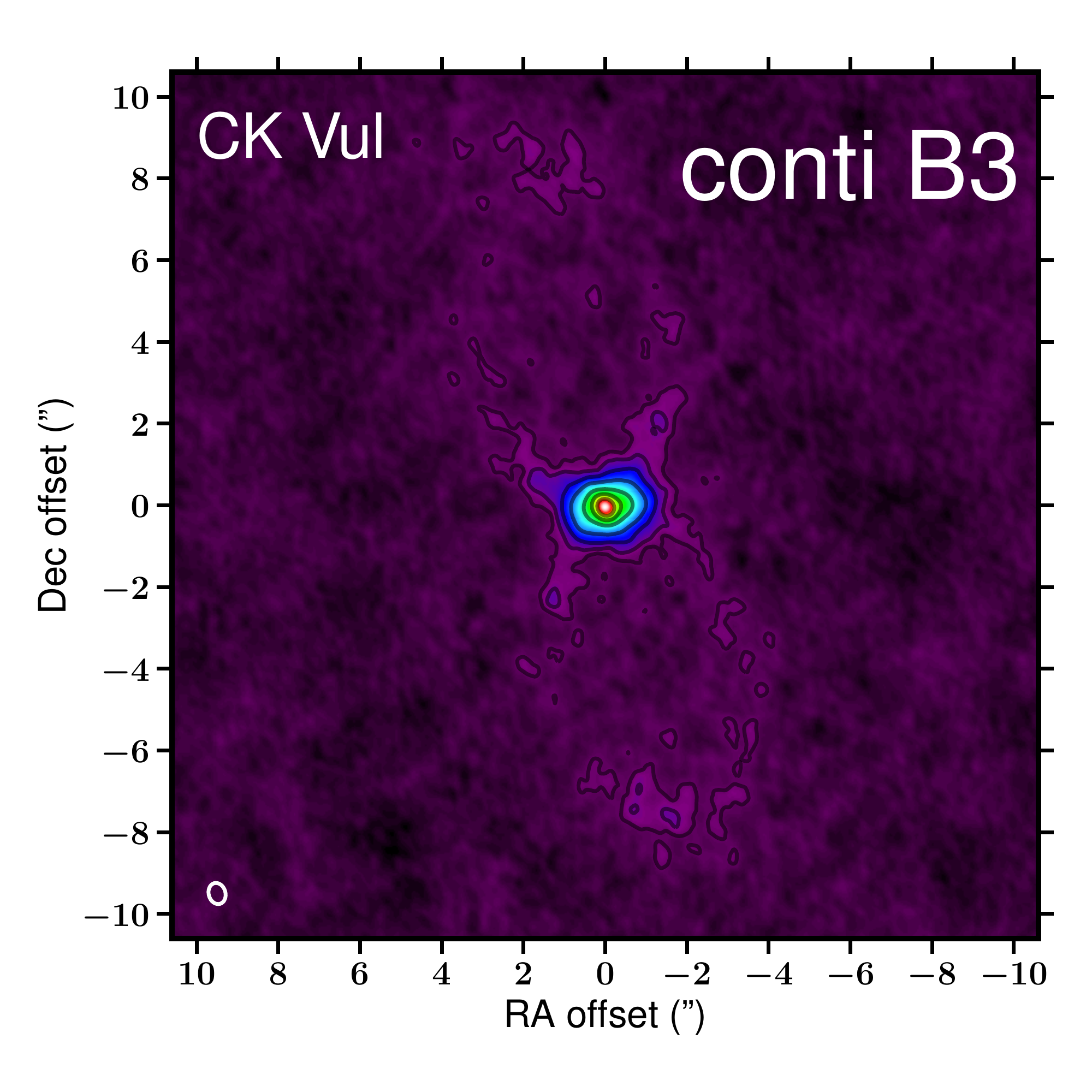}
  \includegraphics[trim=0 35 35 0, width=.225\textwidth]{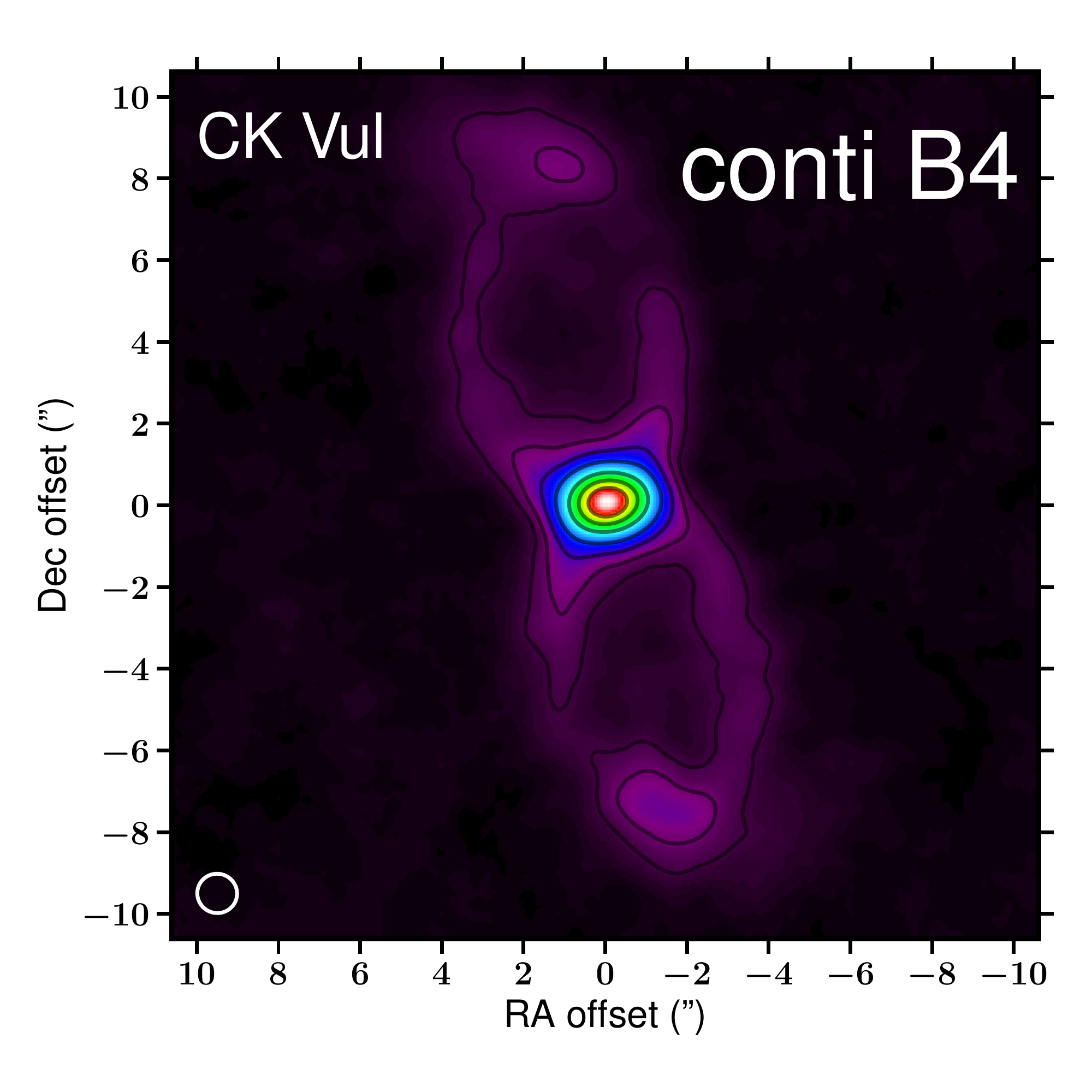}
  \includegraphics[trim=0 35 35 0, width=.225\textwidth]{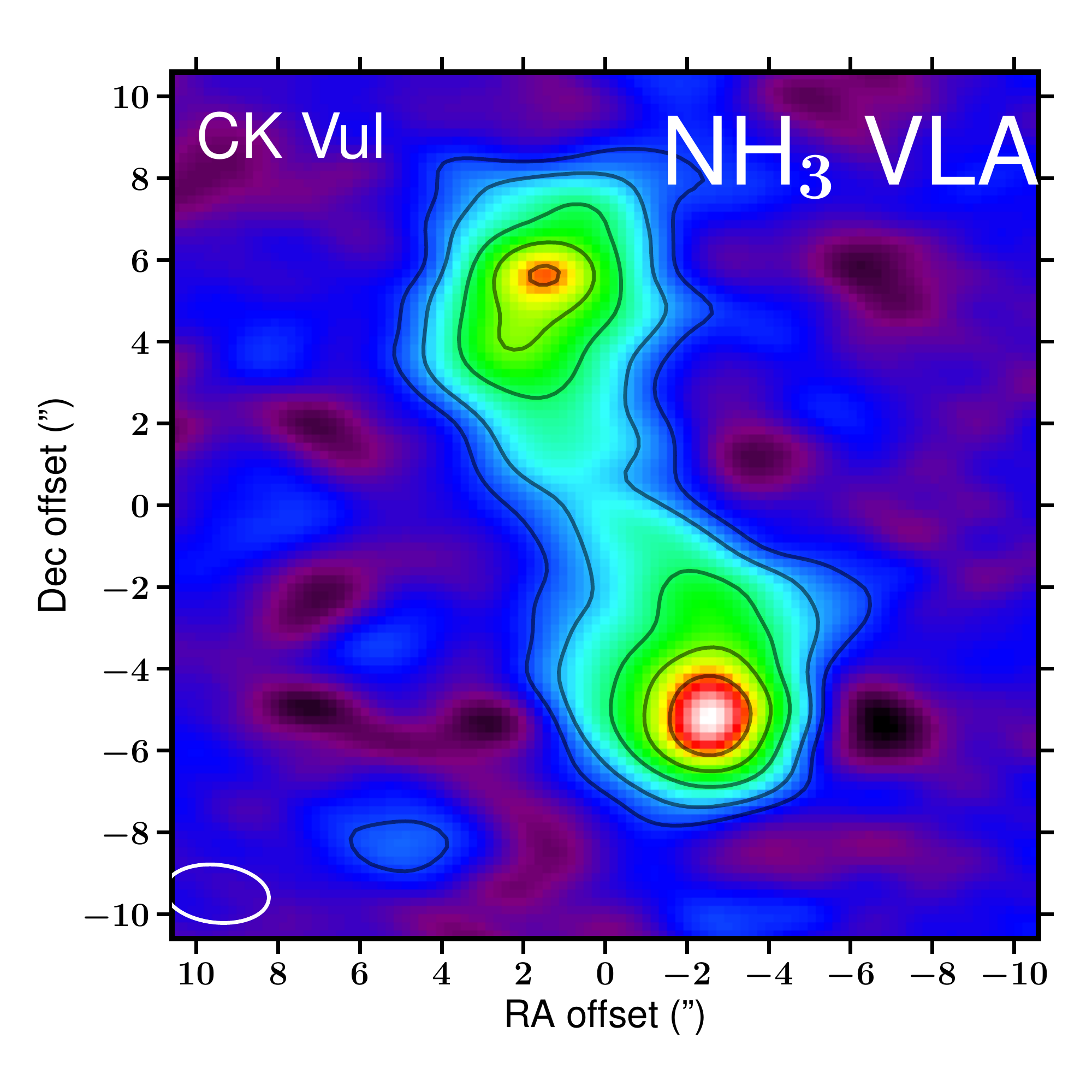}
  \caption{Continuation of Fig.\,\ref{fig-gallery1}. The map of H$_2$CS (left panel) has a low S/N but shows that the dominant component is at the end points of the lobes. Continuum emission in bands 3 and 4 is imaged at a very high S/N, but is presented in the middle panels in the same linear color scale as the other maps. For completeness, we add a map of NH$_3$ obtained with VLA (right panel). Unlike in other maps, the restoring beam is non-circular.}\label{fig-lowres}
\end{figure*}

We define the ``central region'' as the compact component visible in most molecular species and in dust emission within $\sim$2\arcsec\ from the mm/submm continuum peak. All mapped species, except CH$_2$NH, CH$_3$NH$_2$, CH$_3$OH, NH$_3$ and perhaps H$_2$CS, have a significant part of their emission concentrated in this region. Several species, including SO, SO$_2$, and AlF, appear to have emission almost exclusively in this region. For HNCO and HC$_3$N, the images are too noisy to conclude on their spatial extent. The continuum emission is also very strong in this region. Unlike the emission from the molecules, the more sensitive continuum images show an almost continuous loop in the shape of the number ``8'', demarcating the more extended bipolar lobes. This is due to a very high sensitivity of these images, which represent a much larger ``effective'' bandwidth than the spectral-line images. It is possible that AlF and SO have a weak component in the lobes that would be more apparent at a higher S/N. However, with a very deep observations of SO$_2$ averaged over several transitions, we are confident that this species is indeed present exclusively in the central region, strongly suggesting that the region is oxygen-rich and giving physical basis that the central region is a chemically distinctive part of the remnant. 

\begin{figure}[!ht]\centering
  \includegraphics[trim=0 35 35 0, width=.49\columnwidth]{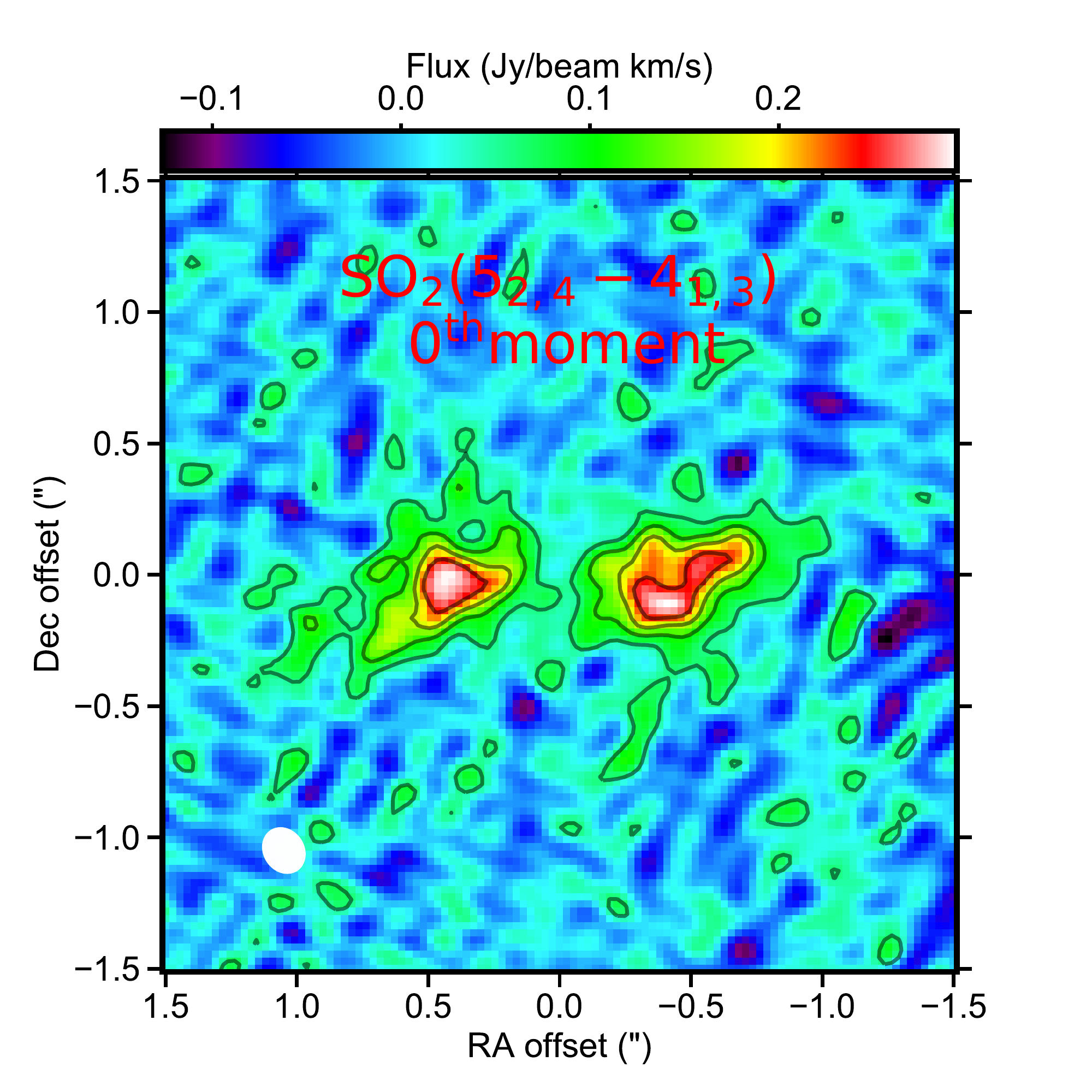}
  \includegraphics[trim=0 35 35 0, width=.49\columnwidth]{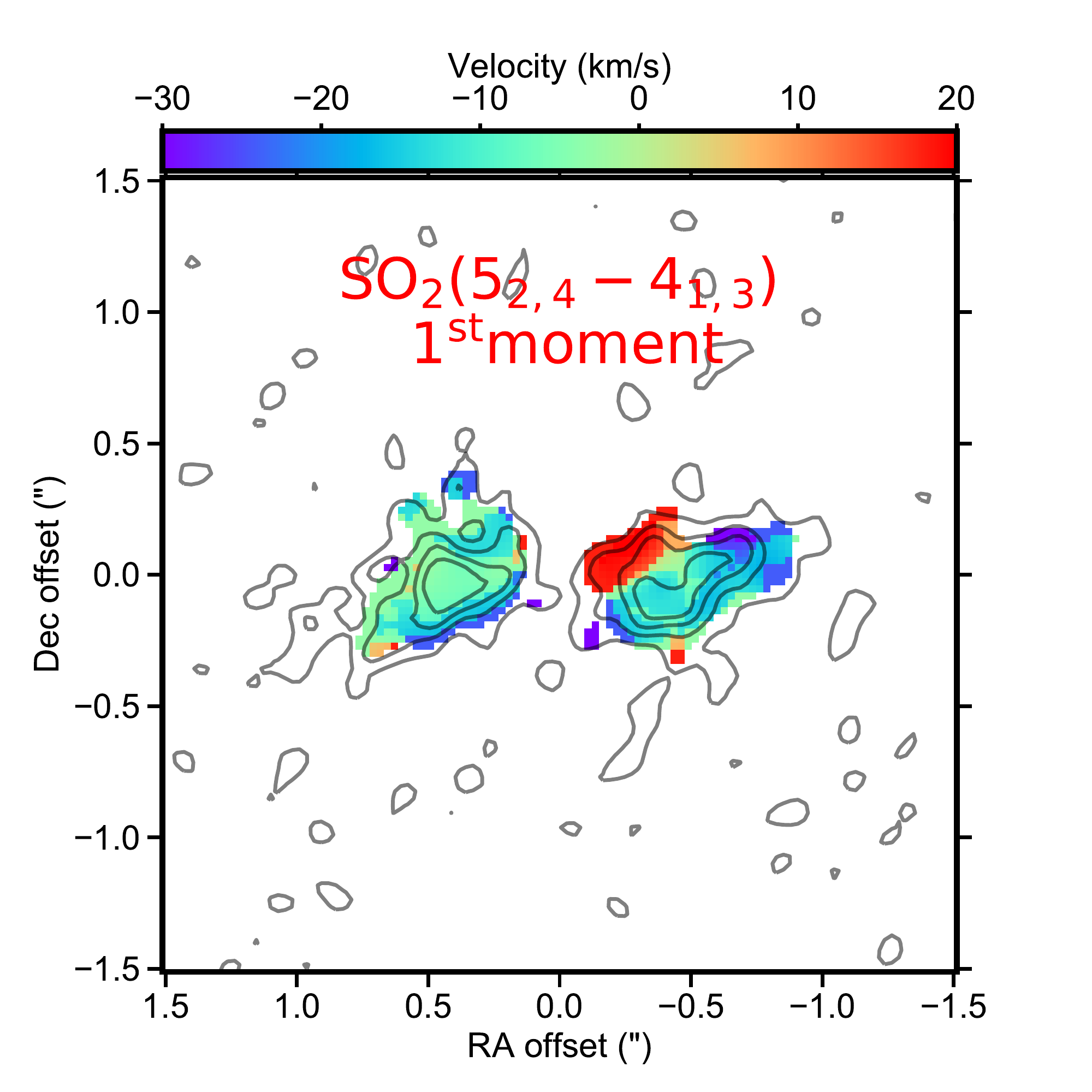}
  \caption{Maps of the SO$_2$ $5_{2,4}-4_{1,3}$ transition. The left panel shows the total intensity map. The right panel shows a first-moment map for emission above the 4$\sigma$ noise level in the original data cube. Both maps present contours of the total intensity at 20, 40, 60, and 80\% of the peak emission. The input data were restored with uniform weighting of the visibilities, resulting in a synthesized beam of size  0\farcs17$\times$0\farcs14 (shown as white ellipse in the left panel).}\label{fig-so2}
\end{figure}

Comparing the appearance of the central region mapped in different species, one can notice variations in position angle (PA) of its major axis. In the continuum, the longer axis is at a PA of 98\fdg2 E of N. In SO$_2$ and SO, the axis changes to 93\fdg5 and 91\fdg2, respectively. In  AlF, the angle decreases to 56\fdg7. Similarly to AlF, most other species (see e.g., HCN in Fig.\,\ref{fig-gallery1}) tend to have the central region elongated along the much more extended ridges of the emission that define the bipolar lobes of the nebula. At the sub-arcsec resolution, the central region splits into multiple clumps (see e.g., SiO and HNC in Fig.\,\ref{fig-gallery1}) and a large portion of the flux in the region may in fact be still associated with the lobes. Perhaps only the continuum and the compact emission of SO, SO$_2$, and AlF represent material that is unique to the central region. The morphology of the compact emission of SO$_2$ mapped at sub-arcsec resolution, as shown in Fig.\,\ref{fig-so2}, is reminiscent of a disk or torus seen edge-on, but its kinematics are not consistent with solid-body or Keplerian disk rotation.

\begin{figure}[!ht]\centering
  \includegraphics[width=.99\columnwidth]{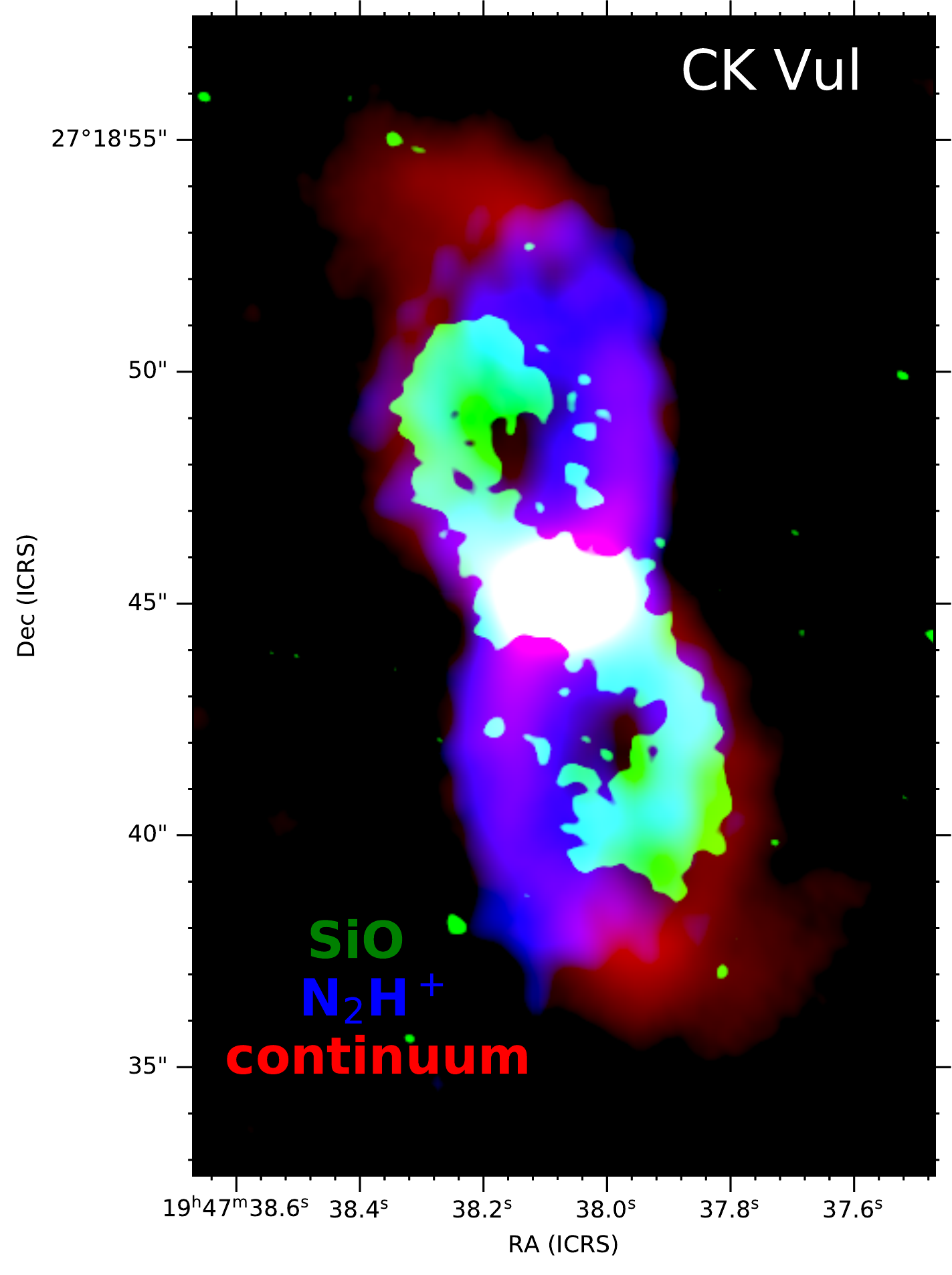}
  \caption{Composite RGB image of dust continuum emission (red), SiO (green), and N$_2$H$^+$ (blue). All three species overlap in the central region.}\label{fig-RGB}
\end{figure}

The extended emission of most species forms a characteristic S-shaped structure with stronger emission in the center and near the apexes of the lobes. Examples include SiS, SiO, HCN, HNC, H$_2$CO, CS and CO. A few complex species show the same S-type overall shape but do not display the central peak. CH$_3$OH, CH$_2$NH, CH$_3$NH$_2$, NH$_3$, and perhaps H$_2$CS belong to this group.  In all cases the extent of the ``S'' is smaller than that of the ``8'' outlined by the continuum emission (cf. Fig.\,\ref{fig-RGB}).

Two molecular ions that  ALMA covered, HCO$^+$ and N$_2$H$^+$, show emission that is more patchy than that of neutral  species and their overall shape appears to be a mirrored image of the S-shaped  emission seen in most other molecules. This \reflectbox{S}-shaped morphology may also be assigned to CN, although the S/N of its image is modest. The S- and \reflectbox{S}-shaped structures complement each other forming a very regular structure in the shape of an "8". This structure is surrounded by an even larger 8-shaped nebula seen in the continuum dust emission, as illustrated in Fig.\,\ref{fig-RGB}. 

Although the molecular nebula displays a high level of a point symmetry, it also shows irregularities. For example, for several species at the northern tip of the S-shaped regions, an extra weak component (a "bump") is present, see HNC or CH$_3$OH in Fig.\,\ref{fig-gallery1}. It has a corresponding weaker feature in the southern  tip that is  apparent only in a few species, e.g. in H$_2$CO.

The most extended emission is observed in the low-$J$ lines of CO which stretches even beyond the continuum lobes, most readily around the waist of the bipolar structure. As noted in Sect.\,\ref{sec-obs}, CO emission may be still partially filtered out by the interferometers due to the missing short spacings. The fact that we only see CO emission beyond the continuum region indicates that the images of all other species are not significantly affected by interferometric filtering.

As shown in \cite{nature} \citep[see also][]{Eyres}, the molecular nebula is embedded within a much larger nebula of optical emission of the extent of 71\arcsec\ \citep{hajduk2007}. The brightest parts of this optical nebula are located in the center of the remnant and have spatial scales comparable to those observed with ALMA and SMA. Projected on the sky plane, the brightest clump, which we call the northern jet and which was analyzed in detail in \citet{xshooter}, overlaps spatially with part of the emission of the molecular northern lobe. Therein, the recombining plasma is seen mainly towards the western half of the lobe, where molecular ions dominate. Other optical clumps are located chiefly outside the molecular nebula: one is seen east of the northern lobe and in the direct vicinity of its western "wall"; another one is seen directly east of the southern molecular lobe (see Fig.\,2 in \citealt{nature}). Weak atomic emission forms a cocoon which surrounds the molecular emission. This morphology of the  atomic emission has remained unchanged over the past decades and was confirmed in Aug 2015, i.e. three years prior to the ALMA survey, by Hubble Space Telescope imaging (H. Bond, priv. comm.).

The morphology and kinematics of the nebula in three dimensions will be analyzed in a forthcoming dedicated paper.

\section{Radiative-transfer model of molecular emission}\label{sec-rad-transf}
To learn more about the molecular remnant, we performed an advanced radiative-transfer analysis of the observed molecular emission. 

\subsection{The modeling procedure}\label{sec-cassis}
We used the CASSIS software (version 5.0) to perform radiative-transfer modeling of the observed molecular transitions detected in the ALMA and SMA surveys. Our modeling procedures are similar to those used in KMT17, but we took an additional non-LTE approach after an initial LTE treatment and, very importantly,  were able to individually analyze different parts of the resolved molecular nebula. Whereas in KMT17 the excitation analysis was mainly used to get an overall overview of the molecular composition and to verify the line identifications, the main motivation here is to constrain the physical conditions and relative molecular abundances in the different regions.

\begin{figure}[!ht]
  \includegraphics[width=.99\columnwidth]{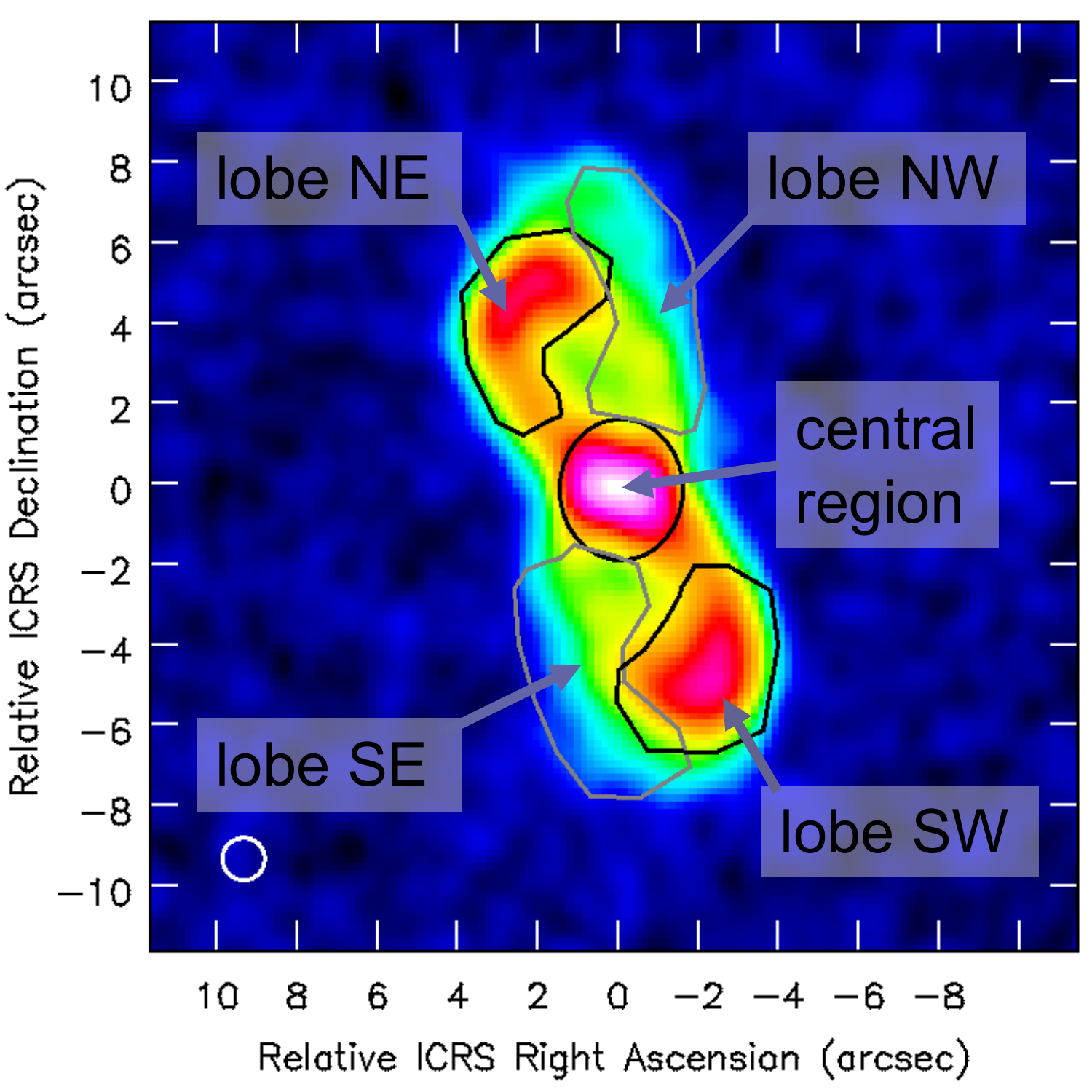}
\caption{Outlined are five regions (central, and the NE, SE, SW, NW lobes) defined for our excitation study. The background image shows the combined emission of multiple transitions and molecules.}\label{fig-regions}
\end{figure}

The molecular nebula was divided into five regions outlined in Fig.\,\ref{fig-regions}. They reflect the morphological differences in the most relevant species (Sect.\,\ref{sec-morph}). This spatial division also allowed us to better separate spatio-kinematic components of the gas and thus model them with a better accuracy than what was possible in the single-dish composite spectrum of KMT17. We call the regions hereafter: central region (the inner region within a diameter of about 3\farcs5), lobe NE, lobe SW, lobe NW, and lobe SE. In CASA, we restored the image cubes with a common beam size of 1\arcsec\ and for each of these regions extracted an averaged spectrum in brightness temperature units. 

The beam filling factor was set to 1.0 for almost all models. The actual beam filling factors are likely less than one but at this stage they could not be reliably and independently constrained for most molecules. One notable exception is SO$_2$ whose emission is more compact than the arbitrarily-defined area of the {\it central region}. We ran extra models adopting the observed size of the SO$_2$ emission in order to calculate its beam filling factor (see Fig.\,\ref{fig-SO2spectra}), but we present in this section the result with the filling factor of 1.0 to use consistently the same scaling of column densities for all species. Not accounting for the real beam filling factors affects chiefly the estimates of the optical depth and consequently may affect temperature constraints for species with very saturated emission.  

Each velocity component was represented by a Gaussian profile with line widths and line centers constrained in a fit independent of the excitation models. Our errors in physical gas quantities typically do not account for uncertainties in line positions, widths, and their deviations from idealized Gaussian shapes. In most cases, they are, however, negligible compared to other sources of uncertainty. Several molecules (e.g., CH$_3$CN or CN) required adding line position and width as free parameters because the kinematic parameters are strongly coupled to the excitation conditions owing to close blends of transitions from different energy levels or owing to a hyperfine structure.  

We first modeled all the observed species assuming they are excited under LTE conditions and in homogeneous isothermal gas, similar to what has been done for the entire source in KMT17. Only uncontaminated transitions were selected for the comparison with the modeled spectra. Physical parameters were identified within a regular grid of models through finding the $\chi^2$ minimum. In the simplest case, the grid free parameters were the excitation temperature and column density. More elaborate models included additionally the isotopic or spin form ratio (e.g., for CH$_3$OH or H$_2$CO), line position, or line-width as free parameters. Some molecules and locations required simultaneous simulation of two or even three overlapping velocity components. 
Despite the best effort, many of the LTE models turned out to be inadequate, i.e., the solutions were producing spectra that were in very poor agreement with the observed ones. Most often, LTE models failed in reproducing strong saturated lines from the lowest energy levels or converged at suspiciously low excitation temperatures and unrealistically high column densities. In other cases, like for CH$_3$OH or CH$_2$NH, some lines were not produced in the simulations at all, that is their fluxes were zero or nearly zero even though they are readily observed (this had been already noted for some species in KMT17). 

More successful was a non-LTE approach. We used CASSIS and its $\chi^2$ minimization procedures plugged into RADEX \citep{radex}.\footnote{\url{http://home.strw.leidenuniv.nl/~moldata/radex.html}} This allowed us to simulate spectra taking into account collisions of a given molecule with a dominant collisional partner. RADEX is a one-dimensional radiative-transfer code that  solves the statistical equilibrium equations with the escape probability parameter calculated for a uniform isothermal sphere.\footnote{cf. \url{http://home.strw.leidenuniv.nl/~moldata/radex_manual.pdf}} Most of the predictions for line intensities calculated with RADEX reproduce the  observed values better than calculated under the LTE assumption and imply slightly higher temperatures than LTE. One exception was H$_2$CS, whose emission was too weak to produce well-defined $\chi^2$ minima in our non-LTE analysis and for which we use LTE constraints. As a non-LTE tool, RADEX provides constraints on the {\it kinetic} temperature. For simplicity, we will refer hereafter to the all the forms of gas temperatures derived from an excitation analysis (kinetic and excitation) as simply ``temperature''.     

The collisional rate coefficients of the modeled molecules were taken from the LAMDA database\footnote{\url{http://home.strw.leidenuniv.nl/~moldata/}} \citep{lamda} and from CASSIS repositories. The tabulated rates were based on the following original sources: \citet{ratesSO2} and \citet{SO2-He} (SO$_2$); \citet{ratesSO} (SO); \citet{ratesSiS} (SiS and SiO); \citep{ratesHCN} (HCN and HNC); \citep{ratesHNCO} (HNCO); \citep{ratesCH3CN} (CH$_3$CN); \citep{ratesH2CO} (H$_2$CO and H$_2$CS); \citep{ratesCH3OH} (CH$_3$OH); \citep{ratesHC3N} (HC$_3$N); \citep{ratesCS} (CS); \citep{ratesHCOplus} (HCO$^+$ and N$_2$H$^+$); \citep{lique2009} (NO); \citep{CN} (CN); and \citep{ratesCO} (CO). For CH$_2$NH, the tabulated rates were acquired from A. Faure \citep[priv. com.; see also][]{Faure}. For AlF and $^{26}$AlF, we used the rates of collisions with para-H$_2$ from \citet{AlFcollisions}. Excitation of PN by para-H$_2$ was modeled using the data from \citet{PN}. For NS and CH$_3$NH$_2$, no collisional rate coefficients were available, thus leaving us with LTE constraints only. The lack of collisional rate coefficients, especially in low gas temperatures or in terms of the number of state-to-state transitions,  often strongly limited our modeling of some molecules.

We assume that throughout the entire molecular nebula H$_2$ is the main collisional partner for all molecules. For some of the molecules analyzed with RADEX, the H$_2$ collisional rates are re-scaled from the rates derived for collisions with He. For some other molecules, rates for either ortho or para versions of H$_2$ are available and in a few cases data for both spin forms of H$_2$ are tabulated. The observations in hand do not allow us to differentiate between the spin forms of H$_2$ and thus we often used rates of one of the H$_2$ spin forms to represent all forms of H$_2$. If formed at high temperatures (which is very likely in our case), H$_2$ should exist at an equilibrium ortho to para ratio of 3. The near-infrared X-shooter observations of excited H$_2$ in the northern lobe (Sect.\,\ref{sec:xshooter}) are consistent with this equilibrium ratio of the ortho and para forms. Unfortunately, it is unclear if the near-infrared emission comes from the same volume of gas that is bright in molecular mm-wave lines. 

The role of collisions with \ion{H}{I} and free electrons in excitation is currently unclear. Both types of particles are abundant in the remnant \citep{xshooter}.  \citet{Goldsmith2017} have studied the importance of the electrons on the excitation of high dipole moment molecules. For HCN, they note that the collisional rates for this process are $10^5$ times  higher than for collision with H$_2$ and find  electron excitation to be of practical importance if the H$_2$ density is  $\leq 10^{5.5}$ cm$^{-3}$ and the electron abundance $\ge 10^{-5}$. As discussed below, such conditions may pertain in large parts of the envelope, except for the central region. Nevertheless, collisions with electrons (and with atomic hydrogen) are not considered in our analysis mainly because of the 
general non-availability or rate coefficients for these processes. 
We \textit{do} consider helium, which is particularly abundant in the optical nebula. With a He/H abundance of 0.26 (or a mass ratio of 0.5), He may play as important a role in the excitation of the molecules as para-H$_2$. With these shortcomings in mind, the density of a collisional partner we obtain can be roughly thought of as the sum of densities of H$_2$ and He. 

In some models, we included rare isotopologues (e.g., $^{26}$AlF, $^{13}$CO, Si$^{18}$O, HN$^{13}$C, H$^{13}$CN, and H$^{13}$CO$^+$) to improve our coverage of $E_u$ and provide a correction for optical depth effects. A full analysis of isotopic ratios  and their spatial variations will be discussed in a dedicated paper. Collision rates were assumed to be the same for all isotopologues of a given molecule. Also, for an analysis of SiO emission, we included archival observations from the VLA (Sect.\,\ref{sec-obs}). 

The only external radiation taken into account in the radiative transfer was the cosmic microwave background (CMB) of 2.73\,K. Dust radiation within the entire nebula is negligible for the radiative transfer in mm lines discussed here \citep[c.f.][]{ncrit}. For instance, for the central region, the ALMA band 3 continuum flux density is only 160\,mJy
and has no practical effect on the analyzed lines, independent of whether the radiation is internal or external. 

If for a given molecule too few lines were observed or the emission was too weak in a given region, that molecule, even when considered as detected, was dropped from the analysis.


\subsection{Kinematics of the remnant}
\begin{figure*}[!ht]\centering
  \includegraphics[width=0.93\columnwidth,page=3]{plotSpectraAll.pdf}
  \includegraphics[width=0.93\columnwidth,page=4]{plotSpectraAll.pdf}
  \includegraphics[width=0.93\columnwidth,page=1]{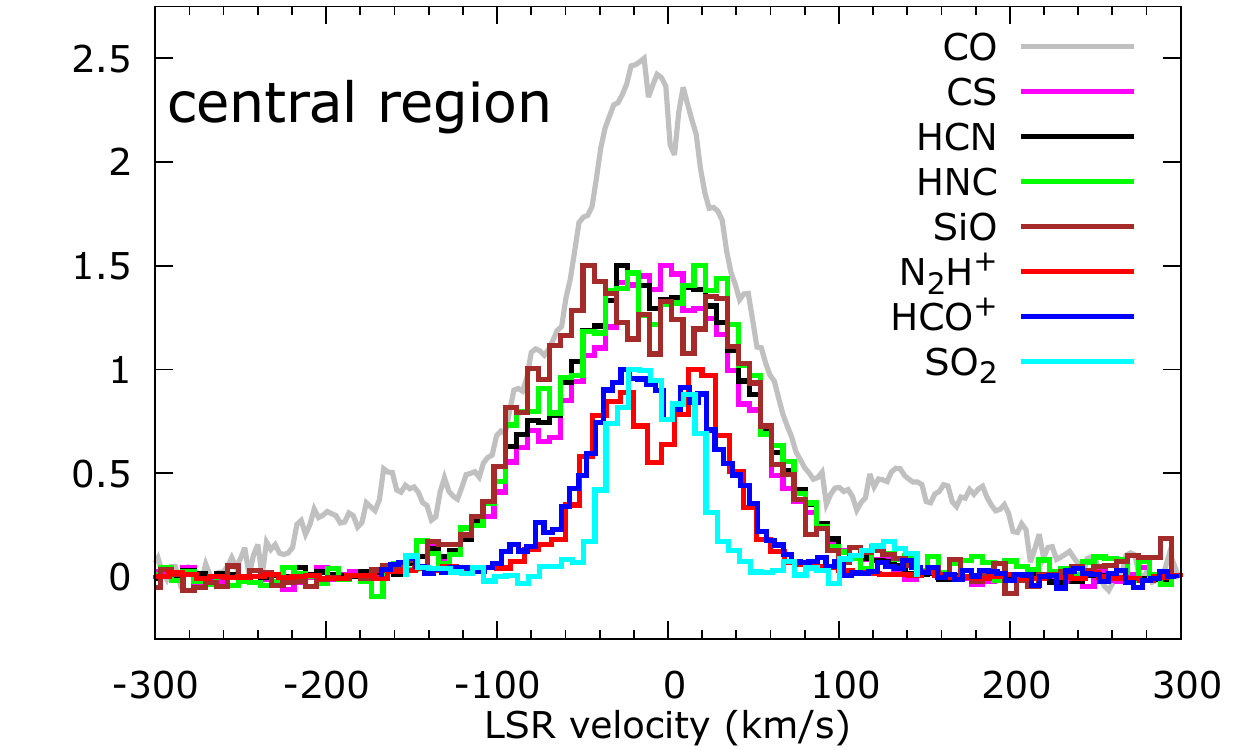}\\
  \includegraphics[width=0.93\columnwidth,page=6]{plotSpectraAll.pdf}
  \includegraphics[width=0.93\columnwidth,page=5]{plotSpectraAll.pdf}
\caption{Spectra of selected species for five regions of the remnant. The spectra represent a weighted mean of multiple transitions of the species and its isotopologues. Some of the lines are optically thick (see text). The intensity scale is arbitrary.}\label{fig-regions-spectra}
\end{figure*}

Sample spectra of the analyzed regions, shown in Fig.\,\ref{fig-regions-spectra}, illustrate the complex kinematics of the remnant. 
The systemic LSR velocity of CK\,Vul found by KMT17 is of about --10\,\kms\ (corresponding to a heliocentric value of 8\,\kms) and hereafter we refer to the observed absolute LSR velocities. 

The central region displays both the narrowest (SO$_2$) and broadest (CO) line profiles present in the remnant. The narrow lines of molecular ions and of SO$_2$ show line splitting suggestive of at least two slow-moving ($\lesssim$20\,\kms) spectrally blended sub-components. The emission of CO, on the other hand, shows three separate components. Two are fast, >100\,\kms, and seen exclusively in this molecule (Fig.\,\ref{fig-regions-spectra}). The central region contains several spatio-kinematic components  which may have different excitation conditions and compositions. 
To simplify our excitation analysis, we considered only one component in the central region 
at LSR velocities between about --120 and 110\,\kms.

The regions defined within the lobes also show multiple peaks in their spectra. In general, the strongest emission is redshifted in the northern lobe and blueshifted in the southern lobe but all regions have both blue- and redshifted emission. As in the total intensity maps, an interesting pattern can be seen when comparing the emission of ions with emission of most neutral species: where the strongest neutral component is redshifted, ions peak at symmetrically blueshifted velocities and vice versa, as if ions were avoiding the regions with the highest molecular diversity. Admittedly, though, the primary and secondary peaks of ions are not very different. To ease our analysis, we define the \emph{primary} component of each lobe as the one that is the strongest in neutral species: for velocities higher than about --60\,\kms\ in NE; lower than 40\,\kms\ in SW; higher than --20\,\kms\ in NW; and lower than --20\,\kms\ in SE. As \emph{secondary}, we define components with velocities close to those of the emission peaks of the ions. It is apparent in Fig.\,\ref{fig-regions-spectra} that fitting a single Gaussian spectral profile to these components is not fully adequate but it is required by our analysis tools.   

Although a full analysis will be performed in a dedicated paper, for the sake of further discussions, we propose here a simple interpretation of the complex kinematics in the lobes. By analogy to pre-planetary nebulae and the red nova V4332\,Sgr \citep{submmRN}, we hypothesize that the molecular remnant has a bipolar form, for example an hour-glass structure, whose long axis lies almost in the plane of the sky. We assume a velocity field in the form of a Hubble-flow, where the material moves the faster the farther away it is located from the center of the remnant. The opening angle of the structure is much wider than its inclination angle and the slight asymmetry in the velocities covered by emission in the northern and southern lobes indicates that the northern lobe is slightly closer to us than the southern one. In a lobe entirely filled by emitting gas, the extreme radial velocities seen in spectra would then correspond to the walls of the structure. For instance, in the SW lobe the sharp redshifted peaks of CO and ions (Fig.\,\ref{fig-regions-spectra}) may correspond to the far wall of the southern lobe, whereas the emission at moderate and negative velocities, which is dominated by neutral species, may correspond to the inner parts and the near wall of this lobe. The hypothesized hourglass structure of CK\,Vul is not fully filled by emitting gas.

\subsection{Excitation and abundances of central region}\label{sec-excit-central}
\begin{figure}[!ht]
  \includegraphics[width=\columnwidth]{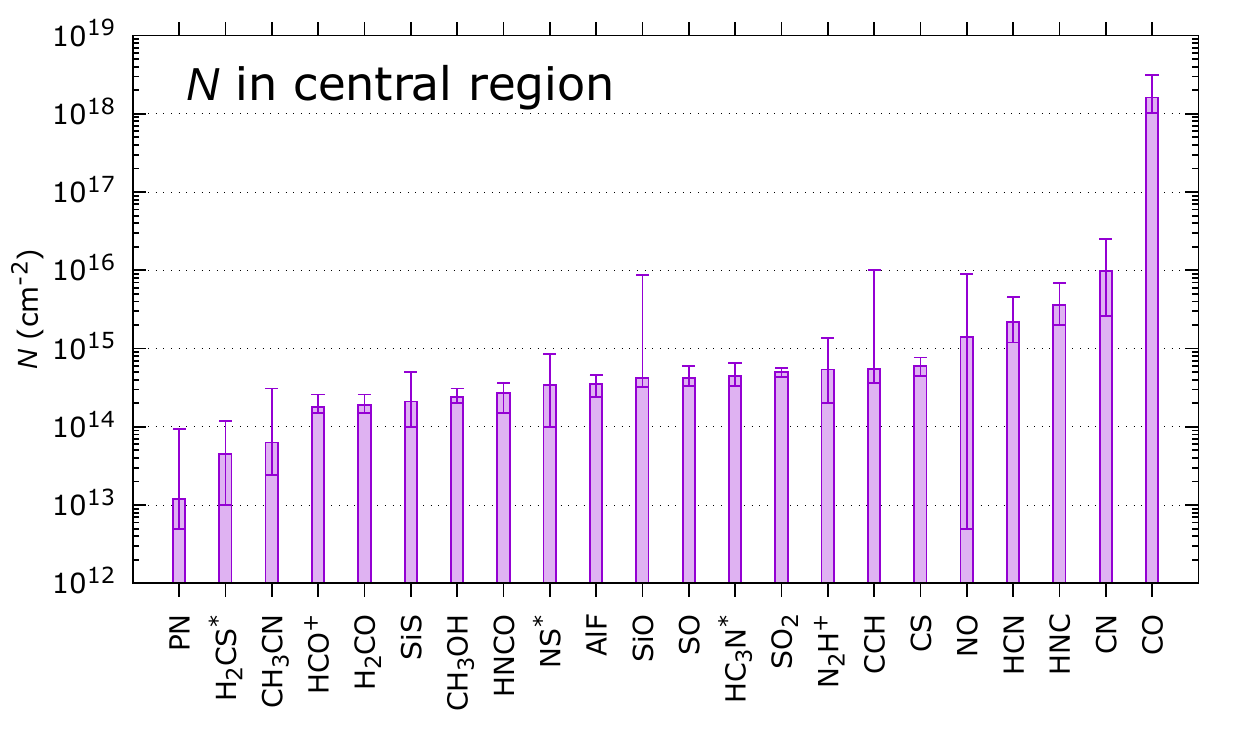}
  \includegraphics[width=\columnwidth]{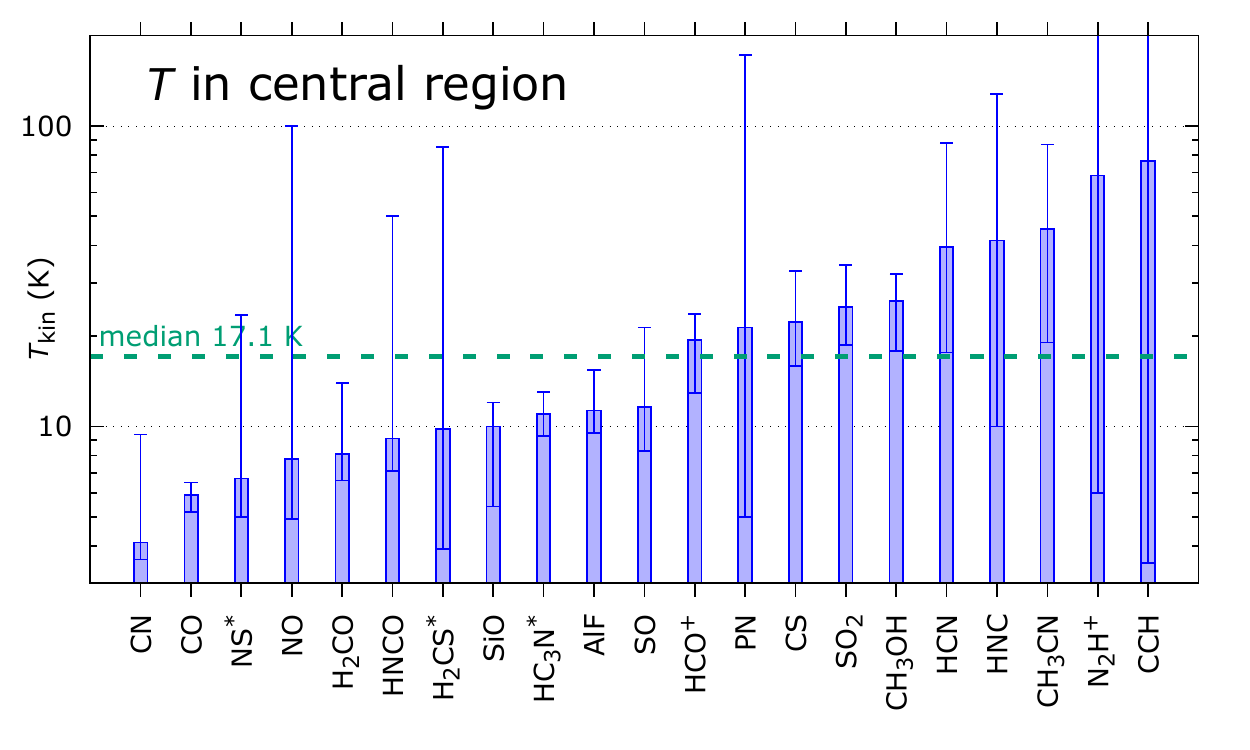}
\caption{Column densities (top) and temperatures (bottom) of molecules in the central region. The error-bars represent 3$\sigma$ uncertainties derived by the $\chi^2$ minimization procedure. The errors are largely underestimated. The column densities represent the main isotopologues of the species. A correction of about 20\% and 30\% is necessary for O- and C-bearing molecules to account for the presence of $^{17}$O and $^{13}$C, respectively. The vertical dashed line shows the median temperature of 17.1\,K. For molecules labeled with an asterisk, LTE results instead of non-LTE are presented.}\label{fig-centr35-NT}
\end{figure}

The column densities ($N$) and temperatures ($T_{\rm kin,ex}$) derived 
for the main velocity component in the central region are shown in Fig.\,\ref{fig-centr35-NT}. The errors are highly underestimated as they do not account for systematic errors, but only incorporate uncertainties from random noise and calibration errors. One of the systematic errors that is not accounted for is related to the break-down of the assumption of an isothermal gas. Another is a bias introduced by the selection of transitions taken into the fitting procedure. 


\paragraph{Temperatures} The median temperature for species we modeled with the highest confidence is 17.1\,K. Our results for CN, CO, and H$_2$CO suggest slightly lower temperatures which may be the result of too few transitions observed, severe saturation, modest S/N, or complex line profiles. Much higher temperatures are possible for several species, including CH$_3$CN, HCN, and HNC, but these solutions have very low confidence levels, too. 

 Many  molecules observed in the central region display line profiles that become narrower with increasing $E_u$. We interpret this as a sign of temperature variations coupled to a velocity field within the region. Models based on fits to a high number (>5) of transitions often under-represent fluxes of lines from the lowest energy levels. The region is thus not isothermal and the derived temperature can be thought of as a source-averaged temperature or one that is close to the temperature of the component with the largest volume. 
 

\paragraph{Column densities} We interpret the column densities shown in Fig.\,\ref{fig-centr35-NT} as representative of {\it relative} abundance patterns. A critical assessment of the uncertainties shows that errors are much larger than the error-bars representing formal 3$\sigma$ confidence levels shown in the figure. Nevertheless, it is safe to say that CO is the most abundant of the species observed at submm wavelengths. CN, HCN, and HNC (and perhaps NO, SiO, HC$_3$N, and CCH), after correcting for their respective isotopologues as constrained in \citet{kami-singledish}, have similar column densities and hence abundances. Their column densities at $\gtrsim$10$^{15}$\,cm$^{-2}$ are about two orders of magnitude lower than of CO. CN may be a few times more abundant than the rest of these species. The rest of the molecules analyzed in the central region ---  CS, N$_2$H$^+$, SO$_2$, CH$_3$CN, SO, AlF, NS, HNCO, CH$_3$OH, SiS, H$_2$CO, and HCO$^+$  --- have column densities on the order of $10^{14}$\,cm$^{-2}$. The sulfur-bearing molecules may be slightly more abundant than the rest of the species within this group. The least abundant molecules we were able to analyze in the central region at our sensitivity levels are PN and H$_2$CS with $N$ somewhat below $10^{14}$\,cm$^{-2}$, but even values lower than $10^{13}$\,cm$^{-2}$ are possible. No reliable constraints were possible for CH$_3$NH$_2$ but, assuming LTE, we found a conservative upper limit on its column density of $2 \times 10^{14}$\,cm$^{-2}$. 

\paragraph{Optical depth} The emission of almost all observed transitions of $^{12}$CO, $^{13}$CO, $^{12}$CN, SiO, H$^{12}$CN, H$^{13}$CN, HN$^{12}$C, and HN$^{13}$C is optically thick with $6 \gtrsim \tau \gtrsim 1$. The line saturation was partially taken into account in our modeling, but adds uncertainty to the model predictions for these molecules. The optical depths are likely larger than in our models, considering that the realistic beam billing factors are smaller than assumed, i.e., $<1$. It is likely that our analysis yields slightly underestimated column densities for the species with most severely saturated emission.

\paragraph{Total density} Constraints on the total density (H$_2$ and He) are rather poor for the central part of the nebula. Most spectra are consistent with a density of 10$^{4\pm1}$\,cm$^{-3}$ but models of AlF and H$_2$CO are indicative of an even higher density, $n \gtrsim 10^6$\,cm$^{-3}$. The central region may have a substructure with a considerable range of densities.

\begin{figure*}[!ht]\centering
  \includegraphics[width=0.63\columnwidth,page=1]{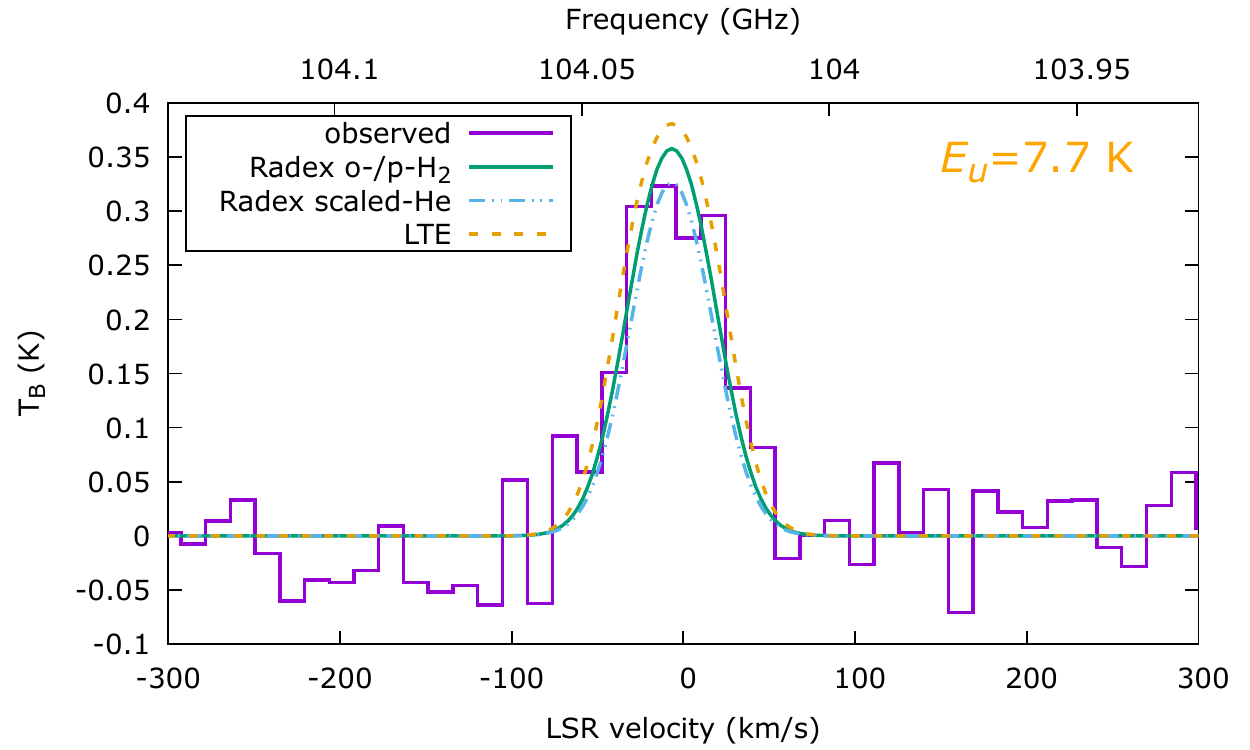}
  \includegraphics[width=0.63\columnwidth,page=2]{plotSO2panels.pdf}
  \includegraphics[width=0.63\columnwidth,page=3]{plotSO2panels.pdf}
  \includegraphics[width=0.63\columnwidth,page=4]{plotSO2panels.pdf}
  \includegraphics[width=0.63\columnwidth,page=5]{plotSO2panels.pdf}
  \includegraphics[width=0.63\columnwidth,page=7]{plotSO2panels.pdf}
  \includegraphics[width=0.63\columnwidth,page=8]{plotSO2panels.pdf}
  \includegraphics[width=0.63\columnwidth,page=10]{plotSO2panels.pdf}
  \includegraphics[width=0.63\columnwidth,page=11]{plotSO2panels.pdf}
  \includegraphics[width=0.63\columnwidth,page=12]{plotSO2panels.pdf}
  \includegraphics[width=0.63\columnwidth,page=13]{plotSO2panels.pdf}
  \includegraphics[width=0.63\columnwidth,page=14]{plotSO2panels.pdf}
  \includegraphics[width=0.63\columnwidth,page=15]{plotSO2panels.pdf}
  \includegraphics[width=0.63\columnwidth,page=16]{plotSO2panels.pdf}
  \includegraphics[width=0.63\columnwidth,page=18]{plotSO2panels.pdf}
  \includegraphics[width=0.63\columnwidth,page=19]{plotSO2panels.pdf}\hspace{0.05cm}
  \includegraphics[width=0.63\columnwidth,page=20]{plotSO2panels.pdf}\hspace{0.05cm}
  \includegraphics[width=0.63\columnwidth,page=21]{plotSO2panels.pdf}
\caption{Observed (histogram) and simulated spectra (lines) of selected SO$_2$ transitions in the central region of the remnant. The simulations include the best LTE model (dashed; $T_{\rm ex}$=7.7\,K and $N=8.3\times 10^{15}$\,cm$^{-2}$), the best RADEX model with collisional rates of ortho and para H$_2$ (solid; $T_{\rm kin}$=10.7\,K and $N=5.4\times 10^{15}$\,cm$^{-2}$), and the best RADEX model with collisional rates of H$_2$ scaled from calculations of collisions with He (dashed-dotted; $T_{\rm kin}$=19.4\,K and $N=5.1\times 10^{15}$\,cm$^{-2}$). Since both collision partners are abundant in CK\,Vul, we are not able to indicate which of the two models is more adequate. This illustrates one of limitations of the non-LTE excitation analysis.  Column densities are given for a beam filling factor of 0.41. Main species whose lines are contaminating the spectra of SO$_2$ are indicated. Panels are ordered with increasing $E_u$ from left to right and from top to bottom.}\label{fig-SO2spectra}
\end{figure*}

The radiative-transfer modeling of the central region turned out to be unexpectedly challenging and we noticed systematic effects which cannot be explained within our simple model. This is especially apparent for SO$_2$. With over twenty SO$_2$ transitions detected at a good S/N and over a hundred transitions covered within a broad range of $E_u$ values and frequencies, we expected to be able to constrain the physical conditions of the central region to a much better accuracy than presented here. In Fig.\,\ref{fig-SO2spectra}, we display selected spectra of SO$_2$ and the corresponding best-fit models generated under the assumption of LTE and using RADEX with H$_2$ or scaled-He collisional rates. The mismatch between the observations and the models, especially for lines weaker than 0.1\,K, is highly unsatisfactory, but cannot be explained by incorrect temperature, opacity effects, or line contamination. We ascribe the problem to an oversimplified treatment of the source thermal and density structure, and -- to a lesser degree -- to uncertainty in the collisional rates.  


\subsection{Excitation and abundances in the lobes}\label{sec-excit-lobes}
\begin{figure}[!ht]
  \includegraphics[width=.99\columnwidth]{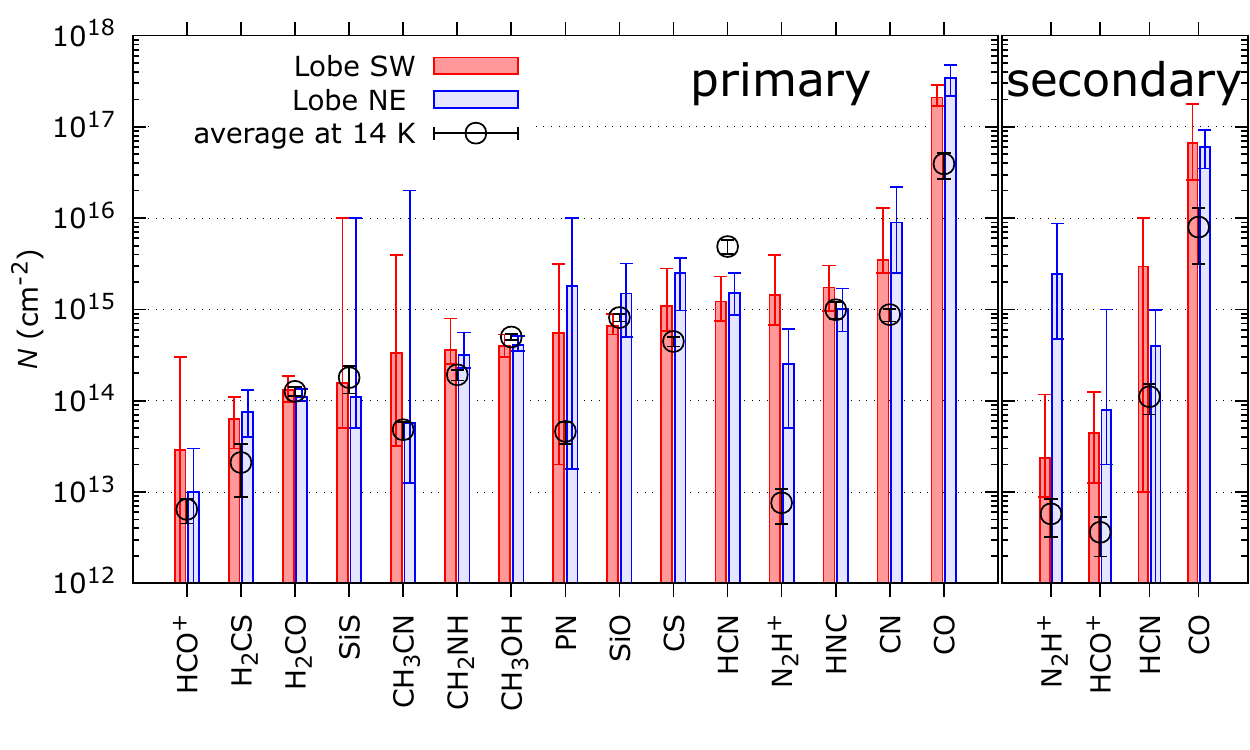}
  \includegraphics[width=.99\columnwidth]{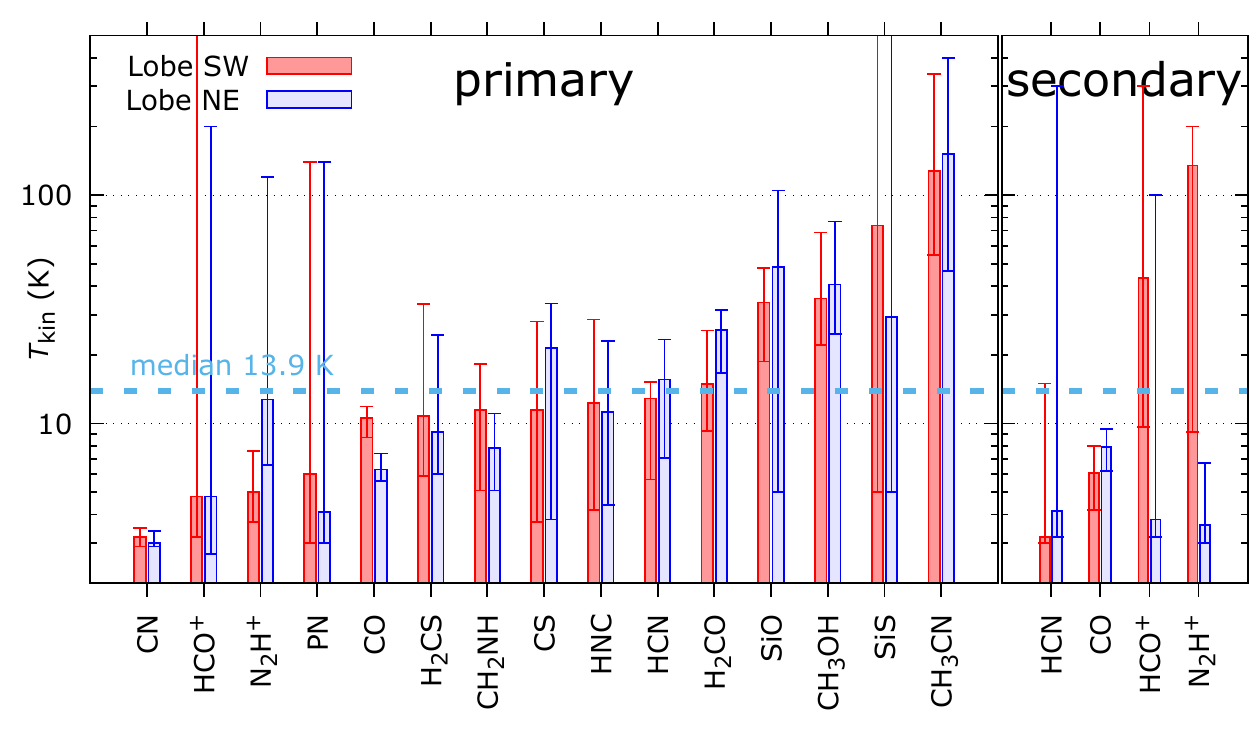}
\caption{Similar to Fig.\,\ref{fig-centr35-NT} but column densities (top) and temperatures (bottom) are shown for the SW and NE regions within the lobes. The primary (neutral) and secondary (ionized) components are defined in the text. The black circles with error-bars show molecular column densities calculated at a fixed temperature and density. }\label{fig-lobeSW-NE-NT}
\end{figure}

\begin{figure}[!ht]
  \includegraphics[width=.99\columnwidth]{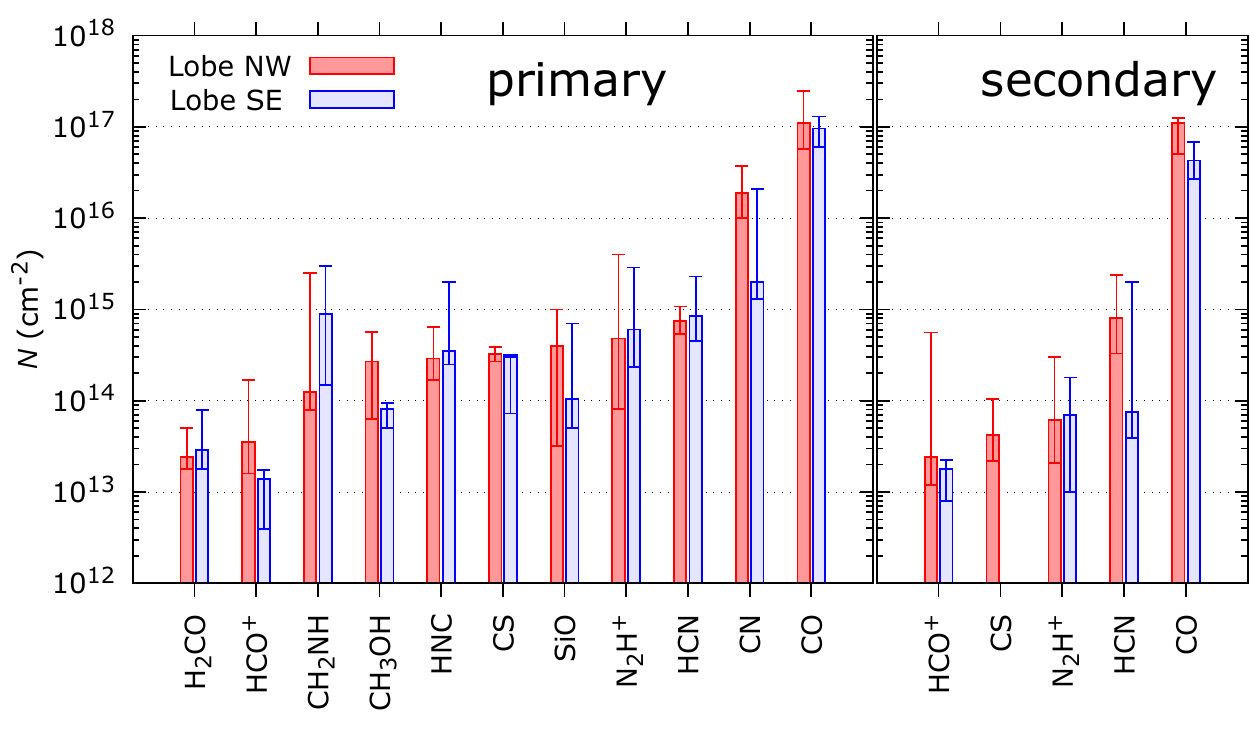}
  \includegraphics[width=.99\columnwidth]{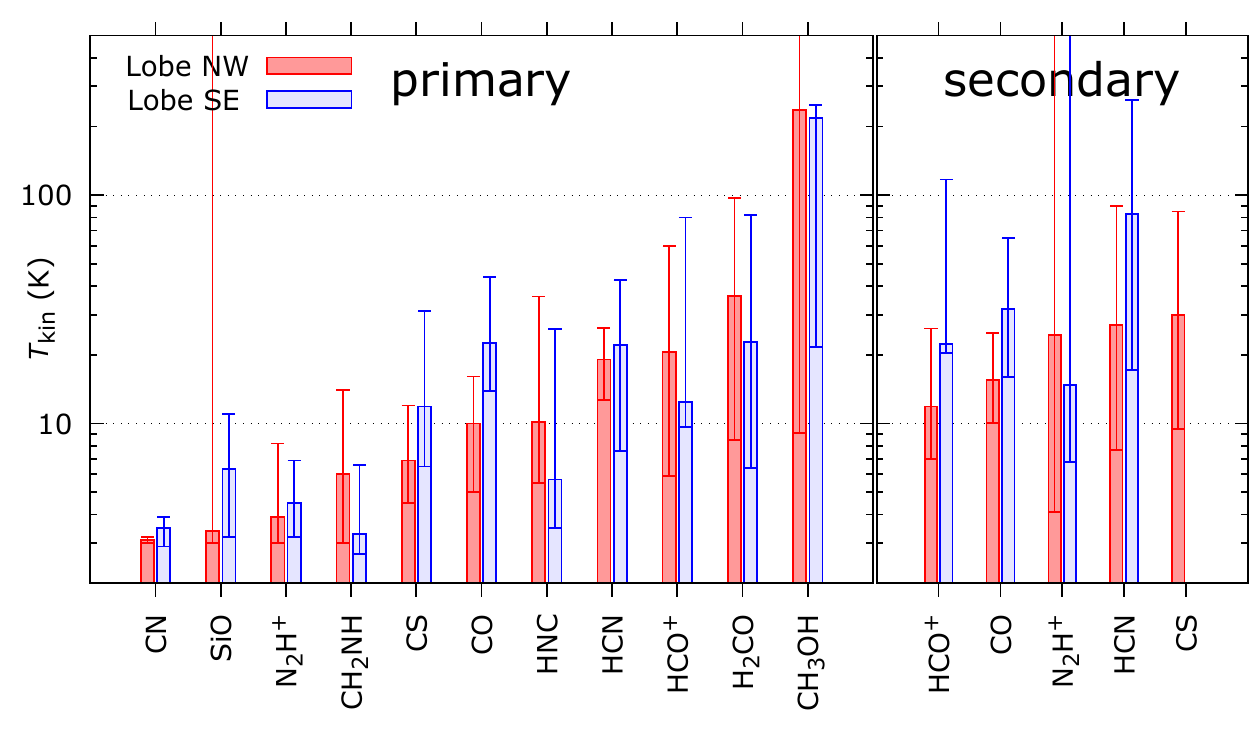}
\caption{Similar to Fig.\,\ref{fig-lobeSW-NE-NT}, but for the SE and NW regions.}\label{fig-lobeSE-NW-NT}
\end{figure}

Column densities and temperatures derived for the four regions within the lobes are presented in Figs.\,\ref{fig-lobeSW-NE-NT} and \ref{fig-lobeSE-NW-NT}. 
%
%
%
The assumption of isothermal gas worked better for the lobes than for the central region. 
A limiting factor in the analysis were complex line profiles, especially in SE and NW, which often were only very crudely approximated by a Gaussian.

\begin{figure*}[!ht]
  \includegraphics[width=0.63\columnwidth,page=1]{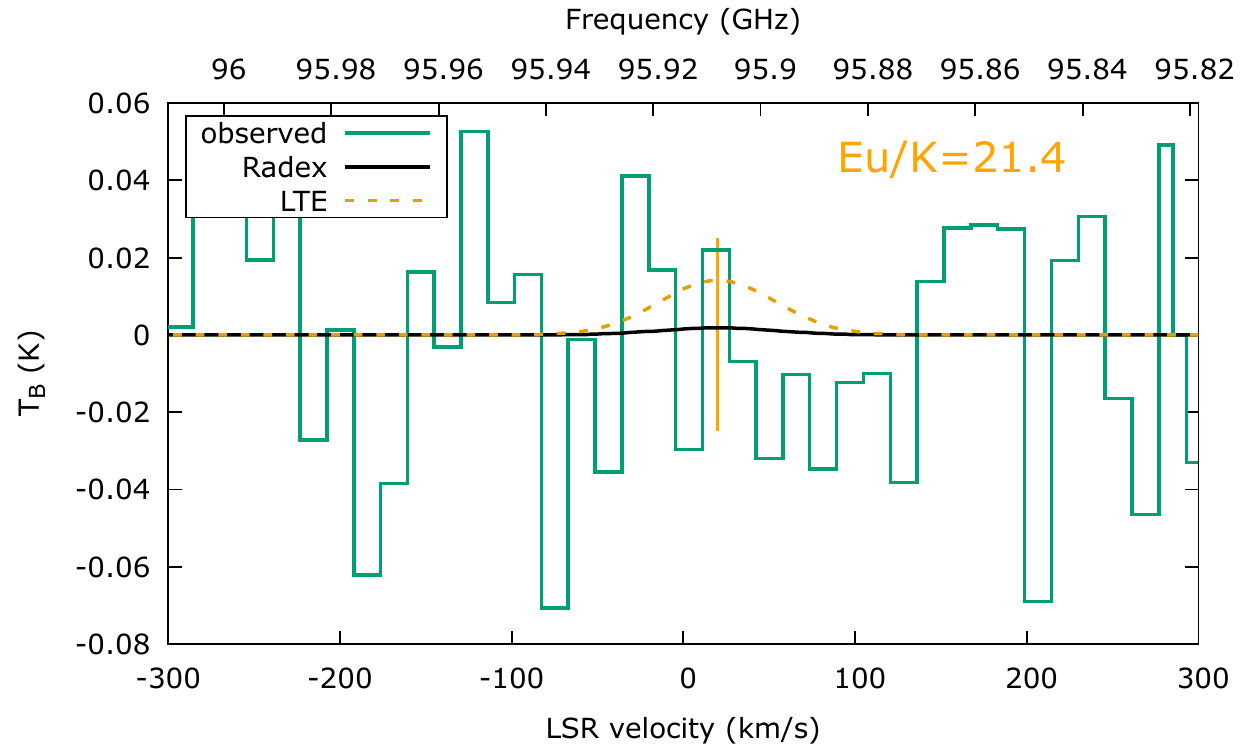}
  \includegraphics[width=0.63\columnwidth,page=2]{plotCH3OHpanels.pdf}
  \includegraphics[width=0.63\columnwidth,page=3]{plotCH3OHpanels.pdf}
  \includegraphics[width=0.63\columnwidth,page=4]{plotCH3OHpanels.pdf}
  \includegraphics[width=0.63\columnwidth,page=5]{plotCH3OHpanels.pdf}
  \includegraphics[width=0.63\columnwidth,page=6]{plotCH3OHpanels.pdf}
  \includegraphics[width=0.63\columnwidth,page=7]{plotCH3OHpanels.pdf}
  \includegraphics[width=0.63\columnwidth,page=8]{plotCH3OHpanels.pdf}
  \includegraphics[width=0.63\columnwidth,page=9]{plotCH3OHpanels.pdf}
  \includegraphics[width=0.63\columnwidth,page=10]{plotCH3OHpanels.pdf}
  \includegraphics[width=0.63\columnwidth,page=11]{plotCH3OHpanels.pdf}
  \includegraphics[width=0.63\columnwidth,page=12]{plotCH3OHpanels.pdf}
  \includegraphics[width=0.63\columnwidth,page=13]{plotCH3OHpanels.pdf}\hspace{0.65cm}
  \includegraphics[width=0.63\columnwidth,page=14]{plotCH3OHpanels.pdf}\hspace{0.65cm}
  \includegraphics[width=0.63\columnwidth,page=15]{plotCH3OHpanels.pdf}
\caption{Sample lines of CH$_3$OH in the observed (histogram) and simulated spectra of the NE lobe. The simulations under non-LTE conditions in RADEX (black solid line) were performed with the best-fit gas parameters: a kinetic temperature of 40.7\,K, column density of $4.1\times10^{14}$\,cm$^{-2}$, and H$_2$ density of $5.6\times10^4$\,cm$^{-3}$. The LTE simulations (dashed orange line) were performed for the same temperature and column density.}\label{fig-methanol-spectra}
\end{figure*}

Non-LTE effects in the lobes are particularly noticeable in the excitation of methanol and methanimine. Simple LTE models of CH$_2$NH over-predict fluxes of the $3_{1,2} \to 2_{1,1}$ transition near 199.8\,GHz for a wide range of physical conditions; but in RADEX simulations, the line is well reproduced, along with all other observed lines. For CH$_3$OH, best-fit LTE models point towards solutions with a very low temperature ($\sim$3.5\,K) and an enormous column density. However, the corresponding best-fit spectra do not satisfactorily reproduce most of the observed lines. In particular, several CH$_3$OH transitions, including those near 156.6 and 107.0\,GHz, were predicted to have detectable fluxes under LTE conditions, but our observations show weak absorption at these frequencies. As shown in Fig.\,\ref{fig-methanol-spectra}, RADEX models much better reproduce all the observed transitions and indeed even predict the weak absorption in the aforementioned transitions. The absorption is produced against the CMB and is a consequence of an anti-inversion in level populations. It is related to the structure of energy levels of CH$_3$OH and has been reported towards other astronomical sources, typically at lower radio frequencies \citep[e.g.,][]{walmsley1988,pandian}. In CK\,Vul, this non-LTE effect makes this molecule especially useful for constraining local densities of H$_2$ and He in the CH$_3$OH-bearing gas. Our models yield $n$(H$_2$) of $(1-20)\times10^4$\,cm$^{-3}$. Gas densities obtained for other molecules are consistent with those derived for methanol but have typically much larger uncertainties. There are indications that densities in the SE and NW regions are still an order of magnitude lower, $10^3$\,cm$^{-3}$, but at a very low confidence level.


The physical conditions and column densities of a given molecule derived for regions SW and NE, which trace the S-shaped structure as mentioned in Sect.~\ref{sec-morph}, are very similar. This is also true for regions SE and NW, which trace the \reflectbox{S}-shaped structure. Hence, we group the regions into two pairs, SW--NE and SE--NW, in the following discussions. 

\subsubsection{Lobes SW and NE}

The median temperature in the SW and NE lobes is about 13.9\,K, only slightly below 17.1\,K found in the central region. However, if our constraints on the excitation of CH$_3$OH and CH$_3$CN are correct, it is likely that parts of the lobes or some molecules are excited at higher temperatures of $\approx$50\,K (and even up to 400\,K at the 3$\sigma$ significance level). This result may be important for understanding the origin of these and other complex species in the lobes. Note, however, that CH$_2$NH, also a large species, is excited at a much lower temperature that is consistent with the median temperature of 13.9\,K. Our grid of models for CN and CO suggests somewhat lower temperatures for these two molecules. This is consistent with the emission from CO being more extended than that from most other species.
However, both CN and CO have very saturated lines, and therefore the lower temperature may be a result of a poor fit. 
Temperature constraints in the weaker secondary components of SW and NE are very uncertain and much more scarce, but the obtained results agree with the values found in the primary emission components.   


The emerging patterns of relative abundances in the primary components of the SW and NE regions are very similar to those found for the central region. The dominant species is still CO, which is followed by CN, HCN, HNC, CS (all C-bearing), and SiO molecules with abundances that are two orders of magnitude lower than CO. N$_2$H$^+$, SiS, CH$_3$CN, and PN may have similar abundances, but the uncertainties are large. Column densities a few times lower, $<$10$^{15}$\,cm$^{-2}$, characterize H$_2$CO, CH$_2$NH, and CH$_3$OH. The least abundant species observed are H$_2$CS and HCO$^+$, but we can barely place any constraints on the column density of the latter. Some species present in the center of the remnant are missing in the lobes, in particular SO, SO$_2$, and AlF. For SO$_2$, if we assume a temperature of 13.9\,K, an H$_2$ density of 10$^{4}$--10$^{5}$\,cm$^{-3}$, and a FWHM of 95\,\kms, we calculate a conservative upper limit on its column density in the SW lobe of 10$^{14}$\,cm$^{-2}$ [or 10$^{13}$\,cm$^{-2}$ for $n$(H$_2$)>10$^6$\,cm$^{-3}$]. We derive a similar upper limit on the column density of SO.

Emission of the majority of neutral species is much weaker in the secondary component of both regions. We were able to model a few neutral species in this component, but their temperatures are very poorly constrained, affecting our column density determinations. Many species have column densities well below 10$^{14}$\,cm$^{-2}$. Ionic species seem to have slightly higher abundances in the secondary components than in the main ones. The average HCO$^+$ to CO column density ratio for both regions is twice higher in the secondary component than in the primary (but both ratios are consistent within 3$\sigma$ uncertainties). Although our measurements are very uncertain, we conclude that the secondary components have a much lower content of most neutral molecules at nearly the same abundances of ionic species. 


The strongest emission lines observed in the SW and NE lobes, although typically weaker than in the central region, are still severely saturated. While our modeling procedure takes saturation into account, the applied correction may be insufficient for the strongest lines. For instance, our models of SiO and $^{29}$SiO converge at a column-density ratio of about 9, while the ratio should be close to the solar value of 19.7 without Si processing \citep[as expected in this object;][see also KMT17]{kamiAlF}. Future improvements on the modeling approach should implement a more realistic kinematic structure of the nebula to study the excitation and abundances in these regions at better accuracy. 

\subsubsection{Lobes SE and NW}
In the SE and NW regions, most species only show very weak emission, often with irregular line profiles. In consequence, only few species could be fully analyzed. The derived temperatures and column densities, shown in Fig.\,\ref{fig-lobeSE-NW-NT}, are consistent with the constraints derived for the same molecules in the brighter SW and NE regions. Although HCO$^+$ and N$_2$H$^+$ are the only species that are significantly brighter in SE/NW than in the SW/NE lobes, their derived column densities, both for the primary and secondary components, are consistent throughout the lobes within the large uncertainties. 


\subsubsection{A generalized model of the lobes}
 Overall, we found that the temperature does not vary drastically throughout the different spatio-kinematic components of the lobes, although there is a considerable temperature gradient in the central region. To investigate whether relative molecular abundances can be studied more clearly, we repeated the analysis of the brighter SW and NE lobes by fixing the temperature at 14\,K and the total density at $10^5$\,cm$^{-3}$ for all molecules. Column densities were averaged for both lobes and are presented with black symbols in Fig.\,\ref{fig-lobeSW-NE-NT}. They suggest that the detected ions have similar column densities in the primary and secondary components whereas neutral species are about one order of magnitude less abundant in the secondary components compared to the main ones. Within the primary component, the molecular ions are among the least abundant detected species.

It is therefore reasonable to conclude that there is a chemical differentiation between the primary and secondary components: the near and far sides of each of the lobes are chemically different. However the conditions and composition for the region pairs in the north and south lobes are essentially the same. The difference in morphological types displayed by molecular emission in Fig.\,\ref{fig-gallery1} is mainly a consequence of chemistry, not excitation. 
 The chemistry of the central region must be different from that in the lobes, with the central parts being much richer in O-bearing molecules. 

The results of our modeling can be compared to those presented in KMT17. The analysis of the interferometric data provided us with basic parameters within different parts of the nebula whereas the single-dish data of KMT17 allowed us to derive basic parameters for the entire source. The ``integrated'' column densities reported in KMT17 cannot be directly compared, but they are in reasonable quantitative agreement with our results. Temperatures derived in the LTE analysis in KMT17 are generally lower than the values presented here. This is chiefly due to the non-LTE effects that are accounted for in the current study.

The overall characteristics of the molecular remnant, including the range of excitation temperatures, the range of total densities, chemical composition and its complexity make it very similar to the prototypical envelope of OH231.8+4.2 and some pre-planetary nebulae. We highlighted some of these similarities in KMT17, but our improved analysis and the updated view on OH231.8+4.2 based on the new ALMA imagery of \citet{oh231} reinforce the resemblance. The purely phenomenological similarities may be related to similar dynamical histories of these objects.

\section{Equilibrium chemistry and elemental abundances}\label{sec-CE}
One would desire to use the observed molecular abundances to infer elemental abundances in CK\,Vul. However, currently no astrochemical framework exists for circumstellar chemistry of eruptive objects, whether driven by shocks or otherwise \citep[but see, e.g.,][]{andrew}. Historically it has been instructive to compare abundances of circumstellar molecules to predictions of equilibrium chemistry, i.e., chemistry in thermal equilibrium (TE). In particular, such TE models, which have a very long tradition \citep{russell1934,tsuji1964}, have been used to explain abundances seen around cool evolved stars and to differentiate between carbon- and oxygen-rich asymptotic giant branch (AGB) stars \citep{tsuji1973}. Admittedly, TE chemistry is known to fail in explaining many of the chemical characteristics of circumstellar envelopes \citep{cherchneff} but nevertheless the TE approach is still used in the field \citep{agundez} and might be expected to provide first clues.

We used the chemical equilibrium code GGChem of \cite{woitke} to simulate molecular gas abundances for different physical conditions. Although the code has dust condensation implemented, we do not use it in our models because it would complicate the model too much. The production of solids may alter the gas abundances of some species but the condensation sequence is very poorly understood, even in ordinary AGB stars. GGchem is lacking several molecules of our interest, e.g., N$_2$H$^+$ and HNCO. Molecular ions have insignificant abundances in TE  and thus HCO$^+$ was omitted in our analysis.   

The GGchem code is only applicable to temperatures higher than 100\,K. We derived much lower temperatures for CK\,Vul, but these do not necessarily correspond to the temperatures in which these molecules formed. One possibility is that most of the molecules we observe today formed when the remnant was smaller, denser, and warmer. At some point after the eruption, fast adiabatic expansion might have stopped chemical reactions, leaving molecular abundances almost unchanged while the expansion continued. This hypothetical evolution of molecular abundances is, starting with \cite{mccabe1979},  sometimes called ``frozen chemistry'' in the context of AGB stars. Within this working hypothesis, we found that the observed relative abundances of the four or five major species are qualitatively close to these expected at TE for temperatures of $\sim$700--2500\,K and densities of the order of 10$^{10}$--10$^{14}$\,cm$^{-3}$. Indeed, such densities and temperatures would be expected if the observed gas seen today at 10$^{5}$\,cm$^{-3}$ and 14\,K has adiabatically expanded from a much smaller remnant, assuming an adiabatic exponent of $\gamma$=1.4. 

The comparison of the observed abundances to TE predictions is not straightforward because even the older and warmer remnant was not isothermal at the time of the hypothetical ``freeze out''. We compared our derived abundances in the lobes and in the central region to \emph{peak} molecular abundances predicted in TE in a \emph{range} of temperatures from $\sim$700--2500\,K. We first searched with a $\chi^2$ test for the set of CNO elemental abundances that would best reproduce the observed molecular abundances relative to CO. Our grid was constructed for CNO abundances every 0.1 or 0.2\,dex. For all other metals we assumed solar abundances, except for sulfur which was found to be slightly less abundant in CK\,Vul (S=6.76) than in the Sun (S=7.20) \citep{xshooter}. The [C, N, O] abundances, expressed on the scale where the abundance of hydrogen is 12.0, are [7.8, 9.2, 7.6] in the lobes and [10.0, 8.0, 9.6] in the central region. In linear relative scale, these correspond to (1.6 : 39.8 : 1.0) in the lobes and (100.0 : 1.0 : 39.8) in the central region. 
For reference, the solar values are [8.4, 7.8, 8.7] or (4.0 : 1.0 : 7.9). We are more interested here in the relative CNO abundances than in their absolute values. Although highly uncertain, these derived abundances suggest very high abundance of N compared to C and O in the lobes, as we suggested in KMT17 based only on the inventory of the observed molecules; in the central region, N is the least abundant among the CNO elements while C and O have comparable abundances. Extra models where the abundance of S was an additional free parameter suggest an even $\sim$2 times lower sulfur content than in Tylenda et al., of S=6.4. Based on the abundances of AlF, we obtained also the abundance of fluorine at F=5.0, which hints at an enhancement in CK\,Vul compared to the solar composition (F=4.89). If confirmed, it could provide additional evidence (cf. KMT17) for nuclear processing of the remnant. 

It is difficult to assess uncertainties of such a brute chemical analysis, in particular because equilibrium conditions are unlikely to have occurred at any stage of the remnant's evolution. We were surprised, however, that this procedure resulted in relatively reasonable results. For instance, N and O abundances within the lobes are in reasonable agreement (i.e., within an order of magnitude) with those derived from the optical nebular "jet" seen toward lobe NE, where [N, O] are [8.2, 8.1] \citep{xshooter}. To investigate the scale of the uncertainties, we tested the models against observations using Monte Carlo Markov Chain (MCMC) simulations. We used the {\it emcee} implementation \citep{emcee} of the MCMC method and compared observed abundances with their 1$\sigma$ uncertainties to these expected in TE. We used 32 starting sets of parameters, 64 walkers, and draw 4000 samples. As priors, we considered only solutions with 5.0<C<12.0 and 3.5<N<12.0 and 4.0<O<12.0 and the sum of the three abundances <27. 
Taking the most likely value as the 50th percentile and uncertainties as 16th to 84th percentile, for the central region we obtain
$
\mathrm{C} = 9.2_{-1.3}^{1.0}, 
\mathrm{N} = 6.9_{-0.3}^{0.6}, \mathrm{and}~
\mathrm{O} = 6.3_{-0.1}^{0.1}.
$
With these large (1$\sigma$-like) errors, the derived abundances do not provide strong constraints on the relative CNO abundances, but it is likely that C is more abundant than N and O in the central region, with (C : N : O)=(794.0 : 4.0 : 1.0).

%

A more advanced analysis of the derived molecular abundances awaits chemical models that take into account high-temperature reactions and shock-induced effects. 



\section{Shocked molecular gas}\label{sec-shocks}
\begin{figure*}[!ht]
  \includegraphics[height=5.5cm]{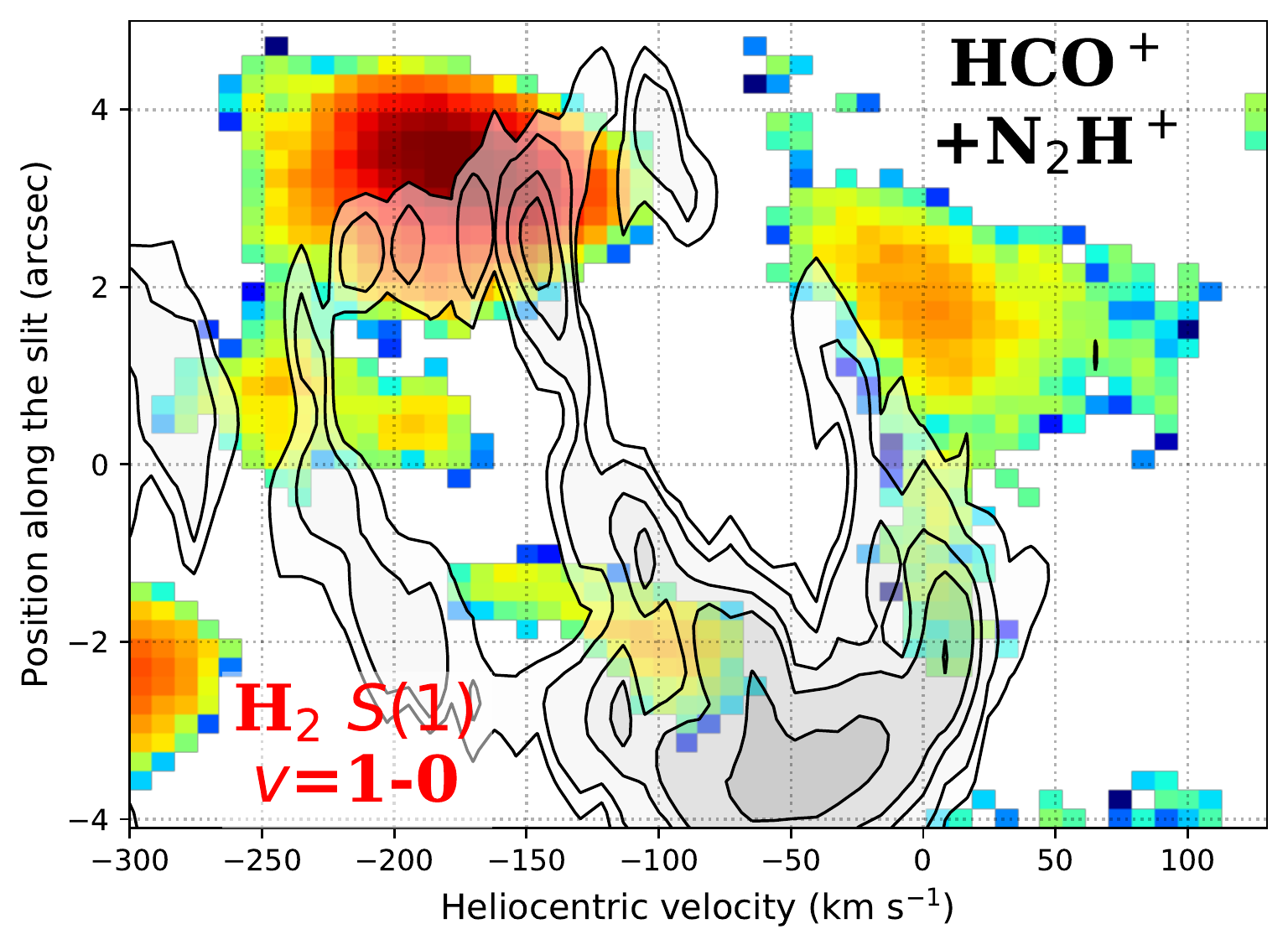}
  \includegraphics[height=5.5cm]{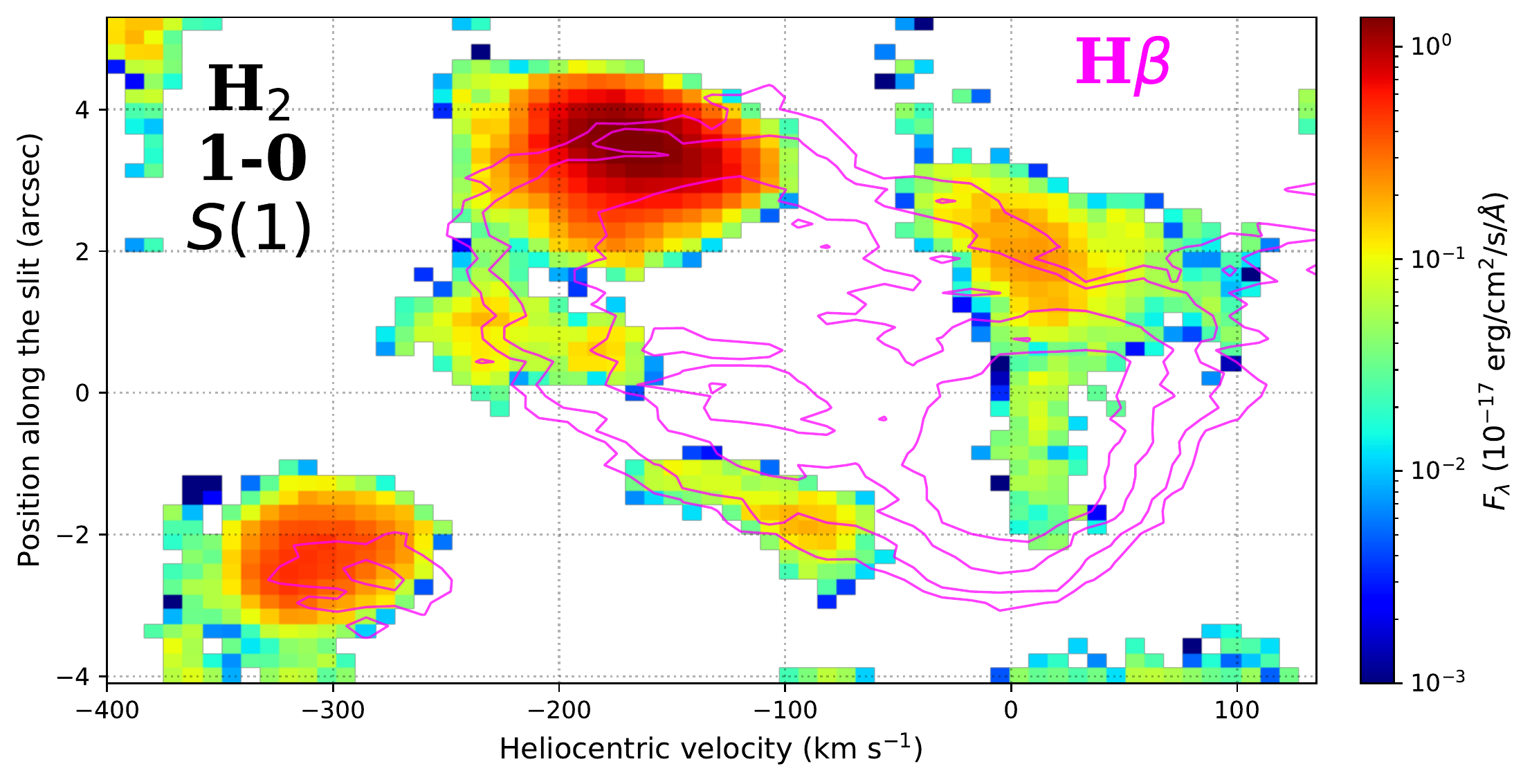}\\
  \includegraphics[height=5.5cm]{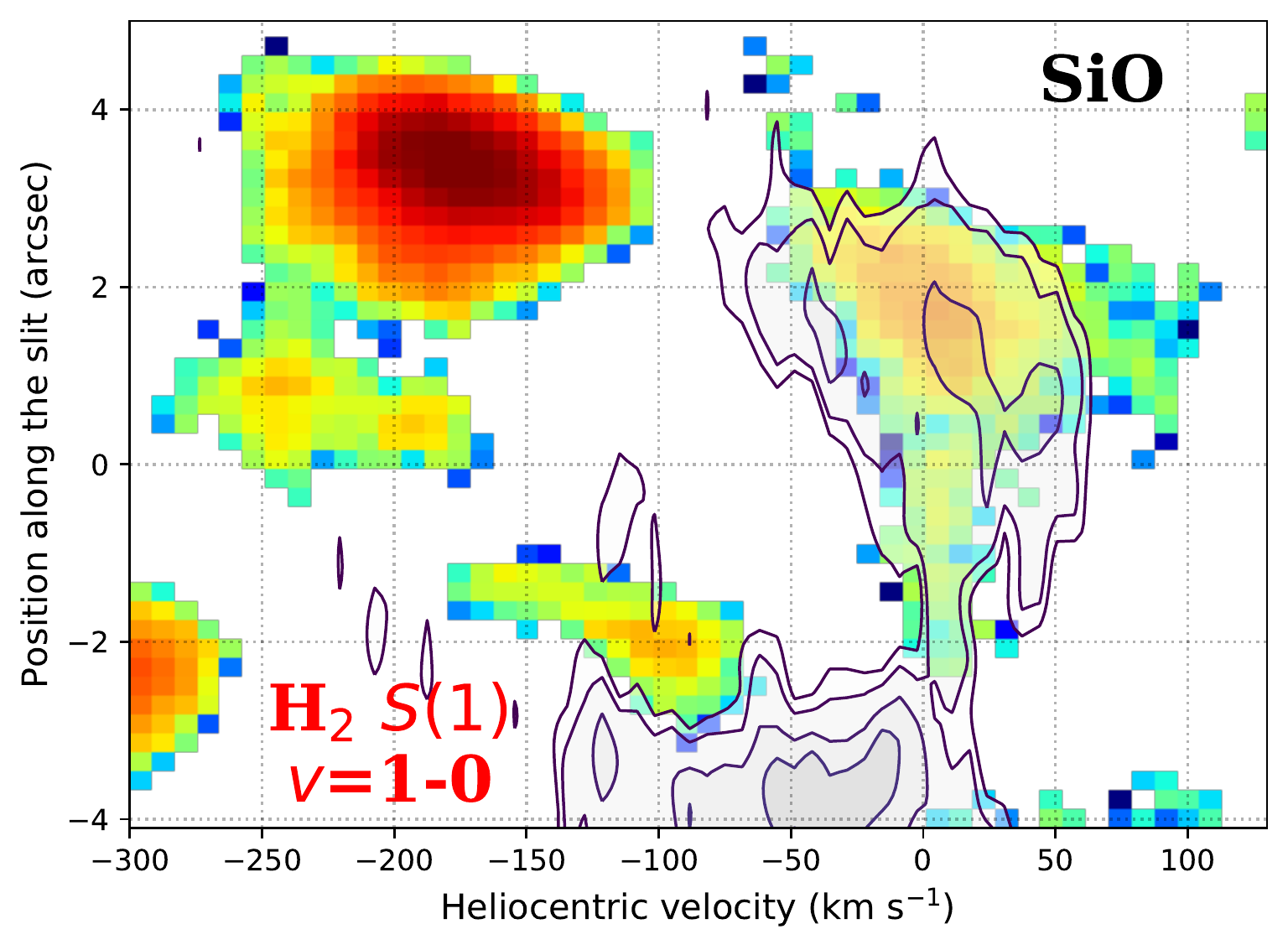}
  \includegraphics[width=0.095\textwidth]{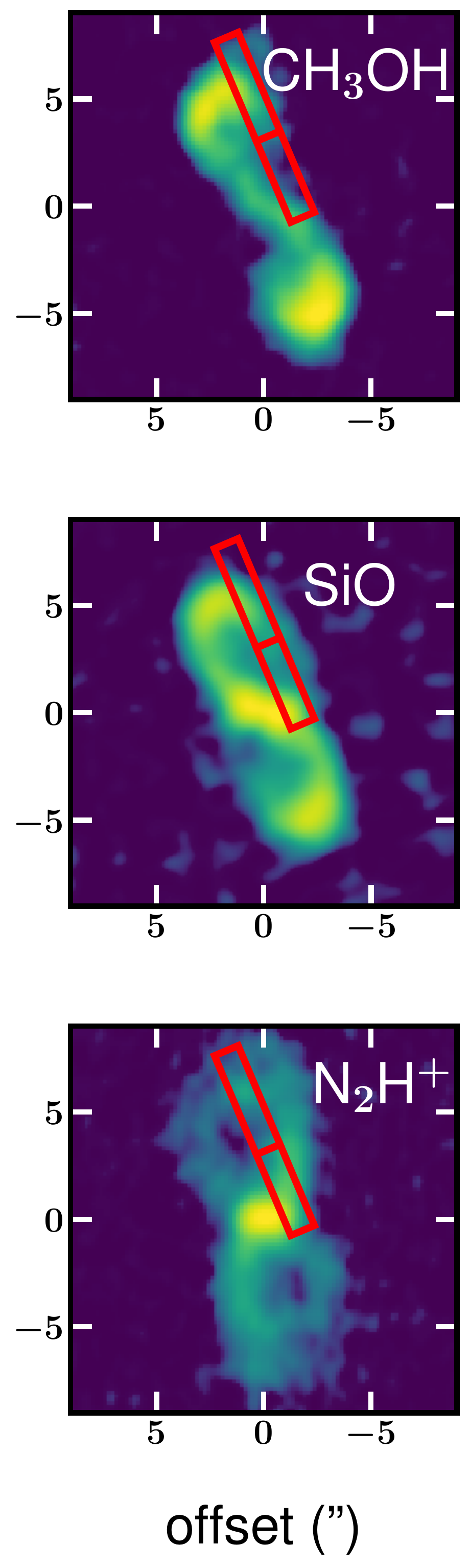}
  \includegraphics[height=5.5cm]{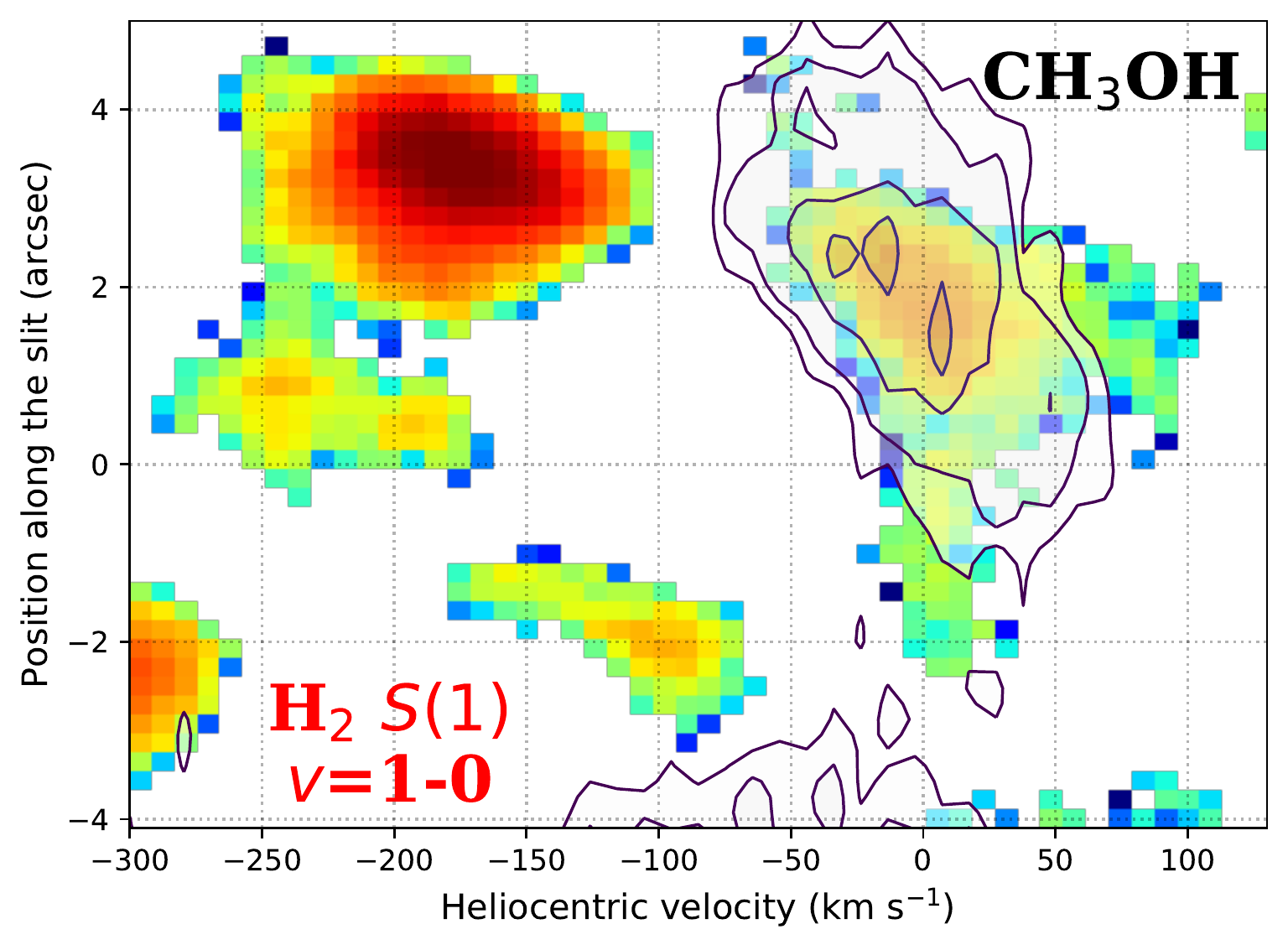}
\caption{Comparison of different phases of gas emission along the X-shooter slit. The main four corner panels show position-velocity diagrams with the emission of the H$_2$ $\varv$=1--0 $S(1)$ line shown in color. The emission of SiO (bottom left), CH$_3$OH (bottom right), and molecular ions (top left) was extracted from ALMA cubes representing average of multiple transitions of the species. (The emission at velocities $<$250\,\kms\ in the map of the ions comes from a nearby line of other species and should be ignored.) The emission in the optical Balmer H$\beta$ line (top right, magenta) and that of H$_2$ come directly from the  X-shooter observations \citep[Sect.\,\ref{sec:xshooter}; see also][]{xshooter}. Note the span of radial velocities is wider in the panel presenting data for H$\beta$. The contours are drawn at 3, 5, 10, 15, and 30, times the rms noise of the respective map. The inset in the middle of the bottom raw shows as a red rectangle the region probed in the position-velocity diagrams. it is overlaid on maps the emission of CH$_3$OH, SiO, and N$_2$H$^+$ (the same as in Fig.\,\ref{fig-gallery1}). The RA and Dec offsets in the inset are given in arcsec. Position along the slit is shown negative for regions south of the line dividing the rectangle. The position-velocity diagrams show a reduced range of offsets compared to the original extent of the X-shooter slit to avoid a contamination from a background star at offsets $<$--4\arcsec\ \citep[cf.][]{xshooter}.}\label{fig-H2}
\end{figure*}

Among the detected species, the presence of the complex organic molecules, such as methanol and methylamine, is the most puzzling. Their formation in CK\,Vul has been already discussed in KMT17. There, we could not uniquely identify which of several possible formation mechanisms has been active in CK\,Vul. The interferometric data presented here show that some of the complex species, including methanol, are almost exclusively seen in the lobes, which have been created in an energetic dynamical event associated with fast shocks. Shocks may be responsible for gas-phase methanol (and species alike) in two ways: ({\it i}) as already discussed by KMT17, through its direct creation in endothermic reactions in the hot gas phase whereby shocks provide the necessary energy for the reaction \citep{Hartquist} or ({\it ii}) with shocks providing means via sputtering or evaporation or desorption to release methanol from icy grain mantles to the gas phase. As mentioned, optical and NIR spectroscopy of regions overlapping with the molecular nebula indicates the presence of active shocks. They are manifested through atomic excitation and the presence of vibrationally-excited H$_2$ emission \citep[Sect.\,\ref{sec:xshooter};][]{xshooter}. Admittedly, in the position-velocity diagrams shown in Fig.\,\ref{fig-H2}, the different phases -- ionic, neutral, and complex molecules observed at mm wavelengths; vibrationally-excited H$_2$; and recombining plasma -- are not perfectly coincident and all partially overlap only in region B of \citet{xshooter}. That, however, may be expected depending on the range of shock velocities, viewing angle, and shock geometry \citep[][and references therein]{Hollenbach89,HM,DopitaSutherland}. Detailed modeling of shocks with CK\,Vul's peculiar elemental composition would be necessary to verify the possible shock structures. One problem to address is the cooling time of molecular gas seen at millimeter wavelengths down to 14\,K at the simultaneous presence of hot plasma at an electron temperature of 10--15\,kK and H$_2$ gas of an excitation (vibrational) temperature of $\approx$2600\,K. 

Whereas methanol emission is commonly found toward young stars and star-forming regions, this molecule is hardly ever observed in objects that are in advanced stages of stellar evolution. In recent years, however, evidence has been obtained for the presence of CH$_3$OH in a diverse  but yet small sample, of evolutionary-advanced stars, all of which are not ordinary AGB stars.  \cite{IRC10420surv} tentatively found weak signatures of methanol in the emission spectrum of the yellow hypergiant IRC+10420, a nitrogen-rich descendant of a massive star. \citet{olofsson} discovered methanol in the circumstellar environment of a low-mass star HD\,101584, which is thought to be in a post-AGB phase and some 600\,yr after a dynamical event (possibly a common-envelope ejection) that produced a bipolar envelope. Methanol is found only in the tips of jet-like features of HD\,101584, not much dissimilar to methanol emission in CK\,Vul. \cite{oh231} found methanol near the AGB star QX\,Pup close to the center of the pre-planetary nebula OH231.8+4.2, the ``Rotten Egg Nebula'', whose kinematic age is of a few hundred years. Also in this object methanol is found in bipolar lobes and is interpreted as a signature of circumstellar shocks. 
Most recently, methanol was tentatively found by \cite{morris} in the massive interacting binary $\eta$\,Car, which has a nitrogen-rich molecular environment \citep{loinard}, has undergone an eruption nearly two centuries ago, and has circumstellar material which has been repeatedly shocked in binary wind interactions. In the case of $\eta$\,Car, the detection was made in an absorption spectrum so that the overall distribution of methanol in the envelope is yet unknown. 
All these methanol-hosting objects (omitting perhaps the uncertain case of IRC+10420) experienced a violent event centuries ago and display a bipolar nebula which at some point of time was shocked. There are likely other objects with similar characteristics that have not been detected in methanol emission but no systematic searches exist. Based on the similarity of CK\,Vul to other eruptive sources with regard to methanol, we advocate shock-induced chemistry as the origin of the complex species in the molecular remnant of CK\,Vul. Although we focused our attention on the well-observed methanol, according to some authors other simpler species may be produced by shocks too, notably ammonia \citep[cf.][]{cherchneff,ammonia-shocks,wong}.

Examining our maps in the hope to find  clues about its formation, we note that in studies of the dense ISM methanol abundances are often compared to those of H$_2$CO, as this latter molecule is an intermediate step in the hydrogenation of CO to CH$_3$OH on icy dust grain mantles
\citep{charnley1997}. 
Our maps show two types of regions: (1) the central one where the abundances of both molecules seem to be nearly the same and (2) the lobes where methanol is about four times more abundant than H$_2$CO (Fig.\,\ref{fig-lobeSW-NE-NT}). Nearly equal abundances of the two molecules were found in HD\,101584. Star-forming regions have typically a broad range of the ratios, between 0.1--20 \citep[e.g.,][]{vanderTak2000}. Therefore, our results do not provide conclusive clues on the origin of the complex species in CK\,Vul, but may indicate that at least two formation or processing mechanisms are active.

Shocks play a major role during the outbursts of red novae and may be responsible for multiple peaks in their light curves \citep{MetzgerPejcha2017}. Theoretical \citep[e.g.,][]{pejcha16,pejcha17,morgan} and observational studies \citep[e.g.,][]{tylenda2011,TK2016} suggest that mass loss is taking place in the orbital plane of the binary years prior to the merger. This phase is typically short (a few tens of orbital periods) but the mass-loss history may last much longer in systems involving red giant stars \citep{MetzgerPejcha2017}, such as the progenitor of CK\,Vul postulated by \citet{kamiAlF}. This phase may lead to the accumulation of cool dust in the circumstellar environment, mainly in the orbital plane of the system. Provided the local densities are high enough, $>10^6$\,cm$^{-3}$ in CK\,Vul, ices can form on those grains (cf. KMT17). The total mass of this material is expected to be comparable to that ejected during the merger (i.e., the red-nova outburst). The merger-burst ejecta may interact with the material accumulated earlier in the orbital plane. Inevitable shocks can then release species locked in ice to the gas phase. An outflow associated with the merger event may next disperse the gas containing now species like methanol to large distances from the coalesced binary. \cite{MetzgerPejcha2017} speculate that multiple peaks in light curves of red novae may be related to sub-structures within the pre-merger envelope and indeed computer simulations of similar events often display spiral-like structures \citep{nandez,morgan,pejcha17}. In the case of CK\,Vul, which displayed at least three peaks in its light curve \citep{shara85}, this interpretation suggests that two large substructures existed within its envelope (the first peak is thought to originate as radiation from the outermost layer of the expanding merger ejecta). However, the formation of icy grains capable of methanol production in such a progenitor system with an RGB star and a circumbinary disk is very unlikely because complex species have never been observed in similar systems, not even in binaries with an AGB star, and the isotopic composition of the gas, at the $^{12}$CH$_3$OH to $^{13}$CH$_3$OH column density ratio of 3.8 (KMT17), indicates that methanol must have formed from material already highly processed by nuclear burning. \cite{kamiAlF} suggest that material from deep layers of the star, highly enhanced in $^{13}$C, was dispersed only after the plunge-in phase of the merger. The formation of the complex species on ice predating the 1670 eruption appears therefore to be problematic and other scenarios involving shocks should be explored. In particular, it is possible that some of the molecules were dispersed to the lobes more recently than 350\,yr ago, as proposed by \cite{Eyres}. 


\section{Molecule survival}\label{sec-surv}
We next discuss the time scales on which the observed molecules can survive in the remnant. While the eruption of CK\,Vul took place 350\,yr ago, it is possible that the molecular remnant is younger and produced by more recent activity. For selected molecules, we calculated the characteristic lifetimes, $t_0$, after which they are dissociated by the standard interstellar radiation field (ISRF). The number of molecules then scales with time as $\exp(-t/t_0)$. The lifetimes under the ISRF are listed in Col.\,(2) of Table\,\ref{tab-dissociation}. In the calculations, we used the photodissociation rates of \cite{dissociation}.\footnote{ \url{https://home.strw.leidenuniv.nl/~ewine/photo/index.html}} Unshielded, some of the molecules of our interest would be able to survive for decades but it is unlikely they would survive in the ISRF for centuries. Molecules most prone to photodestruction -- such as SO, SO$_2$, HC$_3$N, or CH$_3$CN -- are missing or are nearly undetectable in the lobes of the remnant. This may be suggestive of their destruction by an external radiation field. However, the gas in CK\,Vul is shielded by dust and in lines of the abundant molecules such as H$_2$ and CO. Shielding by dust is expected to be effective in the molecular lobes because they are surrounded by a dust cocoon seen in continuum emission by ALMA \citep[Figs.\,\ref{fig-gallery1}--\ref{fig-lowres};][]{kamiAlF,Eyres}. 

Analyzing optical emission lines, emanating from within the northern lobe, \cite{xshooter} found a reddening with $E_{B-V}\approx0.9$\,mag or $A_V \approx 2.8$\,mag, which we assume is mainly circumstellar in origin. \cite{hajduk2013} observed two stars shining through the southern lobe and found $A_V =$3.3--4.4\,mag with unknown contribution from the interstellar component. We recalculated the lifetimes of molecules assuming $A_V=3$\,mag for standard interstellar dust, and from larger and less opaque grains, at the gas-to-dust mass ratio of 124 \cite[see][for more details on the assumed dust properties]{dissociation}. We used shielding functions from Heays et al., which include effects in lines. Results are shown in Cols.\,(3)--(4) of Table\,\ref{tab-dissociation}. The presence of ISM grains makes it possible for the observed molecules to survive for a very long time, longer than 350\,yr. The lifetimes in the presence of the large grains considered by Heays et al. are typically a few times shorter than the age of the remnant. It is uncertain what kind of grains populate the dusty remnant of CK\,Vul, but given its anomalous elemental and molecular compositions and eruptive history, dust may have a peculiar chemical composition and size distribution. In such a case, the total to selective extinction law would also be different and the assumed $A_V$ may not be adequate. Nevertheless, if the molecules formed 350\,yr ago and are shielded by big grains, with the calculated lifetimes a considerable fraction of molecular species would survive, except perhaps for a few most fragile ones which indeed are almost absent in the lobes. We conclude that the lifetimes in Table\,\ref{tab-dissociation} that were calculated with an attenuated ISRF are consistent with the molecule formation 350\,yr ago or more recently. 

\begin{table*}
\caption{Characteristic molecular lifetimes (in years) in the lobes under different radiation fields and shielding conditions.}\label{tab-dissociation}
\centering
\begin{tabular}{c ccc cccc} 
\hline\hline
         &\multicolumn{3}{c}{external interstellar radiation field (ISRF)} & \multicolumn{4}{c}{hypothetical central white dwarf radiation field}\\
         & $A_V=0$ & $A_V=3^{\rm{mag}}$ & $A_V=3^{\rm{mag}}$ & $A_V=0$ & $A_V=3^{\rm{mag}}$ & $A_V=0$ & $A_V=3^{\rm{mag}}$ \\
Molecule & no grains& ISM grains& big grains& $T_{\rm eff}=20$\,kK& $T_{\rm eff}=20$\,kK& $T_{\rm eff}=60$\,kK&$T_{\rm eff}=60$\,kK\\
(1) & (2) & (3) & (4) & (5) & (6) & (7) & (8)\\
\hline\hline
HC$_3$N     & 4.5E+0	&7.9E+3	&5.3E+1	&1.27E-1& 	3.78E+2& 	7.53E-3& 	3.02E+1\\
OCS       	& 6.8E+0	&8.1E+3	&7.4E+1	&2.04E-1& 	3.74E+2& 	1.26E-2& 	3.26E+1\\
SO        	& 7.5E+0	&2.2E+4	&7.8E+1	&3.61E-1& 	1.93E+3& 	1.34E-2& 	1.10E+2\\
H$_2$S      & 1.0E+1	&2.1E+4	&1.1E+2	&3.24E-1& 	1.10E+3& 	1.62E-2& 	8.64E+1\\
CH$_3$CN    & 1.1E+1	&7.4E+4	&1.3E+2	&3.57E-1& 	5.03E+3& 	1.11E-2& 	2.96E+2\\
SO$_2$      & 1.3E+1	&4.0E+4	&1.4E+2	&5.32E-1& 	2.65E+3& 	1.84E-2& 	1.81E+2\\                  
HCN       	& 1.9E+1	&1.6E+5	&2.3E+2	&6.55E-1& 	1.13E+4& 	2.04E-2& 	6.59E+2\\
C$_2$H  	& 2.0E+1	&4.5E+4	&2.4E+2 &1.14E+0& 	2.56E+3& 	3.90E-2& 	1.77E+2\\
SiO       	& 2.0E+1	&4.3E+4	&2.4E+2	&6.09E-1& 	2.20E+3& 	2.81E-2& 	1.64E+2\\						
H$_2$CO     & 2.2E+1	&3.4E+4	&2.4E+2	&7.40E-1& 	1.69E+3& 	3.72E-2& 	1.41E+2\\
NH$_3$      & 2.2E+1	&4.1E+4	&2.0E+2	&7.43E-1& 	2.38E+3& 	3.37E-2& 	2.03E+2\\
CH$_3$OH    & 2.3E+1	&6.5E+4	&2.7E+2	&8.23E-1& 	3.23E+3& 	2.87E-2& 	2.51E+2\\
CH$_3$NH$_2$& 4.4E+1	&4.2E+4	&3.4E+2	&1.31E+0& 	2.27E+3& 	1.10E-1& 	2.04E+2\\
CN        	& 6.1E+1	&1.6E+6	&7.4E+2	&--	   &		--	   & --		&	--     \\
CO        	& 1.3E+2	&1.0E+7	&1.6E+3	&--	   &		--	   & --		&	--     \\
H$_2$    	& 5.6E+2	&1.1E+8	&6.7E+3	&--	   &		--	   & --		&	--     \\
HCO$^+$     & 5.9E+3	&2.6E+8	&7.1E+4	&2.84E+2& 	7.88E+7& 	4.58E+0 & 	1.35E+6\\

\hline
\end{tabular}
\tablefoot{Based on \cite{dissociation}. CO and H$_2$ are subject to significant self-shielding in spectral lines.}
\end{table*}

For our understanding of the origin and outcome of the CK\,Vul eruption, it is relevant to examine what kind of star is currently embedded in the remnant. It is not seen directly because is obscured by a thick column of dust associated with the central region. Along our line of sight, the extinction is very high ($A_V \gg 3$\,mag). If dust emission arises from a ring- or disk-like structure seen almost edge-on, whose presence is supported by some mm observations, the polar regions of the remnant may be effectively irradiated by the central star. Based on the ALMA submm observations, there is no significant dust emission in the lobes, i.e., internal to the dust cocoon mentioned earlier. For the cocoon to be seen at mm wavelengths, the central source has to be heating it through radiation. Thus, the extinction intrinsic to the lobes and for sight lines of particles in the lobes towards the central star must have $A_V\ll2.8$\,mag \citep{xshooter}. 

With these observations, we investigate what would be the response of the molecular gas if a white dwarf (WD) was the central star. A WD was postulated by several authors to be an outcome of the eruption of CK\,Vul \citep{shara85,hajduk2007,evans2016,Eyres}. To investigate the internal stellar radiation field, we first used the synthetic spectra of WDs with $T_{\rm eff}$=20\,kK from \citet{WDspectra}.\footnote{The spectra were acquired through \url{http://svo2.cab.inta-csic.es/theory/newov2/index.php}} Spectra of stars with such an effective temperature are similar to that of the ISRF. We consider the outer-lobe regions located $r$=2\arcsec\ away from the remnant center and assume a distance of 0.7\,kpc to CK\,Vul \citep{hajduk2013}. We find that the unattenuated radiation of a WD in the 912--2000\,\AA\ region, relevant for photodissociation, would be about five times stronger there than the standard ISRF. We then used the photodissociation cross sections of \cite{dissociation} to estimate the lifetimes of molecules at $r$=2\arcsec. Results are shown in Col.\,(5) of Table\,\ref{tab-dissociation}. The lifetimes are very short, typically of a few years, and would be even shorter for regions closer to the WD. We also considered shielding by dust within the remnant by reddening the WD spectra with the {\tt{reddening}} procedure in IRAF and a standard extinction curve used in Heays et al. Taking $A_V$=3\,mag as a conservative upper limit on the extinction intrinsic to the lobes, we obtain very long lifetimes, $\gg350$\,yr (see Col.\,(6) in Table\,\ref{tab-dissociation}). However, in this case our hypothetical WD of 20\,kK would have a luminosity of 0.01\,L$_{\sun}$, whereas the central source of CK\,Vul is certainly more luminous than $\sim$1\,L$_{\sun}$ \citep{nature}. To have that order of luminosity, the central WD would have to be hotter than 55\,kK. If indeed an eruption took place on the surface of the hypothetical WD, the WD should still be very hot because the cooling time is expected to be on the order of 1\,Myr \citep{cooling}. Harder radiation of such a hot and luminous source, say of $T_{\rm eff} = 60$\,kK, is $\gtrsim$\,90 times stronger than the ISRF in the outer lobes and would certainly destroy most molecules throughout the entire bipolar structure (see Col.\,(7)) unless the gas is shielded by large quantities of dust. Only with dust extinction with $A_V\approx3$\,mag, the life-times become comparable to the age of the remnant (Col.\,(8)). As discussed above, such a high intrinsic extinction is very unlikely. Also, the source luminosity may be several times higher than 1\,L$_{\sun}$, as 0.7\,kpc is rather a lower limit on the distance and we might not have accounted for all radiated power of the source.  We can thus exclude WDs with $T_{\rm eff} \gtrsim 60$\,kK as the stellar remnant of CK\,Vul. For analogous reasons, the radiation of the central source is unlikely responsible for the creation of molecular ions, HCO$^+$ and N$_2$H$^+$. We favor a scenario where the stellar remnant is cool ($\sim$3000\,K), as other Galactic red-nova remnants, and does not destroy or ionize molecules through its radiation. 

\cite{evans2016} presented infrared observations which have been interpreted as an argument supporting the presence of a UV-bright and hot ($>$14\,000\,K) inner source in CK\,Vul. A WD as hot as 100\,kK was considered. They presented a {\it Spitzer} spectrum which shows an emission line of [\ion{O}{iv}] at 25.89 $\mu$m and multiple `unidentified infrared features' (UIR), which are typically but somewhat controversially attributed to polycyclic aromatic hydrocarbons (PAHs). While the presence of [\ion{O}{iv}] indeed requires a hot ionizing source, we find that the feature can instead be explained by the line $a^6D$ 7/2$\to$9/2 of [\ion{Fe}{ii}] \citep{feII} which does not require such a hard irradiation spectrum. The UIR bands are often interpreted as a signature of a strong UV source in the system. Both observations and theoretical studies show, however, that this is not always the case \citep[e.g.,][]{LiDraine} and those bands can be effectively excited near sources as cool as 3000\,K through visual photons. 
Active shocks may provide UV radiation capable of destroying on-site molecules in CK\,Vul but not much is yet known about the shocks whose signatures are found in CK\,Vul. Shocks could also be responsible for the observed morphology in emission of molecular ions but chemical models adequate for the case of CK\,Vul do not yet exist.


\section{Summary}\label{sec-final}
We presented first results from interferometric line and imaging surveys obtained with ALMA and SMA towards CK\,Vul. We explored emission morphologies of multiple molecular species and dust. Radiative transfer modeling of the molecular emission, which takes into account non-LTE effects, yields temperatures of 17\,K in the central region and 14\,K in the lobes, total gas densities of $10^{4}$--$10^{6}$\,cm$^{-3}$, and constrains column densities of the observed molecules. We find chemical differences between the compact central region and the extended lobes. The central region appears to be more oxygen-rich than the lobes, but the relative abundance of the CNO elements could not be quantified even under the simplified assumption of equilibrium chemistry. There are also some variations of the ratio of neutral to ionized species within the lobes, but more advanced analysis methods are required to reduce the model uncertainties and better understand the ionization structure of the molecular nebula. We find that shielding by dust provides a good protection of the observed molecules against the interstellar radiation field and therefore the molecular gas might have formed 350\,yr ago (or more recently). However, the molecular emission would not be observed if the bipolar mm-wave nebula were illuminated by a hot central source of UV radiation, such as a white dwarf of 1\,L$_{\sun}$. This is in accord with the expectation that CK\,Vul's stellar remnant is a red novae and thus is a cool star.

In the forthcoming papers based on the ALMA and SMA surveys, we will study the 3D structure of the remnant, its continuum emission, and isotopic ratios.

\begin{acknowledgements}
We thank T. Millar, R. Garrod, and A. Burkhardt for discussing the presence of molecules in CK\,Vul.
T. K. acknowledges funding from grant no 2018/30/E/ST9/00398 from the Polish National Science Center. R. T. acknowledges a support from grant 2017/27/B/ST9/01128 financed by the Polish National Science Center.
M.~R.~S. acknowledges a support from grant 2016/21/B/ST09/01626 financed by the Polish National Science Center. K.~T.~W. was supported for this research through a stipend from the International Max Planck Research School (IMPRS) for Astronomy and Astrophysics at the Universities of Bonn and Cologne and also by the Bonn-Cologne Graduate School of Physics and Astronomy (BCGS).
This paper makes use of the following ALMA data: ADS/JAO.ALMA\#2017.1.00999.S, ADS/JAO.ALMA\#2017.A.00030.S, ADS/JAO.ALMA\#2016.1.00448.S, and ADS/JAO.ALMA\#2015.A.00013.S. ALMA is a partnership of ESO (representing its member states), NSF (USA) and NINS (Japan), together with NRC (Canada), MOST and ASIAA (Taiwan), and KASI (Republic of Korea), in cooperation with the Republic of Chile. The Joint ALMA Observatory is operated by ESO, AUI/NRAO and NAOJ. The National Radio Astronomy Observatory is a facility of the National Science Foundation operated under cooperative agreement by Associated Universities, Inc.
The Submillimeter Array is a joint project between the Smithsonian Astrophysical Observatory and the Academia Sinica Institute of Astronomy and Astrophysics and is funded by the Smithsonian Institution and the Academia Sinica.
Based on observations collected at the European Organisation for Astronomical Research in the Southern Hemisphere under ESO programme 099.D-0010(A).
Based on analysis carried out with the CASSIS software. 
\end{acknowledgements}

\end{document}